\documentclass[authoryear,preprint,12pt]{elsarticle}

\usepackage{graphicx}

\usepackage{amssymb,amsthm}
\usepackage{graphicx} 
\usepackage{epsfig} 
\usepackage{amsmath,amsfonts} 
\usepackage{bm}
\usepackage[margin=0.85in]{geometry}
\usepackage{lineno}

\usepackage{psfrag}
\usepackage{hyperref}
\hypersetup{
     colorlinks   = true,
     citecolor    = gray
}

\renewcommand*{\contentsline}[3]{\csname l@#1\endcsname{#2}{}}
\usepackage{tocloft}

\usepackage{color}

\usepackage{enumitem}

\usepackage{epsfig} 
\usepackage{amsmath} 
\usepackage{amssymb}  
\usepackage{microtype}
\usepackage{epstopdf}
\usepackage{booktabs}
\usepackage{breqn}
\usepackage{bm}

\usepackage{multirow}
\usepackage[ruled,vlined,linesnumbered]{algorithm2e}
\usepackage{tikz}
\usetikzlibrary{shapes,arrows,chains}
\tikzset{middlearrow/.style={
        decoration={markings,
            mark= at position 0.6 with {\arrow{#1}} ,
        },
        postaction={decorate}
    }
}

\usepackage{multirow}
\usepackage{url}
\usepackage{makecell}
\usepackage{relsize}

\usepackage{mathtools}
\DeclareMathAlphabet{\pazocal}{OMS}{zplm}{m}{n}
\makeatletter
\newcommand{\vast}{\bBigg@{3}}
\makeatother
\makeatletter
\renewcommand*\env@matrix[1][\arraystretch]{%
  \edef\arraystretch{#1}%
  \hskip -\arraycolsep
  \let\@ifnextchar\new@ifnextchar
  \array{*\c@MaxMatrixCols c}}
\makeatother

\usepackage{subfig}

\usepackage{natbib}
\journal{Transportation Research Part C: Emerging Technologies}

\begin{document}

\begin{frontmatter}

\title{Two-layer adaptive signal control framework for large-scale dynamically-congested networks: Combining efficient Max-Pressure with Perimeter Control}

\author[EPFL]{Dimitrios Tsitsokas}
\ead{dimitrios.tsitsokas@epfl.ch}

\author[ETHZ]{Anastasios Kouvelas}
\ead{kouvelas@ethz.ch}

\author[EPFL]{Nikolas Geroliminis\corref{cor1}}
\ead{nikolas.geroliminis@epfl.ch}%

\cortext[cor1]{Corresponding author. Tel.: +41-21-69-32481; Fax: +41-21-69-35060.}

\address[EPFL]{Urban Transport Systems Laboratory, School of Architecture, Civil and Environmental\\Engineering, EPFL, CH-1015 Lausanne, Switzerland}
\address[ETHZ]{Institute for Transport Planning and Systems, Department of Civil, Environmental and\\Geomatic Engineering, ETH Zurich, CH-8093 Zurich, Switzerland}

\begin{abstract}

Traffic-responsive signal control has received considerable research attention, as a cost-effective and easy-to-implement network management strategy, bearing high potential to improve performance in heavily congested networks with time-dependent traffic characteristics. Max Pressure (MP) distributed controller gained significant popularity due to its theoretically proven ability of queue stabilization and throughput maximization under specific assumptions. However, its effectiveness is questionable under over-saturated conditions and queue spill-backs, while network-wide implementation is often practically limited due to high instrumentation cost that increases with the number of controlled intersections. Perimeter control (PC) based on the concept of Macroscopic Fundamental Diagram (MFD) is a state-of-the-art aggregated control strategy that regulates exchange flows between homogeneously congested regions, with the objective of maintaining maximum regional travel production and prevent over-saturation. However, homogeneity assumption is hardly realistic under congested conditions, thus compromising PC efficiency. In this paper, the effectiveness of network-wide parallel application of PC and MP strategies embedded in a two-layer control framework is assessed in a mesoscopic simulation environment. With the aim of reducing implementation cost of network-wide MP without significantly sacrificing performance gains, we evaluate partial MP deployment to subsets of nodes, indicated as critical by a node classification algorithm that we propose, based on node traffic characteristics. A modified version of Store-and-forward dynamic traffic paradigm incorporating finite queue and spill-back consideration is used to test different configurations of the two-layer framework, as well as of each layer individually, for a real large-scale network in moderate and highly congested scenarios. Results show that: (i) combined control of MP and PC outperforms separate MP and PC applications in both demand scenarios tested; (ii) MP control in reduced critical node sets selected by the proposed strategy leads to similar or even better performance compared to full-network implementation, thus allowing for significant cost reduction; iii) the proposed control schemes improve system performance even under demand fluctuations of up to 20\% of mean. 

\end{abstract}


\end{frontmatter}



\section{Introduction}
\label{S:1-Intro}

Traffic-responsive signal control systems have been the focus of numerous research studies, as a low-cost, easy-to-implement tool for dynamic flow management under rapid and/or random changes in traffic conditions. In contrast to static (fixed-time) signal control plans, which function according to a pre-determined, time-periodic schedule designed to serve a specific demand scenario, traffic-responsive signals dynamically update green/red times of competing streams in order to actively adjust to the prevailing traffic states. Consequently, they carry higher potential of increasing system operating capacity than fixed-time control, by making more efficient road space use, and by preventing excessive demand from causing instability and gridlocks. However, designing such systems in an optimal, cost-effective and practically applicable way is challenging, due to the high complexity and uncertainty involved in describing, modeling and predicting traffic network dynamics and driver decisions.     

Various traffic-responsive signal control systems with different architectures and control algorithms have been proposed in the literature and implemented in the field. Among distributed controllers, Max Pressure (MP) \citep{varaiya2013max,varaiya2013max2} which applies to independent intersections, has gained significant popularity due to its theoretically proven abilities of queue stabilization and throughput maximization, under specific assumptions. Multiple variations of the original algorithm have been proposed and tested, showing promising results. However, its effectiveness in network-scale control in over-saturated conditions is still uncertain, while traffic monitoring equipment requirements cause implementation cost to excessively increase in case of network-wide control. On the other hand, among centralized approaches, perimeter control (PC) based on the concept of the Macroscopic Fundamental Diagram (MFD), has been effective in improving traffic performance of single or multiple neighborhood-sized, homogeneously congested regions. 
PC consists of regulating inter-regional incoming and outgoing flows by adjusting green light duration of the respective approaches at the intersections located on the network perimeter or at the boundaries between regions. The objective is to maintain maximum travel production within the protected regions, according to the specific MFD law that associates regional vehicle accumulation and travel production. However, traffic homogeneity often decreases while congestion starts building, especially at the boundaries between regions where gating can cause queue formation, and as a result, PC effectiveness may drop. Besides, PC control logic is mostly effective in cases of high travel demand, where vehicle accumulation typically goes beyond critical, while it is less effective in cases of lower demand, even though local congestion pockets can still form. 

While both of these control strategies have been extensively studied in an independent implementation framework, little is known about the effectiveness of their parallel/combined use in large-scale networks. Also, MP effect at the network level has not been investigated in detail for high demand conditions. Motivation to combine both strategies in a single control framework derives from their potentially complementing natures, which can act beneficially for the network. On one hand, MP acts on a local level and can prevent or delay the formation of local gridlocks deriving from highly imbalanced queues, which constitute heterogeneity sources that cannot be addressed by region-scale PC. On the other hand, PC can benefit from increased homogeneity owed to MP effects, while it can act on the region-scale against over-saturation, by maintaining vehicle accumulation - and consequently travel production - close to critical. Hence, PC impedes region-extended gridlocks that occur in over-saturated states, which might render MP ineffective. However, MP and PC might also have conflicting goals, as for example MP might provide longer green time for high demand arterial roads towards attractive destinations, while PC might try to decrease inflow to these attractive regions, in order to keep accumulation close to critical which maximizes trip endings. Moreover, one other question regarding MP network-wide implementation is about optimal number and topology of controlled nodes, with regard to the ratio of traffic benefit over implementation cost. Despite the relatively high MP instrumentation requirement per controlled intersection, network-wide MP control typically assumes control of all network intersections. Reducing the number of controlled nodes can lead to significant cost reduction, and, to the best of our knowledge, no studies of partial MP application to network node subsets exist yet in the literature.

In this work, we investigate the effectiveness of a two-layer adaptive signal control framework for network-wide application, combining centralized MFD-based perimeter control with  Max Pressure distributed control. PC is implemented via a Proportional-Integral (PI) regulator, which adjusts the green time of inter-regional approaches on the boundaries of homogeneously congested regions. MP regulator is installed to isolated intersections across the entire network. A method of identifying intersections that are potentially critical for MP control, based on specific node traffic characteristics, is proposed. Several scenarios of different MP node layouts are created based on this method, both for single MP control and as part of the two-layer controller, in combination with PC. The effects that MP might have in the PC control design parameters (such as critical accumulation of maximum network throughput) are also investigated. A modified version of the Store-and-Forward paradigm (see \cite{aboudolas2009store}), with elements of the S-Model (see \cite{lin2011fast}) is used to emulate traffic for different control scenarios. Its modified mathematical structure enables proper modeling of spill-backs and finite capacity queues, while it allows dynamic updates of traffic signal settings 
Simulation tests are performed for a real large-scale network of more than 1500 links and 900 nodes for two demand scenarios, leading to moderate and high congestion in the fixed-time control case, respectively. The two control strategies are applied both independently and in parallel, as embedded in a two-layer framework. Results are compared to the benchmark case of fixed-time control. Regarding MP configuration, full-network implementation to all eligible nodes is compared to partial implementation to subsets of nodes selected both randomly and through the proposed selection algorithm, for different network penetration rates. A detailed analysis of the network behavior for all above scenarios is performed and performance of different control configurations and demand cases is discussed. 

The contributions of this work are the following: we assess the combined application of Max Pressure with perimeter control through mesoscopic simulation for a large-scale network; we test partial MP implementations to subsets of network noes; we develop and assess a node classification algorithm based on actual traffic characteristics to assist identification of more critical nodes for MP control; we show through sensitivity analysis that MP node schemes generated by the proposed method are not sensitive to small demand fluctuations; finally we gain valuable insights regarding the design and behavior of combined application of PC and MP for different-scale demand scenarios. 

The rest of the paper is structured as follows. At first, a review of the most relevant PC and MP studies is done in order to justify the motivation and expected contributions of this work. The theoretical framework of MP and PC control is introduced, together with the required notation explanation. The relations transforming the controller outputs to adjusted traffic signal plans are described for both control strategies and characteristics of the traffic simulation model are briefly discussed. Furthermore, the experiment setup is described and the methodological framework of node classification and selection is discussed. Simulation results and analysis follow, together with an analysis of the system's behaviour for the different scenarios, compared to the fixed-time control case. Finally, the paper is concluded by summarizing the main findings and discussing possible future research.     

\section{Background}
\label{S:1b-Literature}

Several types of traffic-responsive signal control systems have been proposed and applied in the field, following different optimization methods, as well as modeling and control approaches. Some of the most commonly used are SCOOT \cite{hunt1981scoot}, OPAC \cite{gartner1983opac}, PRODYN \cite{henry1984prodyn}, SCATS \cite{lowrie1990scats}, UTOPIA \cite{mauro1990utopia}, and the more recent ones RHODES \cite{mirchandani2001real} and TUC \cite{diakaki2002multivariable}. These systems employ a centralized control logic, in the sense that traffic information from the entire controlled region is required to be transferred to a central processing unit, where the respective control algorithm will process it and determine the control actions that need to be taken, which must be communicated back to every single intersection. This architecture, although it constitutes the state-of-the-art in the field of adaptive signal control due to higher number of degrees of freedom, is characterized by low applicability. This is due to its requirements in communication and computing infrastructure, which imposes a high installation, maintenance and operational cost. Also they provide relatively low scalability, in the sense that gradual installation or expansion of existing system to cover a larger region is challenging. Another implementation difficulty relates to the exponential complexity optimization algorithms that many of these systems employ, which prohibits network-wide, real-time central application due to computational cost. On the contrary, decentralized approaches, based on local control of isolated or coupled intersections, have significantly lower infrastructure requirements and algorithmic complexity, increased scalability, and yet high potential to improve network performance in dynamically-changing traffic conditions, even by starting from local scale. An overview of the different aspects of centralized and decentralized control strategies is found in \cite{chow2020centralised}, while simulation experiments on such strategies are performed by \cite{manolis2018centralised}. A detailed review of existing research in decentralized adaptive signal control can be found in \cite{noaeen2021real}.

Max Pressure (MP) is probably the most popular decentralized, distributed feedback-based controller (also known as Back Pressure) proposed for isolated traffic intersections. Its function is based on a simple algorithm that adapts the right-of-way assignment between competing approaches in real time, according to feedback information of queues forming around the intersection (upstream and downstream), in a periodic cyclic process. In the core of MP lies a pressure component, which quantifies the actual queue difference between upstream and downstream links of every phase, and determines phase activation or green time assignment during the following time slot. Initially proposed for packet scheduling in wireless communication networks by \cite{tassiulas1990stability}, MP was formulated as a signalized intersection controller through the works of \cite{varaiya2013max}, \cite{varaiya2013max2} \cite{wongpiromsarn2012distributed}, \cite{zhang2012traffic} and was theoretically proven by \cite{varaiya2013max} to stabilize queues and maximize network throughput for a controllable demand, based on specific assumptions, such as point-queues with unlimited capacity and separate queues for all approaches. The theoretical stability proof, despite the constraining assumptions on which is was based, together with the element of independence from any demand knowledge, and the decentralized and scalable layout, render MP a promising control strategy, especially for signalized networks facing unstable, excessive or dynamically variable congestion. These traits have motivated a vast amount of research works focusing on MP controller, which generated numerous enhanced and/or improved versions, some of which are listed here. \cite{gregoire2014capacity} and  \cite{kouvelas2014maximum} introduced normalized queues in the pressure calculation, thus implicitly taking into account link size and spill-back probability, while taking into account queue capacity, which was considered infinite in the initial MP of \cite{varaiya2013max}.  \cite{gregoire2014back} presented an MP version with unknown turn ratios but existing loop detectors for all directions at the exit line. \cite{xiao2015throughput} and \cite{xiao2015further} proposed extended MP versions able to address bounded queue length estimation errors and incorporate on-line turn ratio estimation, as well as dynamically update control settings over space according to demand. \cite{le2015decentralized} applied a strict cyclic phase policy, in contrast to phase activation based on pressure which can induce long waiting time to some drivers, and provided stability proof, which applied also for non-biased turn ratios. \cite{zaidi2015traffic} integrated rerouting of vehicles in MP algorithms. \cite{li2019position} proposed a position weighted back pressure control, on the basis of macroscopic traffic flow theory, and integrated spatial distribution of vehicle queues in pressure calculation. \cite{wu2017delay} proposed a delay-based version of MP controller in order to increase equity of waiting time among drivers around the intersection. In \cite{mercader2020max}, pressure is calculated by using travel time estimation instead of queue length, in an attempt of relaxing the need for expensive queue measuring equipment, and findings are supported by simulation and real field experiments. \cite{levin2020max} propose an alternative signal cycle structure with maximum cycle length. However, even though many of the above works provide stability and throughput maximization proofs, they usually refer to moderate and feasible demand sets, while MP performance in highly congested networks is questionable. Unstable behaviour in such conditions can be attributed to the latency of local controllers, in general, in reacting to the rapid forming of congestion, given the lack of knowledge about traffic conditions upstream and out of the proximity of the controlled intersection. Also, in the case of MP, there is little area for improvement if most (or all) controlled queues are saturated, in which case control tends to approximate the fixed-time plan.

Perimeter control (PC), on the other hand, is an aggregated, centralized approach that has been gaining momentum over the last decade, mainly due to significantly reducing complexity in network-wide congestion control. The strategy, also known as gating, is based on the principle of regulating vehicle flows withing a defined, high-demand region (e.g. city center), in order to prevent high congestion, which can lead to gridlocks, delays and service rate decline. Exchange flow regulation happens via dynamical adjustment of the corresponding green times at the intersections located on the regional boundaries (perimeters). \cite{daganzo2007urban} described a theoretical model of adaptive control at an aggregate level, applicable to cities partitioned in neighborhood-size reservoirs. The work discussed the benefits of an inflow-withholding policy to prevent gridlock forming, showing how travel performance can increase for the entire peak period, even for vehicles that are forced to wait before entering the protected region. This policy was based on the hypothesis of a demand-insensitive, uni-modal relationship between vehicle accumulation (or density) and travel production (or space-mean circulating flow), that became later known as the Macroscopic (or Network) Fundamental Diagram (MFD or NFD). MFD concept, firstly introduced by \cite{godfrey1969mechanism}, 
was theoretically founded and described by \cite{daganzo2007urban}, and was empirically validated with real data for the city of Yokohama, Japan, by \cite{geroliminis2008existence}. Similar studies focusing on multi-modality and critical accumulations based on real traffic data are performed by \cite{paipuri2021empirical} and \cite{loder2019understanding}.  Well-shaped low-scatter MFD is observed for homogeneously congested, compact regions (with low link-flow variance) with relatively low scatter, which constitutes an elegant, dynamic modeling and control tool, that can serve as a basis for simplified, aggregated, network-scale, adaptive flow control schemes. Numerous researchers focused on the concept of MFD producing significant contributions regarding its existence preconditions, characteristics, shape, hysteretic behavior (\cite{buisson2009exploring}, \cite{ji2010macroscopic}, \cite{mazloumian2010spatial}, \cite{geroliminis2011properties}, \cite{gayah2011clockwise}, \cite{mahmassani2013urban}), and many more incorporated MFD theory in developing aggregated models for dynamic traffic-responsive urban network control for congestion mitigation (some recent works are \cite{yildirimoglu2018hierarchical}, \cite{mariotte2017macroscopic}, \cite{mariotte2019flow}, \cite{laval2017minimal}, \cite{batista2021identification}). Bi-modal MFDs are utilized in \cite{geroliminis2014three}, \cite{loder2017empirics}, \cite{haitao2019providing}, \cite{paipuri2020bi}, \cite{paipuri2021empirical}. MFD application to ride-sourcing systems in multi-modal networks is proposed by \cite{wei2020modeling} and \cite{beojone2021inefficiency}. Modeling and dynamic control of taxi operations based on MFD is described in \cite{ramezani2018dynamic}.Network partitioning research for the purposes of MFD-based control was also intensified (\cite{ji2012spatial}, \cite{saeedmanesh2016clustering}, \cite{ambuhl2019approximative}). For an overview of MFD research with respect to traffic modeling, the reader could refer to \cite{johari2021macroscopic}.

Perimeter control specifically was intensively studied on the basis of MFD modeling, and a large number of MFD-based PC schemes have been proposed, analyzed and evaluated in the recent years, utilizing different modeling and control methods and focusing on different control aspects.  Proportional-Integral (PI) feedback regulator for single- and multi-regional networks is implemented in \cite{keyvan2012exploiting}, \cite{keyvan2015controller}, \cite{aboudolas2013perimeter}, \cite{ingole2020perimeter}, optimal MPC is implemented in \cite{geroliminis2012optimal}, \cite{haddad2017optimal2}, \cite{haddad2017optimal} with boundary queue consideration, while route guidance is incorporated in PC schemes in \cite{yildirimoglu2015equilibrium}, \cite{sirmatel2017economic}. Stability analysis is done in \cite{haddad2012stability}, \cite{sirmatel2021stabilization}, integrated PC with freeway ramp metering is proposed in \cite{haddad2013cooperative}, robust control is implemented in \cite{ampountolas2017macroscopic}, \cite{mohajerpoor2020h}, adaptive control is implemented in \cite{kouvelas2017enhancing}, \cite{haddad2020adaptive}, \cite{haddad2020resilient}, demand boundary conditions are considered in \cite{zhong2018boundary}, bi-modal MFD is proposed in \cite{geroliminis2014three}, cordon queues impact is assessed in \cite{ni2020city}, remaining travel distance dynamics are integrated in MFD modeling and control in \cite{sirmatel2021modeling}, while data-based model-free PC implementations are proposed in \cite{ren2020data}, \cite{chen2022data}, \cite{zhou2016two}. However, despite the impressive amount of literature and the promising results in terms of network-wide traffic performance, there are still certain concerns regarding MFD-based PC practical application. One is the requirement for homogeneous congestion distribution, which is not common in real congested networks and can even be endogenously compromised, by the PC gating strategy, that tends to create local queues on the boundaries between clusters, which are themselves source of heterogeneity and compromise PC effectiveness. Similarly, PC can have little to no effect in cases of low to medium demand patterns that yet lead to heterogeneous traffic distribution, where gridlocks appear in specific parts/paths of a seemingly uncongested network, creating spill-backs and local congestion pockets, and causing delays.

Based on the above, developing multi-layer control polices comprising of both local and aggregated strategies, seem an intuitive way of achieving network control with multiple objectives. Various multi-layer hierarchical control structures involving perimeter control in the upper layer have been proposed, often with a lower layer focusing on reducing heterogeneity in local scale. \cite{ramezani2015dynamics} propose a two-layer feedback controller, where MPC is utilized to solve the optimal PC problem in the higher layer, while a feedback homogeneity controller is embedded in the lower lever, which acts as PC in a subregional level, aiming at decreasing heterogeneity within regions. Similarly,  \cite{zhou2016two} propose a two-layer framework where demand-balance problem between homogeneous subnetworks is treated via MFD-based PC in the higher layer, and a more detailed traffic model is embedded in the lower layer for optimizing all signals within each subnetwork, with both layers based on MPC optimal control formulations. Based on the same concept, \cite{fu2017hierarchical} propose a three-layer hierarchical control strategy, also including a stability analysis top layer. In all these works, hierarchical communication schemes between layers are required for exchange of information between controllers, which results in increased infrastructure requirements. Furthermore, MPC optimal control formulations increase complexity and computational cost, and performance is affected by the accuracy and fine-tuning of the utilized models.   

Adopting a concept similar to the one of hierarchical schemes,  we design and evaluate a two-layer network control framework consisting of: (i) a PC strategy based on a Proportional-Integral (PI) feedback controller, which manages exchange flows between homogeneously congested regions by adapting the traffic signal plans of intersections on their boundaries; and (ii) independent MP controllers acting on signalized intersections in the interior of the regions, balancing queues in their proximity, at a local scale. Very few studies have examined the behaviour of networks with parallel application of PC and other types of local controllers, and to our knowledge, none includes MP. \cite{yang2017multi} proposed an MPC-based controller incorporating two objectives of network-wide and intersection-scale delay minimization, by formulating and linearizing a complex multi-objective optimal control problem, applied in a connected vehicle environment. However, apart from the increased complexity, model-dependent accuracy and requirement of detailed demand information of MPC, local delay minimization was focused only on intersections on the perimeter of the protected region, while the coupled control scheme is centralized, meaning that communication infrastructure to central controller is required, in contrast to MP which is decentralized. \cite{keyvan2019traffic} examine the effects of combining PC with two different local adaptive controllers, a volume-based strategy and a simplified SCATS strategy. Performance results showed significant improvement in the cases of combined PC and local traffic-responsive control strategies. However, due to the size of the test network, only a small number of intersections were available for the local controllers and no investigation of the spatial control layout was held. Also MP was not tested as local controller. 

In this work, we propose a node selection algorithm to be used for identifying subsets of network signalized intersections that are more critical for MP control, based on variance and mean of normalized queues and spill-back occurrence of adjacent links. To the best of our knowledge, no existing research investigates partial MP application to node subsets. We evaluate and compare several scenarios of independent and combined application of PC and MP, with different MP node layouts and network penetration rates, for a real large-scale network, where traffic is modeled with a modified Store-and-Forward dynamic mesoscopic model. All control scenarios are tested for two trip demand settings, leading to moderate and high network congestion under fixed-time control settings. 

\section{Methodology}
\label{S:2-Methodology}

\begin{figure}[tb]%

\includegraphics[width=1.0\textwidth]{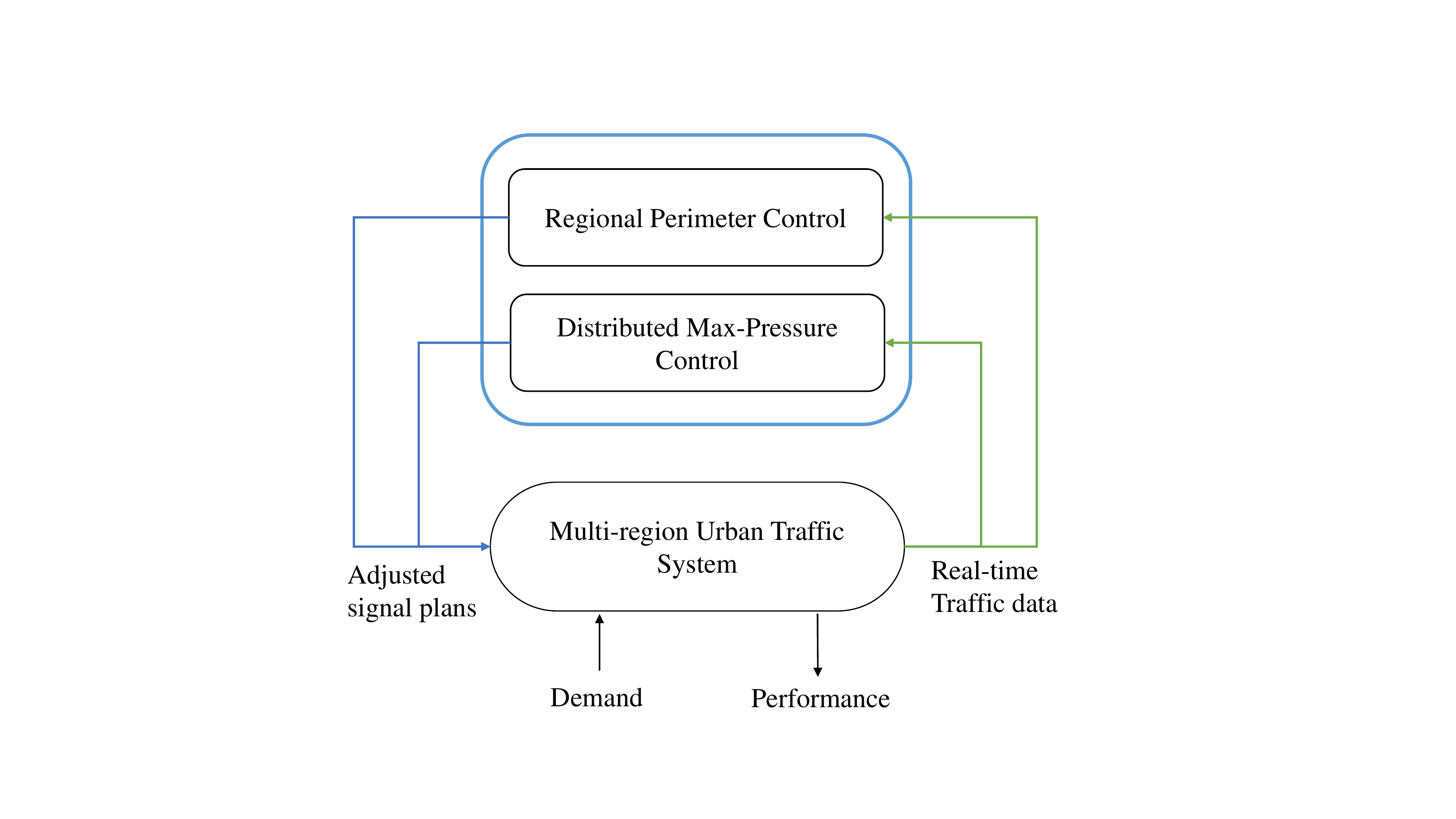}%
\caption{Schematic description of the two-layer controller.}\label{fig:schema2layer}%
\end{figure}

A schematic representation of the proposed two-layer controller is shown in figure \ref{fig:schema2layer}. In the upper layer, perimeter control is applied in an aggregated scale between a set of homogeneously congested regions. At the end of every control cycle, the controller, based on inputs of aggregated regional vehicle accumulation, specifies the target inter-regional exchange flows for the next cycle, which are translated into the respective inter-regional green times between every pair of adjacent regions. The controller-specified interregional green times are then translated to exact green times per approach, for all PC controlled intersections, located on the boundaries between regions, by taking into account the actual boundary queues. In the lower layer, distributed control based on Max Pressure regulator is applied to a set of eligible intersections, in the interior of the regions. This set can contain all or a fraction of signalized intersections of the region, with the exception of those used for PC (if PC is applied in parallel). MP controllers do not communicate with each other or with any central control unit, but operate independently based on queue measurements directly upstream and downstream the controlled intersections, by adjusting green times of the approaches accordingly, at the end of every control cycle. The control layers do not exchange information, however their combined effect is indirectly considered by both controllers through the real-time traffic measurements that they receive as inputs. The mathematical formulations of both controllers are described in the following subsections, followed by a brief description of the utilized traffic model and traffic simulation process.  

\subsection{Max-Pressure Control}

\subsubsection{Review of Max-Pressure feedback controller}

Max-Pressure feedback-based control algorithm was initially formulated for traffic signal control by independently by \cite{wongpiromsarn2012distributed} and \cite{varaiya2013max,varaiya2013max2}, who theoretically proved its stability and throughput maximization properties, though under restrictive assumptions. It is a distributed and scalable control strategy, in the sense that its operation requires no communication with the rest of the network infrastructure and, therefore, it can be introduced to any signalized intersection at any time, without necessitating any readjustment of existing controllers elsewhere. In other words, gradual installation to network is theoretically possible. Moreover, it does not require any knowledge of the actual or expected traffic demand, in contrast to Model-Predictive Control (MPC) approaches, which simplifies practical implementation. The MP version applied in this work only requires real-time queue measurements of the links around controlled intersection and turning ratios of all alternative approaches that traverse the intersection. Both can be measured or estimated with proper instrumentation. Several modified versions of the original control algorithm have been introduced and tested by simulation experiments. The present version is similar to the one described in \cite{kouvelas2014maximum}. The detailed description of MP algorithm together with the necessary notation is given below.

The traffic network is represented as a directed graph $(N,Z)$ consisting of a set links $z \in Z$ and a set of nodes $n \in N$. At any signalized intersection $n$, $I_n$ and $O_n$ denote the set of incoming and outgoing links, respectively. The cycle time $C_n$ and offset - which enables coordination with the neighboring intersections - are pre-defined (or calculated online by a different algorithm) and not modified by MP. Intersection $n$ is controlled on the basis of a pre-timed signal plan (including the fixed total lost time $L_n$), which defines the sequence, configuration and initial timing of a fixed number of phases that belong to set $F_n$. During activation of each phase $j \in F_n$, a set of non-conflicting approaches $v_j$ (i.e.\ connections between pairs of incoming-outgoing links of node $n$) get right-of-way (green light) simultaneously. The saturation flow of every link $z$, denoted as $S_z$, refers to the maximum possible flow that can be transferred to downstream links, depending on link and intersection geometry. The turning ratio of an approach between links $i-w$, where $i \in I_n, w \in O_n$ is denoted as $\beta_{i,w}$ and refers to the fraction of the outflow of upstream link $i$ that will move to downstream link $w$. The present version of MP assumes that turning ratios are known to the controller. However, it has been shown that control effectiveness is not deteriorated if turning ratios are estimated (see \cite{le2015decentralized}). By definition, the following relation stands for every node $n$, 
\begin{equation}
    \sum_{j \in F_n}{g_{n,j}(K_c) + L_n = (\mbox{or} \leq)\hspace{0.2cm} C_n}
    \label{eq:cycle}
\end{equation}
where $k_c = 1, 2, \ldots$ is the discrete-time control interval index, and $g_{n,j}(k_c)$ denotes the green time duration of phase $j$ of node $n$ at control interval $k_c$. The inequality may apply in cases where long all-red phases are imposed for any reason (e.g.\ gating). 

The version of MP controller employed in this work assumes that phase sequencing is given as input (e.g.\ from pre-timed scheduling) and does not change during the control. Consequently, during every cycle, all phases will be activated for a minimum time in the same ordered sequence. As a result, the following constraint also applies for every node $n$, 
\begin{equation}
    g_{n,j}(k_c) \geq g_{n,j,\min},  j \in F_n
\end{equation}
where, $g_{n,j,\min}$ is the minimum green time required for phase $j$ of node $n$, which often matches the required amount of time for the respective pedestrian movements. The control variables of this problem, denoted as $g_{n,j}(k_c)$, represent the duration of the effective green of every stage $j \in F_n$ of all controlled intersections $n \in N$. Assuming that real-time measurements or estimates of the queue lengths (states) and turning ratios of all controlled intersections are available, the pressure $p_z(k_c)$ of every incoming link $z \in I_n$ of node $n$, at the end of control cycle $k_c$, is computed as
\begin{equation}
    p_z(k_c) = \left[\frac{x_z(k_c)}{c_z} - \sum_{w \in O_n}{\frac{\beta_{z,w} x_w(k_c)}{c_w}}\right] S_z, \hspace{0.2cm} z \in I_n
    \label{eq:pressure}
\end{equation}
In equation (\ref{eq:pressure}), $x_z(k_c)$ denotes the average number of vehicles that are present (moving or queuing) in link $z$ during control cycle $k_c$ and $c_z$ denotes the storage capacity (maximum number of vehicles) of link $z$. Queue normalization by the link storage capacity aims at considering the link size, so that pressure of a smaller link is higher than that of a larger one with the same number of vehicles in it. In other words, pressure takes into account the likelihood of link queues - upstream and downstream - to spill-back in the following cycle. Pressures of all incoming links are calculated at the end of every cycle based on the latest queue measurements, which constitute the state feedback variables and are collected through proper instrumentation. Then, pressures are used by the controller for the signal settings update. It should be noted that the above formulation assumes that flow transfer is always possible between $z$ and all $w \in O_n$. Otherwise, the second term of equation \ref{eq:pressure} refers only to downstream links for which flow transfer is allowed from $z$ ($\beta_{z,w} > 0$). In different MP versions, $x_z$ may refer to the instantaneous or the maximum observed queue length at the end of the control cycle. In this work, a preliminary analysis showed that mean queue length values generate better results and this way was adopted. By looking at the two terms, one can notice that, essentially, pressure depicts occupancy difference between upstream and downstream links. Therefore, higher pressure indicates higher potential in traffic production, i.e.\ significant volume waiting to be served and enough available space in downstream links to receive it. Low or close to zero pressure indicates lower need for right-of-way time, due either to small queue upstream, or to lack of space downstream (links close to capacity). We should note that negative pressures are meaningless, so constraint $p_z(k) \geq 0$ must always hold.  

Based on equation \ref{eq:pressure} pressure is calculated for all incoming links $z \in I_n$ of node $n$. Then, the pressure corresponding to every stage $j$ at control cycle $k_c$ is defined as the sum of the pressures of all incoming links that receive right-of-way in stage $j$, as follows.
\begin{equation}
    P_{n,j}(k_c) = \max\left\{0,\sum_{z\in v_j}{p_z(k_c)}\right\},
    \hspace{0.2cm} j \in F_n
\end{equation}
This metric is then used as weight for the distribution of the total available green time between the competing stages of the intersection. 

After pressure values $P_{n,j}$ are available for every phase $j \in F_n$ of intersection $n$, the total amount of effective green time $G_n$, calculated as
\begin{equation}
    G_n = C_n - L_n = \sum_{j\in F_n}{g_{n,j}^\star},
    \hspace{0.2cm} n \in N
    \label{eq:G_n}
\end{equation}
is distributed to the phases of node $n$ in proportion to pressure values. In equation (\ref{eq:G_n}), $g_{n,j}^\star$ denotes the green time assigned to phase $j$ by static fixed-time analysis, using any of the standard algorithms. It holds that $g_{n,j}^\star \geq g_{n,j, \min}, \forall j \in F_n$. 

There are several different approaches that have been proposed regarding green time calculation, some of which also include phase activation based on pressures. In this version, since phases are activated in a strictly defined and non-changing order with a guaranteed minimum green time, green duration, $\tilde{g}_{n,j}(k)$, is assigned to phases proportionately to the computed pressures, as follows
\begin{equation}
    \tilde{g}_{n,j}(k_c) = \frac{P_j(k_c)}{\sum_{i \in F_n}{P_i(k_c)}} G_n,
    \hspace{0.2cm} j \in F_n 
    \label{eq:green_tilde}
\end{equation}

Eq. (\ref{eq:green_tilde}) provides the raw green times calculated according to MP controller. However, these values cannot be applied directly to the intersection signal plan, because it must first be guaranteed that they comply with a set of necessary constraints. Therefore, an additional step is added in the signal update process, whose objective is to translate MP outputs of eq. (\ref{eq:green_tilde}) to practically applicable green times $G_{i,j}$. This is done by solving online, for every control cycle $k_c$, the following optimization problem (similar to \cite{diakaki2002multivariable} but in this case two additional constraints are added): 

\begin{equation}
\begin{aligned}
& \underset{G_{n,j}}{\text{minimize}}
& & \sum_{j \in F_n}{\big(\tilde{g}_{n,j}-G_{n,j}\big)^2} \\
& \text{subject to}
& & \sum_{j \in F_n}{G_{n,j}} + L_n = C_n \\
&&& G_{n,j} \geq g_{n,j,\min}, \; j \in F_n \\
&&&  \big|G_{n,j} - G_{n,j}^p\big| \leq g_{n,j}^R \\
&&& G_{n,j} \in \mathbb{Z}^+ \\ 
&&& \forall j \in F_n \\
\end{aligned}\label{eq:feasible_greens_MP}
\end{equation}


According to the above formulation, the applicable green times for every phase $G_{n,j}$, $j \in F_n$, should be as close to the non-feasible regulator-defined greens $\tilde{g}_{n,j}$ as possible, while satisfying a set of constraints. The first constraint states that eq. (\ref{eq:cycle}) must always hold, therefore the sum of the updated feasible green times plus the total lost time $L_n$ should be equal to cycle $C_n$. The second constraint ensures that all phases get a predefined minimum green duration $g_{n,j,\min}$. In order to avoid potential instability of the system due to large changes in the signal timing happening too fast, we impose a threshold to maximum absolute change of every phase duration between consecutive cycles. This is expressed in the third constraint, where $G_{n,j}^p$ denotes the applied green times of the previous cycle and $g_{n,j}^R$ is the maximum allowed change of the duration of phase $j$ between consecutive cycles. Finally, feasible green times must belong to the positive integers set. This type of integer quadratic-programming problem can easily be solved by any commercial solver fast enough to allow online solution for every control cycle. The solution of this optimization problem, i.e.\ variables $G_{n,j}, \forall j \in F_n$, are the new feasible phase duration for node $n$ which will be applied in the next control cycle. The above process is repeated at the end of the cycle for every controlled intersection, regardless of what is happening to the rest of the network. The controller only requires real-time queue information of the adjacent intersections and respective turning ratios and the algorithm is executed once per cycle. 

\subsubsection{Developing a critical node selection framework for Max-Pressure control}\label{sss:nodeSelection}

While there has been significant research interest in finding more  efficient or less infrastructure-dependent versions of MP controller as part of network-scale signal control systems, little attention has been directed towards defining the optimal number, relative location and traffic characteristics of the intersections that are included in the MP control scheme. In most case studies in literature, either all eligible intersections or only those across important arterial roads are controlled. However, given the high requirements in monitoring equipment that increase proportionately to the number of controlled intersections, it is interesting to investigate how the impact of the control scheme is affected in the following cases: if only a fraction of the eligible intersections are controlled; whether some nodes are more critical than others in the sense of MP control for the same fraction; and what are the characteristics that would allow us to identify them. In an effort to reply to these research questions, we develop a node selection methodology based on current network traffic characteristics, by using principles of traffic engineering combined with an optimization approach. We test several different schemes of MP control, where different fractions of eligible nodes are included in the MP control node set. The controlled nodes are chosen both by the proposed method and randomly, for comparison reasons, and are tested in independent MP schemes as well as combined with specific PC strategies. In this section, the proposed node selection process for MP nodes is described.  

By analyzing the mechanism of the MP controller, we can infer that the process of green time re-assignment among competing phases/approaches is intuitively more beneficial in cases where queues of the competing approaches differ significantly from each other during the day, and thus, there are important pressure differences that can be balanced by the controller. This makes more sense if we think of an intersection where all approaches constantly have similar queues. In such case, pressures of phases would be relatively constant in time and therefore the controller would assign almost unchanged green time to all phases, similar to what the fixed-time control plan would do. In such cases, MP benefits are negligible. Thus, variance of queues of all approaches of an intersection, both incoming and outgoing, is one variable of interest in assessing node criticality. Furthermore, the mean normalized queue of the same approaches during peak period, or in other words the amount of traffic that the node serves with respect to its capacity, are also good indicators, since the impact of a poor or efficient signal plan is intensified if controlled node serves vehicles close to capacity. A node that gets congested in the FTC scenario, in the sense that some or all of its approaches reach jam densities and spill-backs occur upstream, might see higher improvement with MP control, which would affect higher number of trips, compared to a moderately congested node, where spill-backs do not occur. The charging level of each node can be estimated by the average queue over capacity ratio of all node approaches. Finally, the duration of the spill-back occurrence also seems important. For instance, a node might 'see' very high queues for only a short time, which can lead to a medium time-mean average approach queue and the same can happen for a node with medium to high queues that persist for a longer time period. In the former, spill-backs can occur for a short time while the node is relatively empty the rest of the time, while in the latter, spill-backs may not occur but short delays are affecting more drivers for longer time, therefore MP controller may create higher benefit for this node. To differentiate between the two node cases, we can count, for every node, the time during which at least one of the node approaches reaches jam density and spill-backs occur. Through this thinking process, we identify three quantities that should be taken into account in order to accurately assess the overall node significance for MP control. 

Therefore, we define a set of three node assessment criteria, which we linearly combine into a kind of node (dis)utility function, whose coefficients can be determined through a suitable calibration process. In this work, a simple grid-search was performed, which is described in the following subsection. Given traffic information of the current network situation (e.g. FTC), a peak-period $P$ is defined, based on the observed network state, as a set of time steps $T_P$. The selection process can be described, step by step, as follows:

\begin{itemize}
    
    \item For every node $n$ that is eligible to receive MP controller, the following three quantities are estimated: The first, denoted as $m_1^n$, represents the average node congestion level, as the mean over time of the mean occupancy (queue normalized over link capacity) of all incoming links $z \in I_n$ of node $n$, during peak period $P$. It is equal to   
    
\begin{equation}
        m_1^n = \frac{1}{\big\|T_P\big\|}\frac{1}{\big\|I_n\big\|}\sum_{i \in T_P}\sum_{z \in I_n}{\frac{x_z(i)}{c_z}}
    \end{equation}
    
    where $i$ is the simulation time-step index, $T_P$ is the set of time-step indices corresponding to the peak period $P$ and $\big\|T_P\big\|$ is the size of set $T_P$. The second, denoted as $m_2^n$, represents the mean over time of the variance of link occupancy of all incoming links  $z \in I_n$ of node $n$ during peak-period $P$, computed by
    
     \begin{equation}
        m_2^n = \frac{1}{\big\|T_P\big\|}\sum_{i \in T_P}{\mbox{var}\big(X_z^n(i)\big)}
    \end{equation}
    
     where $X_z^n(i) = \big\{x_z(i)/c_z |\forall z \in I_n \big\}$ is the set of normalized queues of all incoming links $z$ of node $n$ at time step $i$. The third quantity, denoted as $N_c^n$,  represents the fraction of the peak period $P$, during which node $n$ is considered `congested'. In this analysis, we assume that a node is `congested' during control cycle $k$ if the average queue of at least one incoming link $z \in I_n$ of node $n$ during $k$ is higher than a preset threshold percentage $p$ of its storage capacity, as shown by binary function $C_n(k)$ below.  
     
     \begin{equation}\label{eq:C_n}
        C_n(k) = \Bigg\{
  \begin{tabular}{ll}
 1, & if \hspace{0.2cm}  $\frac{1}{t_c^n}\sum_{i = (k-1)t_c^n + 1}^{k t_c^n}{x_z(i)} \geq p \hspace{0.1cm} c_z $, \hspace{0.2cm} for any $z \in I_n$   \\
 $ 0, $ & else
  \end{tabular}
    \end{equation}

         \begin{equation}\label{eq:N_c}
        N_c^n = \frac{t_c^n}{\big\|T_P\big\|}\sum_{\forall k \in P}{C_n(k)}, \hspace{0.2cm}  
    \end{equation}
    
    In equation \ref{eq:C_n}, $t_c^n$ denotes the control cycle size in number of simulation time-steps, i.e. $t_c^n = C_n / t$, where $C_n$ denotes the control cycle duration of node $n$ and $t$ the simulation time-step duration. In equation \ref{eq:N_c}, the ratio  $\big\|T_P\big\|/t_c^n$ is the number of control cycles that constitute peak period $P$. In other words, $N_c^n$ represents (in the scale of 0 to 1) the fraction of the peak period $P$ during which, at least one incoming link is congested and causes queue spill-back. In the current analysis, we set $p = 80\%$, since this is shown to significantly increase the probability for spill-back occurrence (see \cite{geroliminis2011identification}).

    \item Then, the level of importance of each node $n$ regarding MP control is estimated as a linear combination of the the above variables, denoted as $R^n$, as follows: 
    
    \begin{equation}\label{eq:Rn}
        R^n = \alpha m_1^n + \beta m_2^n  + \gamma N_c^n
    \end{equation}
    
    Quantity $R^n$ is then used as a base to rank nodes and drive the selection of the most critical ones. The coefficients in equation \ref{eq:Rn} act as weights for the importance of every criterion and their values can be calibrated based on a trial and evaluation grid test, as described below. 
    
    \item Finally, based on a target network penetration rate for MP control (i.e. percentage of eligible network nodes to receive MP controller), nodes are selected in sequence of increasing $R^n$, until the target number is reached.   
    
\end{itemize}

An important step of the above process is finding proper values for the parameters $\alpha, \beta, \gamma$ of classification function $R$ (equation \ref{eq:Rn}), which serve as weights of node variables $m_1$, $m_2$ and $N_c$ that indicate node importance. Since the relative importance of these variables is not straightforward, a trial-and-evaluation test is performed. More specifically, enumeration upon a grid of values of $\alpha, \beta, \gamma$ with subsequent MP simulation is performed. The combination of values leading to minimum total travel time for the same node penetration rate is then found and selected for all experiments. 


    
    
    
    
    


\subsection{Perimeter Control} 

\subsubsection{Proportional-Integral regulator for MFD-based gating}\label{ss:PI_regulator}

The concept of gating in perimeter control strategies for single- or multi-region systems consists of controlling vehicle inflows in the perimeter or the boundaries of the protected regions, in order to prevent vehicle accumulation to rise excessively in their interior and lead to congestion phenomena, such as lower speeds, delays and gridlocks. Flow control can be applied by means of real-time adaptive traffic signals on the perimeter or the boundaries between regions, where green time of the respective approaches is periodically adjusted, based on a control law that takes into account the actual traffic state of the region. State information can be provided in real-time by loop detectors installed properly in the interior of the region, or by other types of traffic measuring equipment. 

The concept of MFD enables the development of reliable feedback control strategies that assess the network state based only on measurements of vehicle accumulation in the system, which is associated to a specific travel production/service rate. Driven by the characteristics of the MFD curve, PC strategies can manipulate perimeter inflows during peak-hours so that vehicular accumulation in high-demand regions is maintained close to critical - for which travel production is maximum - and does not reach higher values belonging to the congested regime of the MFD. When accumulation tends to increase above this level, the allowed perimeter inflow is reduced by means of decreased green time for the approaches on the perimeter leading to the interior of the region. For large scale heterogeneously congested networks, proper clustering into homogeneous regions) that demonstrate a low-scatter MFD is required.

Several approaches for MFD-based perimeter control have been proposed and successfully tested, for single and multi-region systems, and different types of control laws are employed. In many studies, model-predictive control (MPC) schemes are employed for PC implementation, where dynamic aggregated traffic models are used to predict upcoming traffic states, based on which, an optimal control problem is solved for the prediction horizon and control variables are defined. However, in this work we consider no traffic states prediction, as we follow a simpler, less computationally expensive approach, where the system is controlled in real-time through a classical multivariable Proportional-Integral (PI) feedback regulator (see \cite{kouvelas2017enhancing}), as follows: 
\begin{equation}\label{eq:PI}
    \textbf{u}(k_c) = \textbf{u}(k_c-1) - \textbf{K}_P \left[\textbf{n}(k_c) -\textbf{ n}(k_c-1)\right] -\textbf{ K}_I \left[\textbf{n}(k_c) - \hat{\textbf{n}}\right] 
\end{equation}
In the above, $\textbf{u}(k_c)$ denotes the vector of control variables $u_{ij}$ for control interval $k_c$, which, in this work, represent the average green times corresponding to the controlled approaches between adjacent regions $i$ and $j$ (heading from $i$ to $j$), as well as to the external perimeter approaches of every region (if external gates exist), denoted as $u_{ii}$; $\textbf{n}$ is the state vector of aggregated regional accumulations $n_i$; $\hat{\textbf{n}}$ is the vector of regional accumulation set-points $\hat{n_i}$; and $\textbf{K}_P$, $\textbf{K}_I$ are the proportional and integral gain matrices, respectively. If equation \ref{eq:PI} is written in analytical form instead of matrix form, a system of equations will be produced. Every equation specifies the average green time, for the next control interval, for all nodes in specific direction between pairs of adjacent regions (e.g. $u_{ij}$ and $u_{ji}$ for adjacent regions $i,j \in \mathcal{N}$), while the last $\|\mathcal{N}\|$ equations refer to the average green time of all external approaches of each region. The control goal is to maintain accumulation close to critical in all controlled regions, in case of excessive demand, and impede reaching the congested regime of the MFD, where travel production drops significantly. Therefore, the set-points are decided based on the regional MFDs. The number of controlled regions and the respective MFD shapes depend on the network partitioning to a set of homogeneously congested regions $\mathcal{N}$, which can involve real or simulated traffic data. In order to be functional, the PI regulator requires as inputs the real-time regional accumulations, the set-points, as well as the proportional and integral gain matrices. Regional accumulations are supposed to be provided by loop detectors or other measuring equipment, properly distributed in the network. In this work we assume perfect knowledge of regional accumulations, that are averaged over the control interval. 

The PI controller is activated at the end of every control interval and only when real-time regional accumulations are within specific intervals, in the proximity of the specified set-point, i.e. activated when $n_i \geq n_{i,\text{start}}$ and deactivated when $n_i \leq n_{i,\text{stop}}$, for $i \in \mathcal{N}$ and typically $n_{i,\text{stop}} < n_{i,\text{start}}$. This is important as early activation of the PI regulator (i.e. for low accumulation) can lead to signal settings aiming at increasing congestion in the controlled areas, so that production gets closer to critical, which is the target. However, such a policy can accelerate congestion and compromise the system performance. From equation \ref{eq:PI}, the average green time $u_{ij}(k_c)$ of all controlled approaches on the border between regions $i$ and $j$ with direction from $i$ to $j$ is calculated. Based on this average value, the exact green time for every specific intersection is calculated according to the process described in the following subsection. After deactivation of the controller, the FTC signal plan for all PC intersections is gradually restored.

Benefits of PC are more obvious in cases of high travel demand with highly directional flows towards the protected regions, that are usually areas of increased activity levels (e.g. city center). In such cases, the congestion-prone regions are kept close to capacity while queues are forming in the adjacent regions, due to PC. Usually, the adjacent regions are considered less probable to get highly congested and the PC-related queues do not cause significant performance degradation. However, multi-reservoir systems with multiple sources of congestion are more challenging to control with a multi-region PC scheme, as it may be the case that several neighboring regions can get congested at the same time and flow regulation between them is not straightforward. In such cases, the system optimal performance might come from a PC scheme where one region is `sacrificed' and gets congested so that another more critical one is protected. In other words, finding the right values for the proportional and internal gain matrices is not trivial and depends on the actual traffic distribution patterns. In this work, since the focus is on the combination of PC and MP schemes, we calibrate $\textbf{K}_P$ and $\textbf{K}_I$ intuitively, based on a trial-and-error simulation process, in order to achieve some minimum performance improvement.  

\subsubsection{Green time calculation for PC intersections}

After average green time $u_{ij}$ for all PC controlled approaches between adjacent regions $i$ and $j$ (from $i$ to $j$) is defined from equation \ref{eq:PI}, it is used as base to define the exact new green duration for the respective phases containing the approaches leading from region $i$ to $j$. However, since not all intersections serve the same demand, green time assignment is more efficient if decision takes into account current queue lengths of the respective approaches. In other words, assignment should aim at providing longer green to approaches having larger queues, while ascertaining that the mean green of all approaches is as close as possible to the value defined by the PI controller, in an effort to balance queues in the boundaries between regions. Moreover, new greens are subject to a set of constraints, similar to the ones imposed in the case of MP signal update process, i.e. maximum allowed change between consecutive cycles, minimum and maximum green phase duration, constant cycle duration and integer green integrals. 

Hence, for every $u_{ij}$ that is specified by the PI controller, an optimization problem is solved for determining the exact green duration of the primary and secondary phases, denoted as $p$ and $s$, respectively, of all controlled intersections with direction from $i$ to $j$. Primary phase $p$ includes approaches with direction from $i$ to $j$ and secondary phase(s) $s$ include approaches of the same intersection but in the vertical (parallel to the boundary between regions) or the opposite direction (from $j$ to $i$), depending on the traffic characteristics of the nodes. The sum of available green of primary and secondary phase remains constant, in other words, the green time that is removed from the primary phase goes to the secondary phase(s), while duration of any other phases of the node remains unchanged. Assuming that the set of controlled nodes in the direction from $i$ to $j$ is denoted as $\mathcal{M}_{ij}$, $m$ is the node index, $G_{m,p}$ is the final green of the primary phase, $G_{m,s}$ is the final green of the secondary phase, $G_{m,t}$ is the sum of available green time for primary and secondary phases and $Q_{m,p}$ and $Q_{m,s}$ are the sum of the average observed queues of all incoming links belonging to the primary and secondary phases during the last control interval, respectively, the following optimization problem is formulated:     

\begin{equation}\label{eq:feasibleGreensPC}
\begin{aligned}
& \underset{G_{m,p}, G_{m,s}}{\text{minimize}}
& & \sum_{m \in \mathcal{M}_{ij}}{Q_{m,p}\big(G_{m,p}-u_{ij}\big)^2} +  \sum_{m \in \mathcal{M}_{ij}}{Q_{m,s}\big(G_{m,s}-(G_{m,t}-u_{ij})\big)^2}  \\
& \text{subject to}
& & {G_{m,p}} + G_{m,s} = G_{m,t} \\
&&& G_{m,i} \geq g_{m,i,\min}, \;  \text{for} \hspace{0.2cm} i \in \left\{p,s\right\} \\
&&&  \big|G_{m,i} - G_{m,i}^p\big| \leq g_{m,i}^R \\
&&& G_{m,i} \in \mathbb{Z}^+ \\ 
&&& \forall i \in \left\{p,s\right\}, \forall m \in \mathcal{M}_{ij}  \\
\end{aligned}
\end{equation}

In the above, we seek to minimize the weighted sum of the squared differences between the indicated by the PI controller green and the finally assigned green for all controlled intersections $m \in \mathcal{M}_{ij}$ of the approach $i$-$j$, where average queue lengths are used as weights. The first constraint is about maintaining cycle duration, i.e. ensuring that the sum of primary and secondary phases remains constant; the second dictates that minimum green $g_{m,i}^R$ is assigned to all primary and secondary phases; the third ensures that maximum absolute change between new green $G_{m,i}$ and green of the previous control interval $G_{m,i}^p$ is below the preset threshold of $g_{m,i}^R$; and the forth dictates that green time intervals are integer. The new control plans take effect, for every intersection, after the end of their ongoing cycle. 

Regarding the gating of the approaches at the external perimeter of the regions, in the cases where no signal plan exists on the entry node, the control variable $u_{ii}$ is translated to adjustment of the saturation flow of the entry links, i.e. the maximum allowed inflow from external virtual queues to entry links is adjusted accordingly. The constraints of minimum saturation flow and maximum absolute change of saturation flow between consecutive control intervals are also applied. However, in the case of external PC, no optimization problem is solved, but control variables $u_{ii}$ are adjusted accordingly, in order to satisfy the constraints. The same settings are applied to all nodes of the external perimeter of every region. 

\subsection{Traffic simulation details}\label{sbs:simulation}

\subsubsection{Modified Store-and-Forward traffic model}

The effectiveness of the various control scenarios is assessed through simulation experiments performed by a modified version of the mesoscopic, queue-based Store-and-Forward (SaF) paradigm (see \cite{aboudolas2009store}), with enhanced structural properties derived by S-Model (see \cite{lin2011fast}). The exact version of the traffic model used in this work is described in details in \cite{tsitsokas2021modeling} and detailed presentation of its mathematical formulation is omitted here. However, in order to facilitate understanding of this work, a brief qualitative description of the model properties is provided in this section. 

In accordance with the notation of MP mathematical description, the model monitors the state of the network, represented as a directed graph $(N,Z)$, by updating the number of vehicles (or `queue') inside every link $z \in Z$, denoted as $x_z(k)$, according to a time-discretized, flow conservation equation.  Vehicle flow is generated in the network according to a dynamic Origin-Destination demand matrix while routing of the flow is performed via turn ratios received as inputs. Queues are updated at every time-step. Link inflows include the sum of transit flows coming from all upstream links plus the newly generated demand of the link. Outflows consist of the sum of transit flows towards all downstream links plus trip endings inside the link. Backwards propagation of congestion is properly modeled, by replacing the initial assumption of point-queues in SaF by utilizing a more accurate link outflow representation. More specifically, the model assumes zero transfer flow for next time-step if the receiving link is already congested (receiving queue at the current time step is close to capacity). Therefore, in case of gridlock downstream, upstream queue is growing and spill-backs can occur, since no outflow is recorded. Therefore congestion propagation is properly captured by the traffic model. The regional accumulations $n_i$ that appear as elements of vector $\textbf{n}$ in equation \ref{eq:PI}, in real applications is measured through appropriate instrumentation of the network. However, in this work where perfect knowledge of real-time traffic is assumed, $n_i(k)$ is simply the sum of link accumulations $x_z(k)$ of all links assigned to region $i$ at time step $k$, i.e. $n_i(k) = \sum_{z \in i}{x_z(k)}, i \in \mathcal{N}$. 

The inherent inaccuracy in estimating travel time and real queue length of links, owed to the dimensionless nature of the initial SaF version has been decreased by integrating the structure of S-Model (\cite{lin2011fast}), where every link queue $x_z$ consists of two distinct queues, $m_z$ and $w_z$, representing vehicle flows moving and waiting at the intersection at the end of each link, respectively. Transit flow vehicles that enter a new link firstly join the moving part of the link, where they are assumed to move with free flow speed. They transfer to the queuing part after a number of time steps, which is determined by the position of the queue tail at the moment when they entered the link. Moreover, traffic signal settings are properly taken into account. A binary function dictates whether an approach of an intersection gets green light at every time-step, according to the most recent signal plan. If an approach gets red light at any time step, this function will set the current outflow of this approach to zero for this time-step, regardless of the state of the queues upstream or downstream.

By iteratively updating all state variables according to the mathematical relations of the model for a number of time-steps, we get a complete traffic simulation for the specific demand scenario. By knowing the queues of every link of the network at every discrete time-step (state variables $x_z \forall z \in Z$), we can calculate the total travel time of all vehicles as 
\begin{equation}\label{eq:VHT}
    VHT = \sum_{k = 1}^{K}\sum_{z \in Z}{x_z(k)} T + \sum_{k = 1}^{K}\sum_{z \in Z}{x_{VQ,z}(k)} T 
\end{equation}
where the first term represents the total time spent inside the network and the second represents the time spent waiting in virtual queues, i.e.~upstream network links that serve as origins. In equation \ref{eq:VHT}, $x_{VQ,z}(k)$ denotes the virtual queue of link $z$, $T$ denotes the time-step duration and $KT$ is the total simulation time. 

\subsubsection{Dynamic turn ratio update}

The modified version of SaF model that is used as traffic simulator in this work, requires knowledge of turn ratios between every pair of incoming-outgoing links for all network intersections. These values determine how traffic is distributed in the different parts of network, according to the considered origin-destination scenario, and drive the generation of congestion pockets, since they express aggregated driver decisions regarding route selection. It has been shown that adaptive routing decreases the scatter and the hysteresis of MFD as it avoids the development of local gridlocks (\cite{daganzo2011macroscopic}, \cite{mahmassani2013urban}). In order to realistically model traffic distribution in the scope of assessing different control schemes, in this work, turn ratios are estimated and dynamically updated in regular intervals during simulation, on the basis of time-wise shortest path calculation. This is done in order to account for potential rerouting effects, since in reality drivers might react to the enforced control policies by adjusting their route in real time, based on the observed traffic conditions that can be associated to the applied control policy. This process assumes that drivers know or can accurately estimate the average speed of all roads that are included in their alternative paths, and thus select the fastest one among them. Once the final time-wise shortest path is determined for every origin-destination pair of the considered demand matrix, the amount of traffic transferring between all feasible connections of incoming-outgoing links of all network nodes, for a specific future time-horizon, can be counted and turn ratios can be estimated. The process is repeated in regular intervals and turn ratios are recalculated, while accounting for the evolving traffic conditions. Although an equilibrium analysis for path assignment would be a more appropriate approach, it would considerably increase modeling complexity and simulation computational cost, while exceeding the scope of this paper, therefore the simpler shortest path approach was adopted. 

The process of turn ratio recalculation is presented in Algorithm~\ref{alg:alg_1}. Firstly, for a time window of calculation $T_w$, preceding the process, mean speed is estimated for every network link $z \in Z$ based on link outflow and queue values, according to the following relation: 
\begin{equation}\label{eq:v_car}
    v_z(t) = \mbox{min} \left( v_{\textrm{ff}},  \frac{\sum_{j \in T_w(t-1)}{u_z(j)} L_z }{\sum_{j \in T_w(t-1)}{x_z(j)}} \right) \geq v_{\min}  
\end{equation}
where $t$ is the time interval index for the turn ratio calculation, $v_{\textrm{ff}}$ is the free flow speed, $T_w(t)$ is the set of simulation time step indices corresponding to time interval $t$, $L_z$ is the length of link $z$, and $u_z(j)$ and $x_z(j)$ denote the outflow and accumulation of link $z$ at simulation time step $j$, respectively. Since our objective here is to estimate the time to cross a link, a lower threshold $v_{\min}$ is imposed for the cases of gridlocks, for computational reasons, so that no link have infinite traversing time. When mean link speed is calculated for all links, algorithm \ref{alg:alg_1} is called to estimate the new turn rates. Inputs include the origin-destination demand matrix, the simulated inflows to origin links during the previous calculation interval, the ongoing trips from previous calculation interval (in the format of  origin-destination demand), as well as link length and mean speed vectors.   

Firstly, if there are ongoing trips, they are added to the origin-destination demand matrix and ongoing trip stack is initialized. Traversing times for all links are calculated from speed and link lengths. Afterwards, for every origin-destination pair of the demand matrix, the shortest path is calculated by using traversing time as link cost function. If the path's estimated trip duration is longer than the calculation interval $T_w$, the trip is split and a new trip is added to the stack of the ongoing trips, where as origin link is set the link where path is split while destination is the original trip destination. The traffic volume of this path as well as the time already spent in the split link are stored in the stack as well. For the path that is estimated to be traversed entirely in the next interval, connection counters between consecutive links of the shortest path are updated with the respective vehicle volume corresponding to this trip. Trip volume is estimated by the simulated traffic volume that entered the network at the origin links (outflow of virtual queues) during the previous interval $T_w$, which is split to different destinations in proportion with the demand of the original origin-destination matrix. The use of simulated inflow instead of the nominal demand improves the accuracy of the turn ratio calculations by taking into consideration potential delays in trip starting, for example due to external perimeter control or spill-back of queues up to entry links. When all trips of the matrix have been counted, turn ratios for all approaches between upstream and downstream links are estimated by dividing the counted volume of each approach by the total volume of each upstream link. In cases where no vehicles are assigned to an approach, all approaches receive the same turn ratio. At the end, the algorithm returns to the simulator the updated turn rates and a stack of ongoing trips. Dynamic turn ratios determined in this way can reflect possible path adjustments due to congestion-related low speeds. Therefore, the impact of the control schemes under evaluation in terms of traffic distribution in the network is directly taken into account. 

\begin{algorithm}[tb!]
\SetAlgoLined
\KwData{origin-destination pairs $\mathbf{OD}$, inflows of origin links $\mathbf{u}$, ongoing trips o-d stack $\mathbf{OD_s}$, ongoing trips flows stack $\mathbf{u_s}$, time window $T_w$, speed vector \textbf{v}, link length vector \textbf{L}, previous turn ratios $\beta$}
\KwResult{updated turn ratios $\beta$, ongoing trip o-d pairs $\mathbf{OD_s}$ and flow $\mathbf{u_s}$ stacks}
\DontPrintSemicolon
\textup{Initialize connection counters:}  $c_n(z,w) = 0, \hspace{0.2cm} \forall z \in I_n, w \in O_n, n \in N$  \;
Merge ongoing trips stacks $\mathbf{OD_s}$,$\mathbf{u_s}$ with $\mathbf{OD}$,$\mathbf{u}$, respectively\;
Initialize stacks $\mathbf{OD_s} = 0, \mathbf{u_s}$ = 0 \;
Calculate link traversing time $\textbf{t} = \textbf{L} \div \textbf{v}$\;
Create directed graph $G$ with link cost function $\textbf{t}$\;
\For{ all $(o,d) \in $ {\normalfont \textbf{OD}}}{ 
    \If{ $u(o,d)>0 $}{
        find shortest path $p$ from $o$ to $d$ in $G$\;
        \If{duration of $p > T_w$}{
                split $p$ up to link where duration is $< T_w$\;
                store remaining trip in o-d stack $\mathbf{OD_s}$ with the path link after the split as $o$ and same $d$ and volume stack  $\mathbf{u_s}$ \;
                }
            \For{every pair $z$ - $w$ of consecutive links in $p$}{
            update respective connection counter:  $c_n(z,w) = c_n(z,w) + u(o,d)$\;} 
        }
}
\For{ all $n \in N$}{
    \For{all $\{(z,w)| z \in I_n, w \in O_n\}$}{
        \eIf{$\sum_{j \in O_n}{c_n(z,j)}>0$}{
        Calculate turn ratios: $\beta(z,w) = \frac{c_n(z,w)}{\sum_{j \in O_n}{c_n(z,j)}}$
        }{
        keep same turn ratios or set them all equal (if no previous value)\;
        }
    }
}
\Return $\beta, \mathbf{OD_s}, \mathbf{u_s}$
\caption{Turn ratio dynamic adjustment for modified SaF model}\label{alg:alg_1}
\end{algorithm}

\section{Implementation to a large-scale network via simulation}\label{S:3-SetupOfExperiments}

The proposed adaptive signal control schemes are evaluated using the mesoscopic urban traffic model and settings discussed in subsection \ref{sbs:simulation}, which was coded from scratch and executed in Matlab R2020a, while optimization problems \ref{eq:feasible_greens_MP} and  \ref{eq:feasibleGreensPC} are solved by Gurobi 9.1.2 solver called from Matlab via Yalmip toolbox \citep{Lofberg2004}. Real-life large-scale signalized traffic network of Barcelona city center is used as case study and Fixed-Time Control (FTC) settings with no adaptive element, are used as benchmark case. The network model and FTC plans are provided by Aimsun. Both MP and PC schemes are applied separately, as well as in combination, for two different demand scenarios that create moderate and high levels of congestion in the FTC case, respectively. All MP cases are tested in full-network implementation and in node subsets for different penetration rates, selected by the proposed algorithm as well as randomly, for comparison. Detailed description of the network, the simulation settings and the performed experiments are provided in this section.          

\subsection{Case study}\label{ss:Case_study}

\begin{figure}[!b]%
	\centering
    \subfloat[]{\includegraphics[scale=0.70]{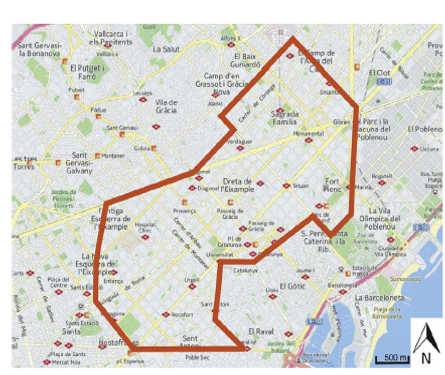}}%
    \qquad
    \subfloat[]{\includegraphics[trim= 2cm 1cm 2cm 1cm, scale=0.45]{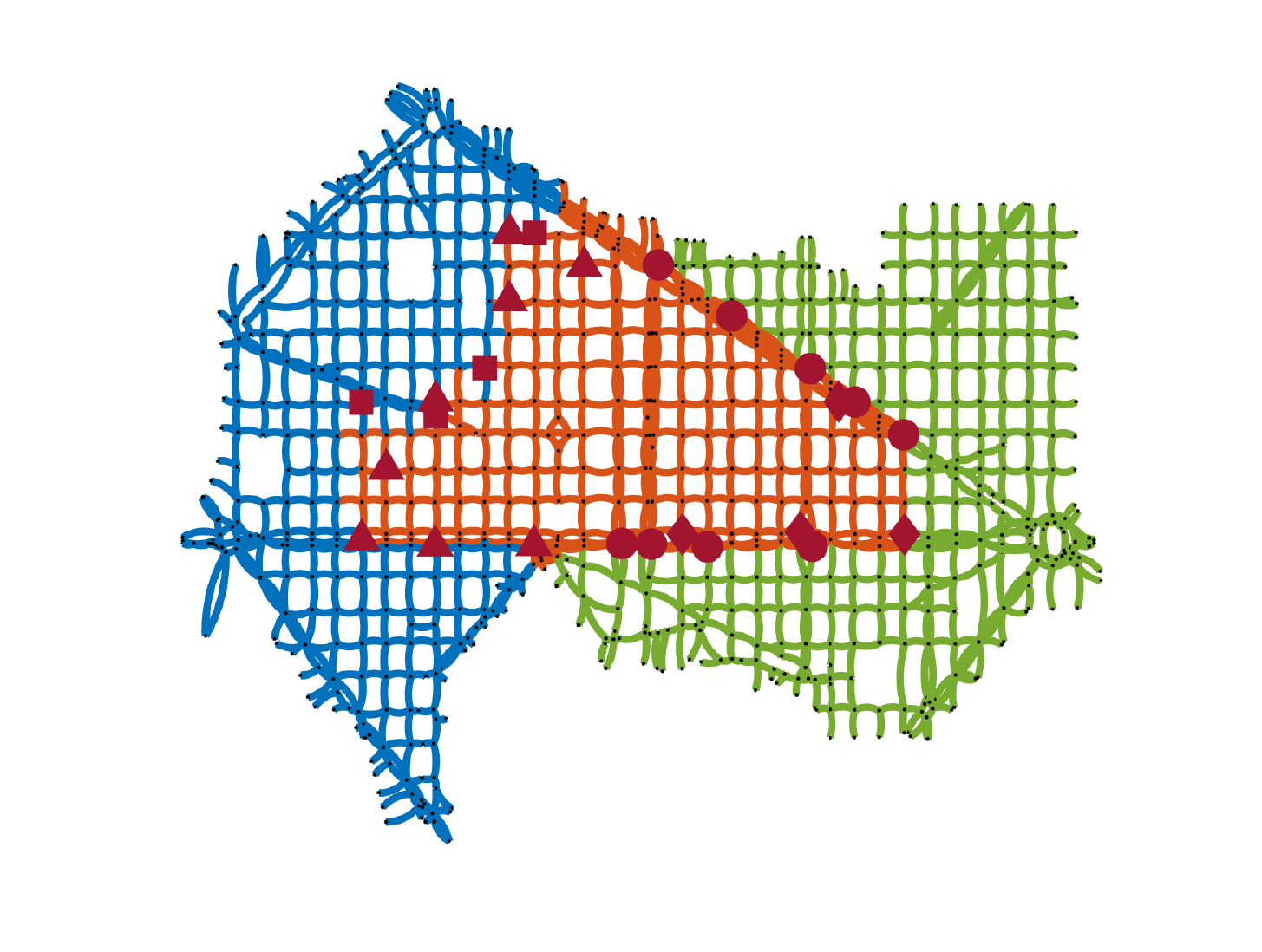}}%
    \qquad
    \subfloat[]{\includegraphics[trim= 8cm 0cm 8cm 0cm, scale=0.35]{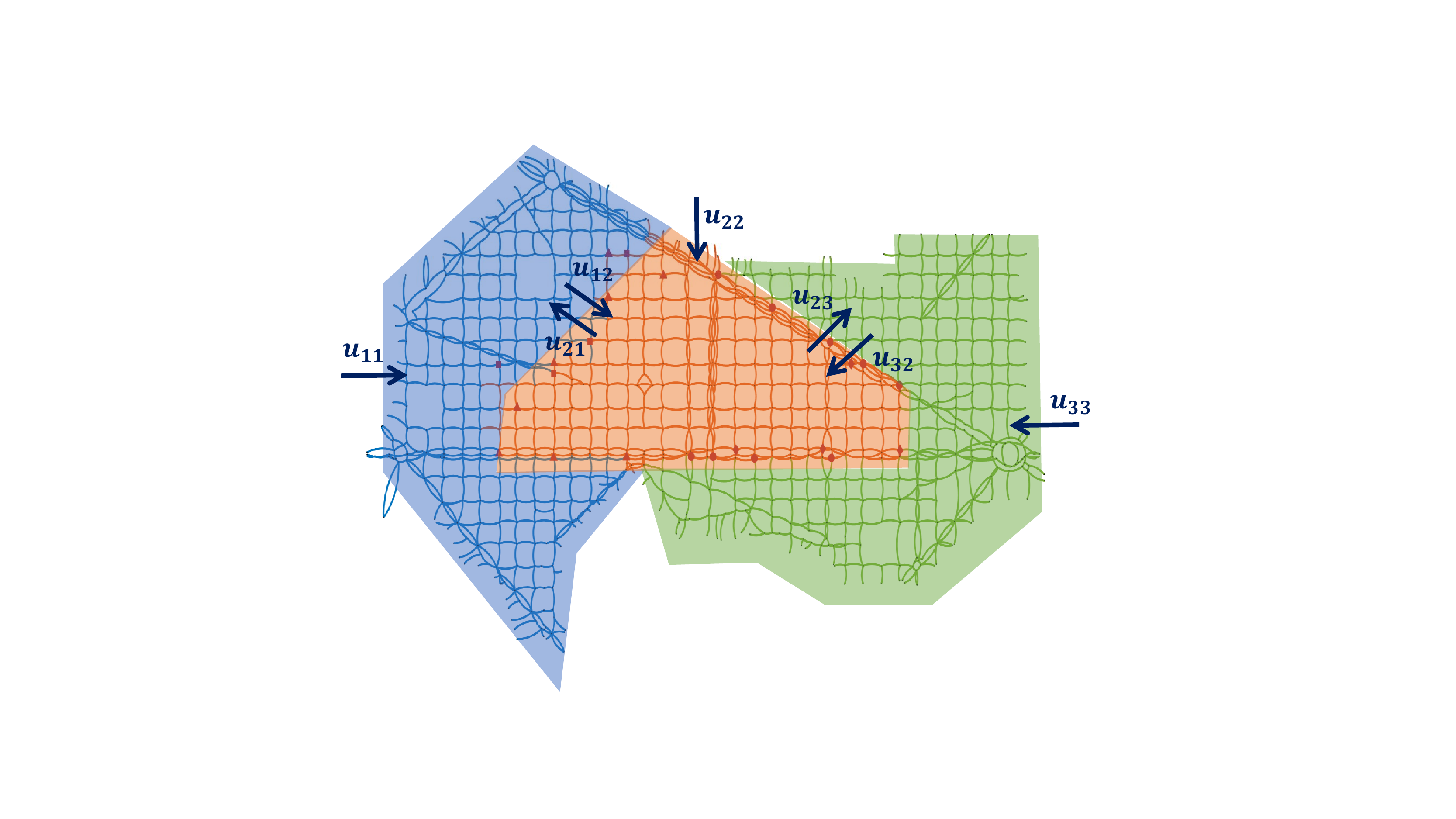}}%
    \caption{(a) Map of the studied network of Barcelona city center; (b) Model of Barcelona network as a directed graph with annotation of nodes used for PC; (c) schematic representation of controlled approaches for perimeter and boundary flow control, with green time per approach as control variable.}%
    \label{fig:maps}
\end{figure}

\begin{figure}[!b]%
	\centering
    \subfloat[]{\includegraphics[scale=0.33]{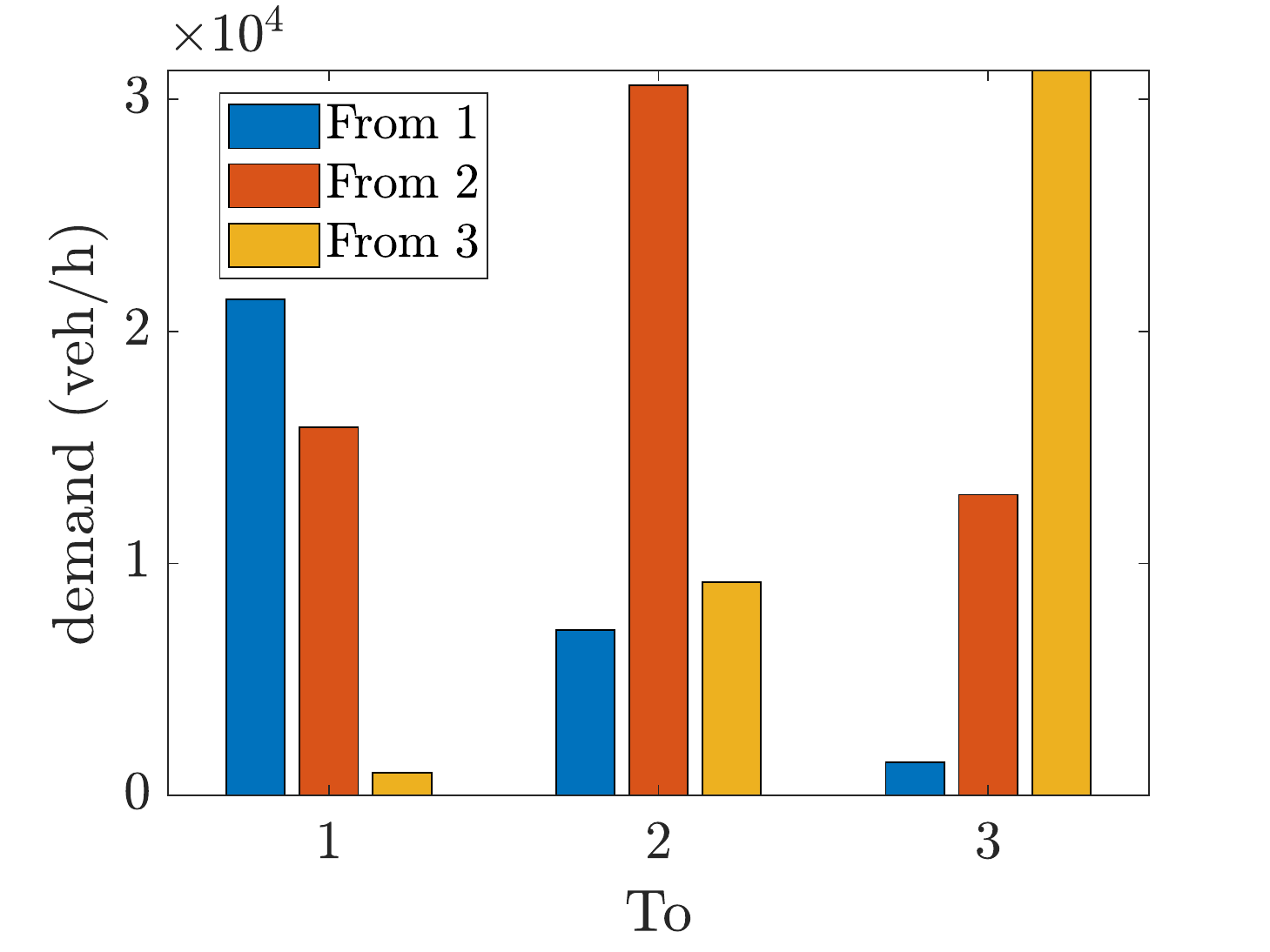}}%
    \qquad
    \subfloat[]{\includegraphics[scale=0.33]{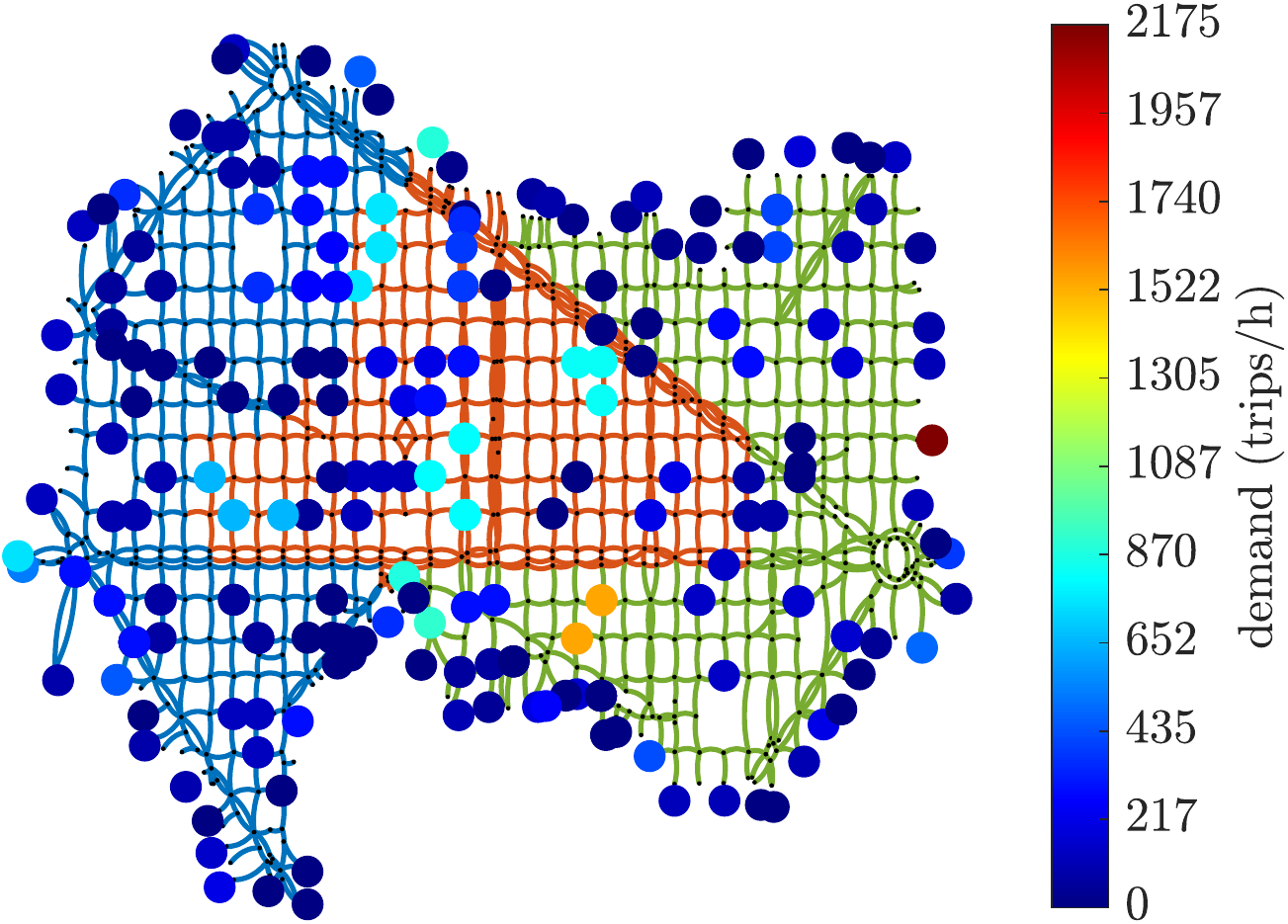}}%
    \qquad
    \subfloat[]{\includegraphics[scale=0.33]{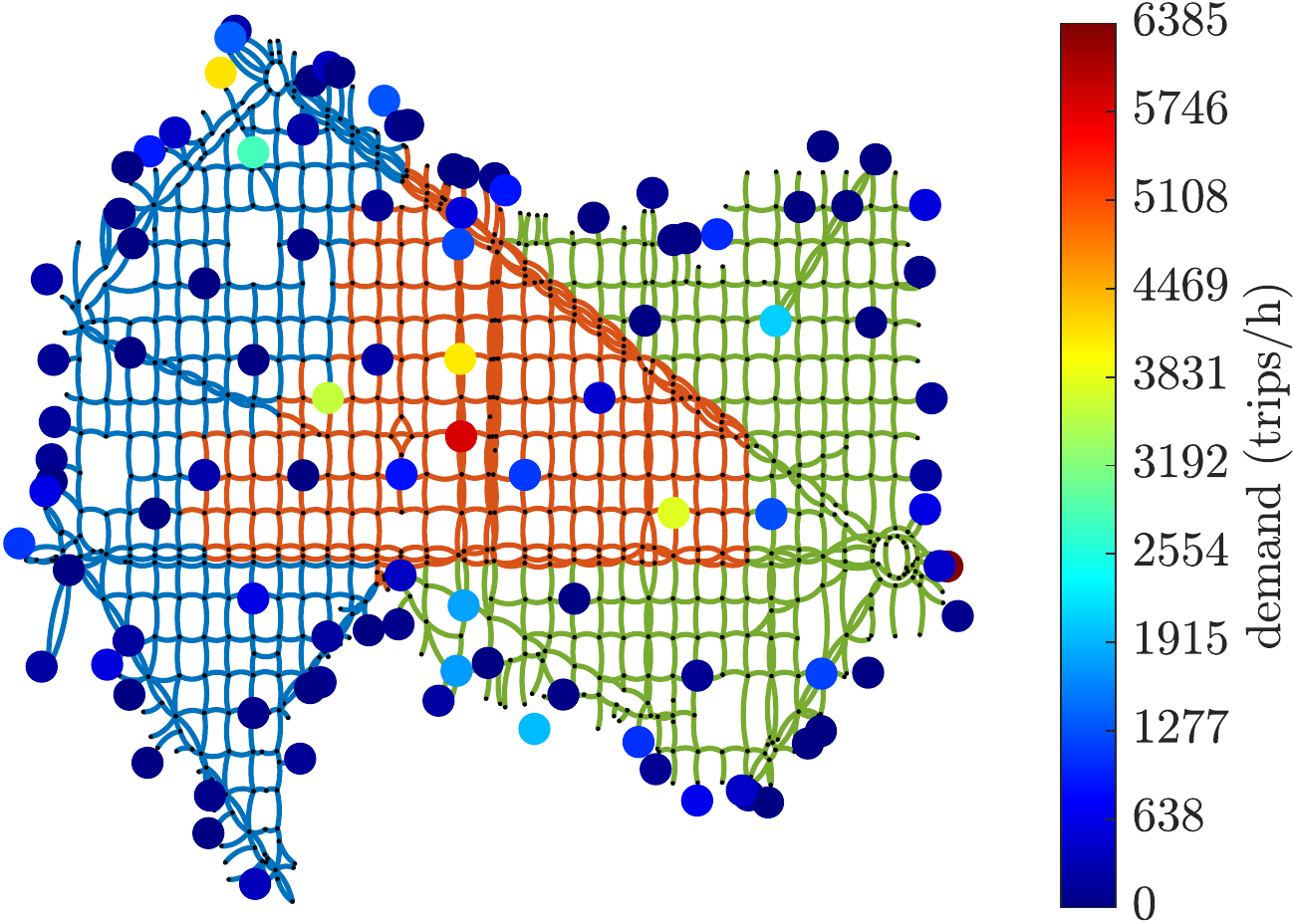}}%
    \qquad
    \subfloat[]{\includegraphics[scale=0.33]{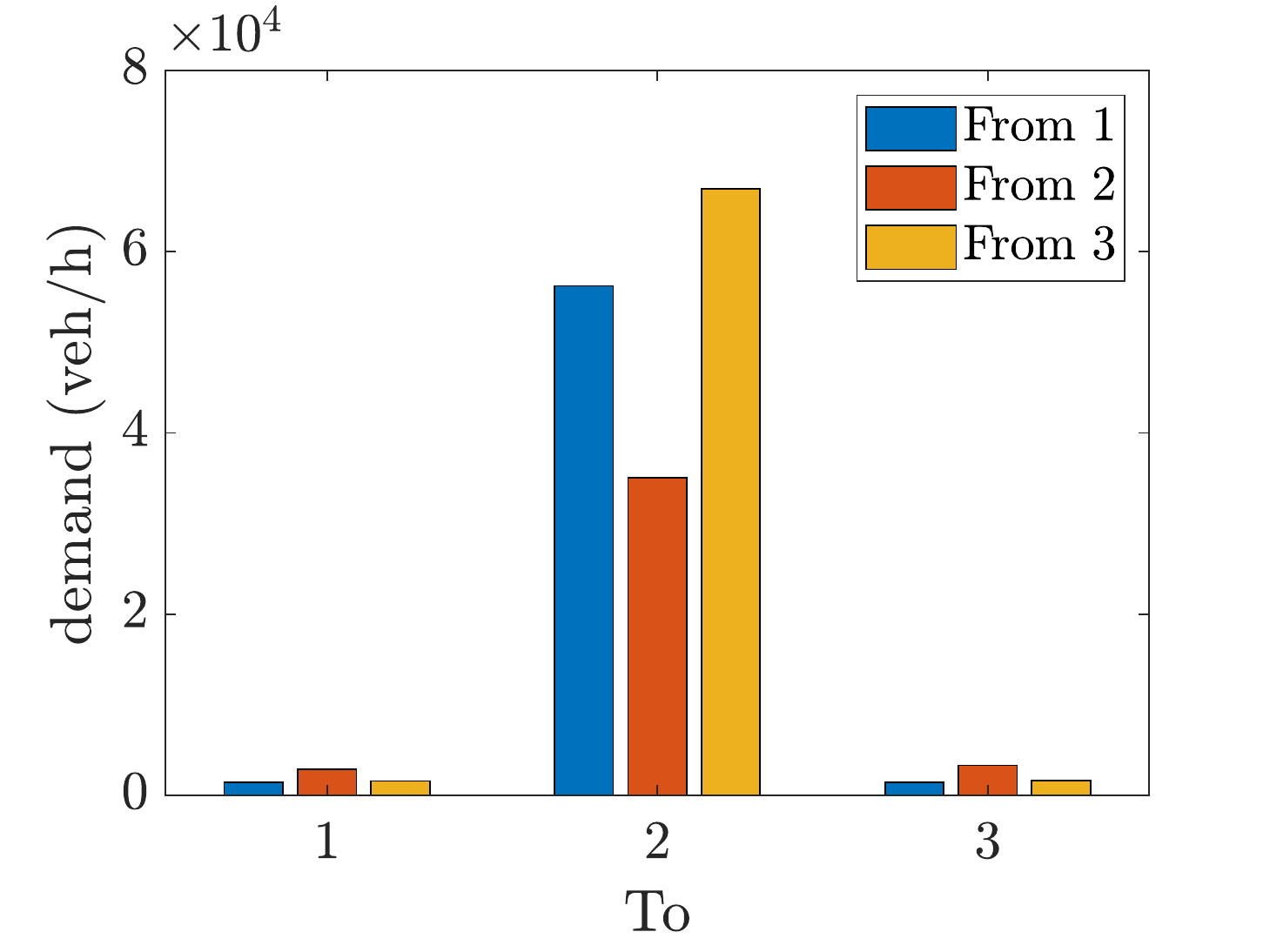}}%
    \qquad
    \subfloat[]{\includegraphics[scale=0.33]{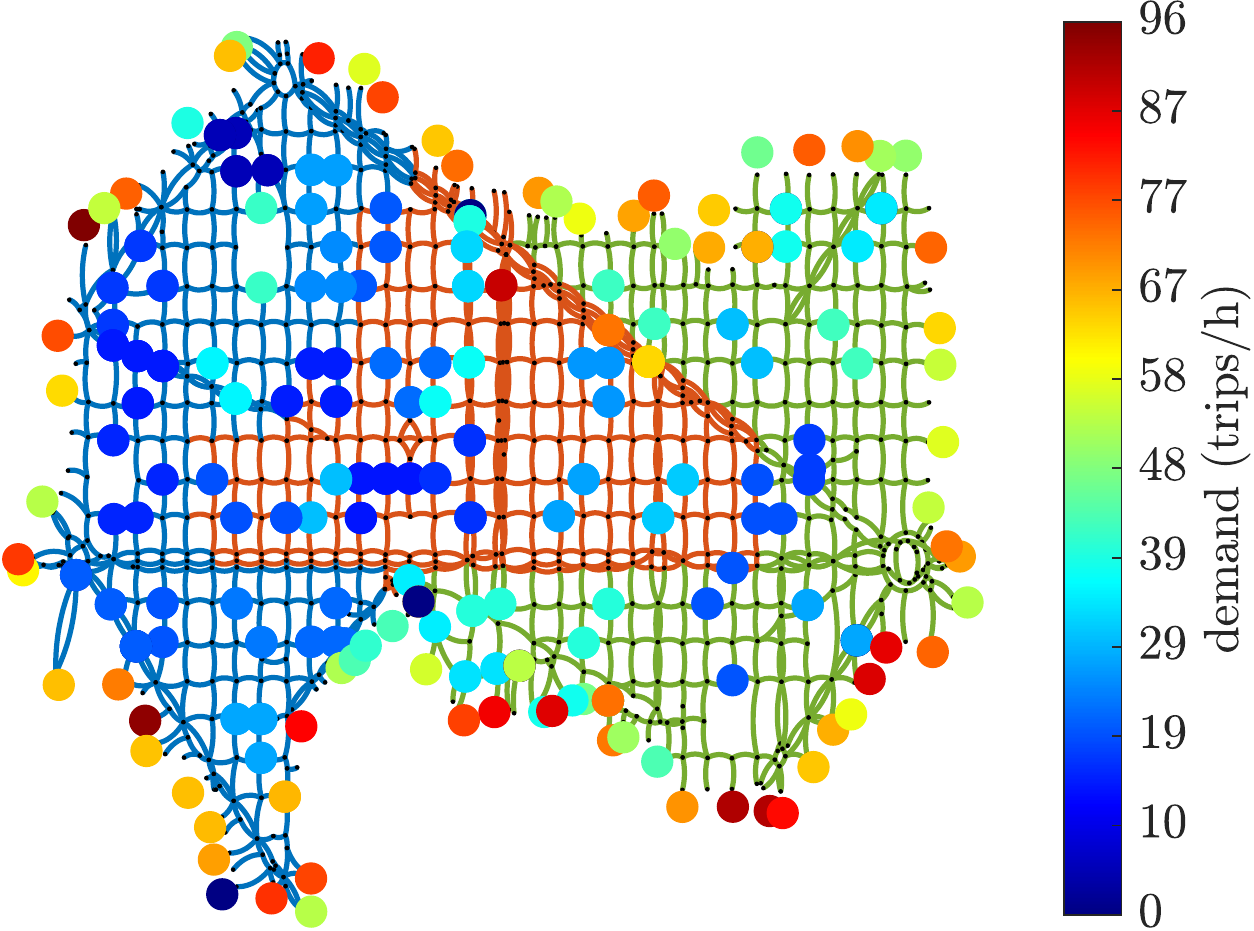}}%
    \qquad
    \subfloat[]{\includegraphics[scale=0.33]{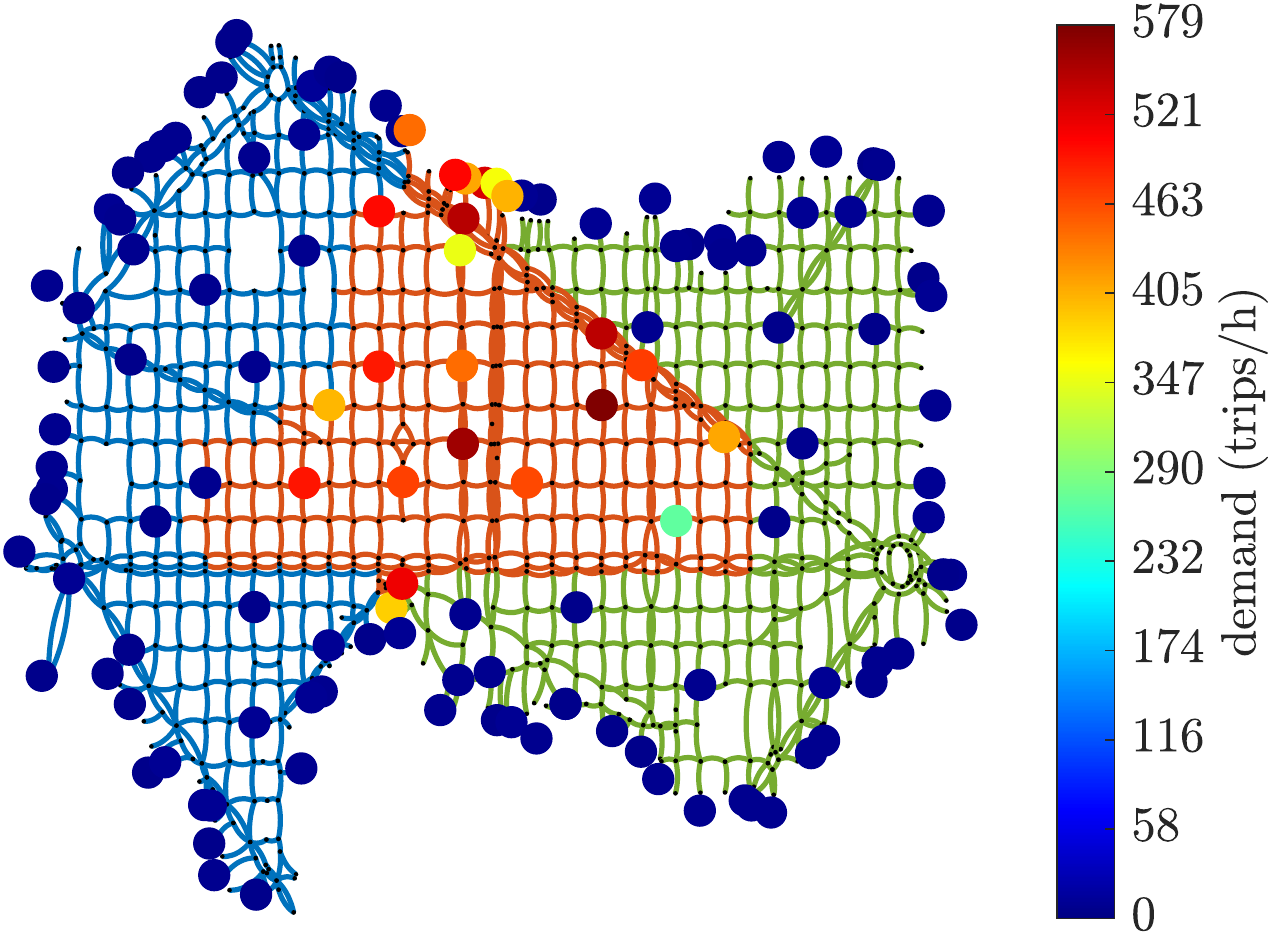}}%
    \caption{Description of the two demand scenarios in peak period: medium (a)-(c) and high (d)-(f). (a) and (d): aggregated trip distribution between regions; (b) and (e): trip origin density; (c) and (f): trip destination density.}%
    \label{fig:demand_description}
\end{figure}

\begin{figure}[!b]%
	\centering
    \subfloat[]{\includegraphics[scale=0.35]{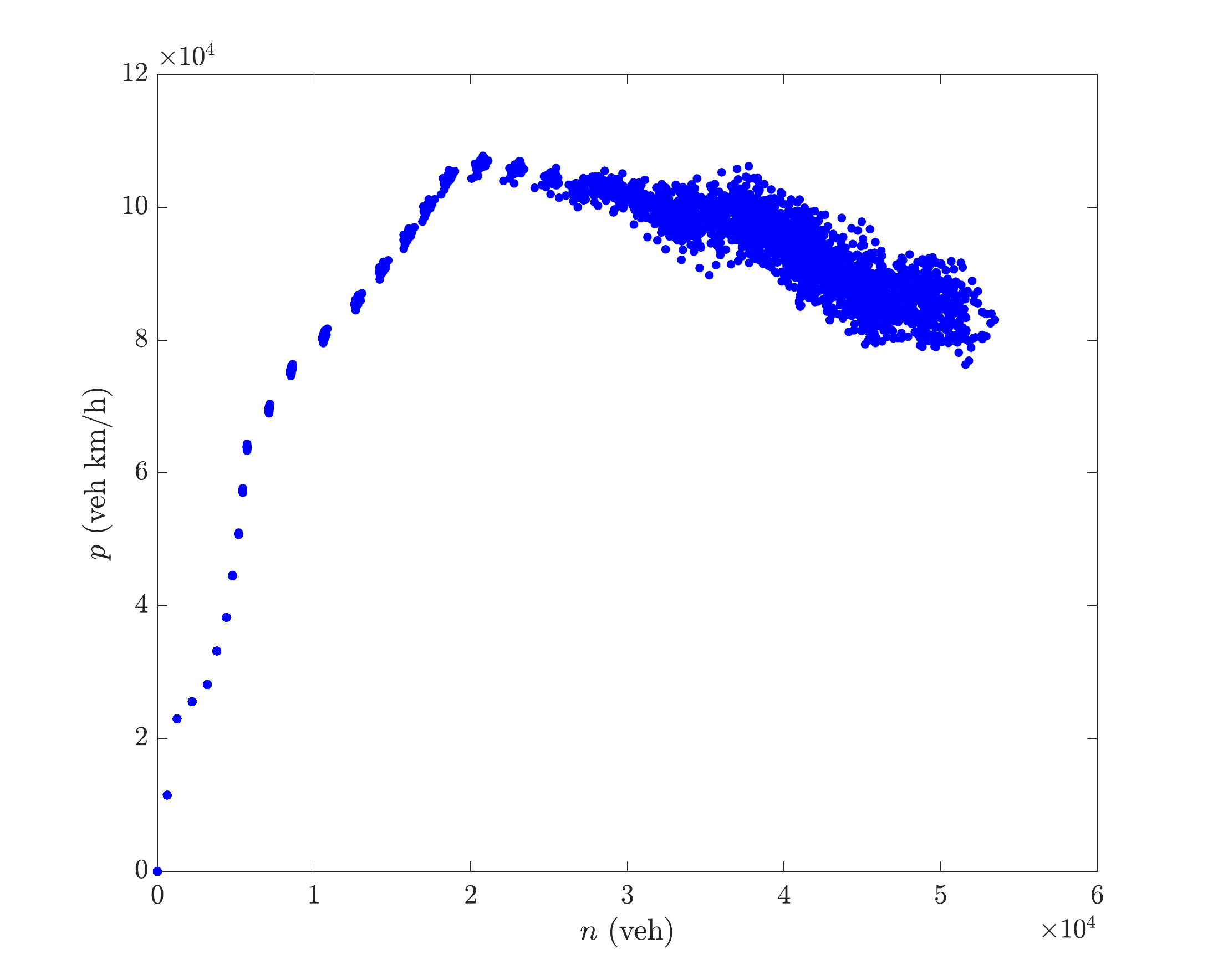}}%
    \qquad
    \subfloat[]{\includegraphics[scale=0.35]{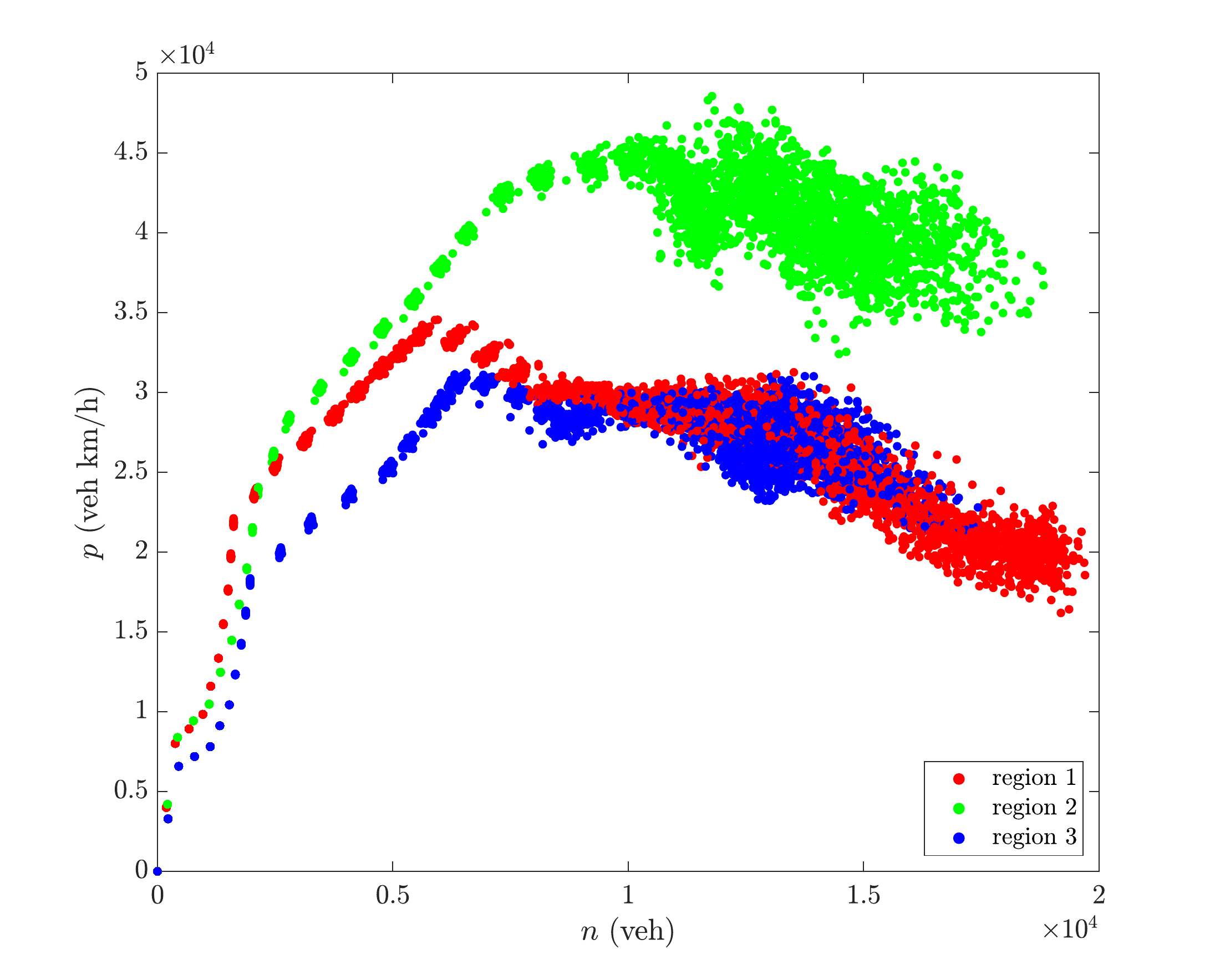}}%
    \caption{Production MFDs for the onload of congestion in the case of FTC: (a) MFD of entire network; (b) MFDs per region after partitioning.}%
    \label{fig:MFDs_NC}
\end{figure}

The traffic network utilised in this study is a replica of Barcelona city center, in Spain, as shown in Figure~\ref{fig:maps}a. It consists of 1570 links and 933 nodes, out of which 565 represent signalized intersections with fixed-length signal control cycles ranging from 90 to 100 sec. Links have from 2 to 5 lanes. All control schemes are tested in two different demand scenarios, one leading to moderate and one to high congestion in FTC settings. For the purpose of PC implementation, the network is partitioned into three regions of similar traffic distribution, according to the clustering method described in \cite{saeedmanesh2016clustering}. The resulting regions are displayed in different colors in figure \ref{fig:maps}b, with region 1 shown in blue, region 2 in orange and region 3 in green. In the same figure, the controlled intersections used for gating through PC are displayed, with different annotation, based on the approach that they control: triangle for 1 to 2 approaches, circle for 3 to 2 approaches, square for 2 to 1 approaches and rhombus for 2 to 3 approaches. Figure \ref{fig:maps}c schematically represents the perimeter and boundary flow control variables $u_{ij}$, which denote the average green time of all approaches in the direction from $i$ to $j$, while $u_{ii}$ denotes the equivalent time for all external approaches of region $i$. All nodes on the external perimeter of all three regions are controlled. This partitioning leads to 4 state and 7 control variables, which are depicted in figure \ref{fig:maps}c. By running a simulation experiment for the FTC case for both demands scenarios, we obtain MFD curves for each of the three regions, which are shown in figure \ref{fig:MFDs_NC}. 

The dynamic profile of total generated demand, for both demand scenarios, consist of a 15-minute warm-up period followed by a 2-hour constant peak demand, for a total simulation period of 6 hours, which is representative of the morning-peak. Medium demand includes 251k trips generating at 88 origin links and heading towards 104 destination links, whereas high demand scenario includes 316k trips, from 123 origins to 130 destinations. Figure \ref{fig:demand_description} graphically describes the spatial distribution of demand for both scenarios: (a) to (c) refer to medium demand while (d) to (f) refer to high demand. The distribution of trips between regions, resulting from suitable clustering process as described below, is presented in the first graph of each row (a and d). Each bar corresponds to the total demand originating from each origin region, each represented by different color, while horizontal axis indicates destination region. Second graph of each row (b and c) presents the spatial distribution  and density of origin points, represented by dots on the network map. Dot color indicates demand volume per origin point, as described by the respective colorbar. Similarly, third graph of each row (c and f) represents the same information for destination points. Apart from difference in total number of trips, the two demand scenarios lead to different traffic patterns, since high demand scenario has a clear directional profile, with trips mainly originating from the periphery (regions 1 and 3) and heading towards the center (region 2), whereas medium demand shows more diverse trip distribution between regions, with more intraregional trips in all regions, plus a less intense directional pattern towards the city center (region 2). The objective of using different demand scenarios is to test the efficiency of the proposed control schemes under different traffic conditions. It should be noted, that no demand information is required by the controllers, which only receive real-time queue measurements and/or turn ratios as inputs. Accuracy and configuration of traffic measurements is not taken into account in this work and all necessary information to controllers is provided directly from the simulator (perfect knowledge assumed).

\subsection{Control scenarios}

The case of fixed-time, static signal control, labelled as 'FTC', is used as benchmark for all tested control schemes. The network performance improvement owed to traffic responsive control schemes based on single MP, single PC and on their combination in a hierarchical framework is calculated with reference to FTC, which is consider as the current network state. Firstly, MP is evaluated as single control strategy (no PC applied simultaneously). With the aim of investigating the performance of MP in relation to the number and location of the controlled nodes, we test the following scenarios: 

\begin{itemize}
    \item MP control of all eligible network nodes. All signalized nodes receive MP controller. 
    
    \item MP control of fraction of network nodes, selected randomly. For each penetration rate of $5\%, 10\%, 15\%, 20\%, 25\%$, 10 randomly created MP node sets are evaluated through simulation.
    
    \item MP control of fraction of network nodes, selected by the proposed algorithm. For the same penetration rates as above, MP node sets are created according to decreasing values of $R$, after suitable parameter calibration. FTC simulation results are used for calculating variables $m_1^n$, $m_2^n$ and $N_c^n$, and thus quantity $R$. 
    
\end{itemize}

Afterwards, MFD-based PC based on the PI controller described in subsection \ref{ss:PI_regulator}, is applied first as a single control scheme and then in combination with distributed MP control, integrated in a two-layer framework. Similar to the case of single MP, the MP layer of the combined scheme is tested for several controlled node layouts, in various penetration rates, as well as in full network implementation (100 \% eligible nodes). The following scenarios are evaluated: 

\begin{itemize}
    \item Single PC for 3-region system  
    
    \item PC for 3-region system combined with MP control to all eligible network nodes
    
    \item PC for 3-region system combined with MP control to fractions of eligible network nodes, selected randomly. For each penetration rate of $5\%, 10\%, 15\%, 20\%, 25\%$, 10 sets of randomly selected eligible nodes are formed (same as in single MP scenarios). MP control is evaluated in parallel with PC scheme.
    
    \item PC for 3-region system combined with MP control to fractions of eligible network nodes, selected by the proposed algorithm. For the same penetration rates as above, MP node sets are created according to decreasing values of $R$, after suitable parameter calibration. FTC simulation results are used for calculating variables $m_1^n$, $m_2^n$ and $N_c^n$, and thus quantity $R$.
\end{itemize}

\subsection{Experiment settings}

The control scenarios under evaluation are simulated for a 6-hour time period for medium demand, and for a 8-hour period for high demand, representative of typical morning peak. High demand required longer simulation time to unload and completely empty the network, due to high number of trips. Simulation time step is set to 1 second. Free-flow speed of vehicles in the moving part of links, $v_{\textrm{ff}}$ is set to 25 km/h and average vehicle length for capacity calculations is set to 5 meters. The time window for turn ratio update is 15 minutes.

MP regulator is active for all controlled intersections during the entire simulation time and signal plans are updated at the end of every cycle,  based on traffic information collected during the last cycle. Only phases lasting longer than 7 seconds in the pre-timed scheduling are eligible for change by the regulator and minimum green time allowed per phase $g_{n,j,\min}$ is also 7 seconds. Maximum allowed fluctuation of green time between consecutive cycles, $g_{n,j}^R$ is set to 5 seconds, for all MP modified phases, to avoid instabilities (similar to \cite{kouvelas2017enhancing}). Inputs for MP regulator are link queues of incoming and outgoing links, as well as estimated turn ratios for all approaches of controlled nodes. For our experiments, this information is provided directed by the simulator, thus assuming that the controller receives perfect real-time traffic information.  

Regarding PC implementation and based on the network partitioning in 3 regions described above, the PI controller of equation \ref{eq:PI} regulates 7 control variables $u_{ij}$, which represent the average green time of all controlled intersections in the approaches between adjacent regions 1-2, 3-2, 2-1, 3-1, as well as those of the external perimeter of each regions 1, 2 and 3 (see figure \ref{fig:maps}c), using 3 state variables, i.e. the regional accumulations $n_i, i=1,2,3$. Therefore, $\textbf{u}$ is a 7x1 vector, $\textbf{n}$ and $\hat{\textbf{n}}$ are 3x1 vectors, while proportional and internal gain matrices $\mathbf{K}_P$ and $\mathbf{K}_I$ are of dimensions 7x3. Four first rows refer to boundary approaches in the order they are listed above and three last rows refer to external perimeter approaches of regions 1 to 3. Similar to MP, for inter-regional approaches, minimum green time per phase $g_{m,i,\min}$ is set to 7 sec and maximum allowed absolute change between consecutive cycles is $g_{m,i}^R$ is set to 5 seconds. For the external perimeter gating, where no signal plan is available and controller regulates saturation flow of entry links, the constraint of allowing at least 15\% of the initial saturation flow holds. Due to different directional patterns of the two demand scenarios, PI parameters differ slightly. For the medium demand scenario, setpoint accumulation for the three regions are $\hat{n_1} = 10000, \hat{n_2} = 12000, \hat{n_3} = 6800$ veh, proportional gain matrix is $\textbf{K}_P = [15, -10, 0; 0, -5, 10; -15, 10, 0; 0, 5, -10; -20, 0, 0; 0, -20, 0; 0, 0, -20]$ and integral gain matrix is $\textbf{K}_I = \textbf{K}_P \times 10$, $n_{i,\textrm{start}} = \hat{n_i}$, $n_{i,\textrm{stop}} = 0.85 \hspace{0.1cm } \hat{n_i}$, for every region $i=1,2,3$. For the high demand scenario, accumulation setpoint is $\hat{n_1} = 4500, \hat{n_2} = 9200, \hat{n_3} = 8000$ veh, proportional gain matrix is $\textbf{K}_P = [18.5, -2.1, 0; 0, -3.3, 6.8; -13.3, 5.6, 0; 0, 4.6, -3.5; -16.4, 0, 0; 0, -9.8, 0; 0, 0, -10.5]$ and integral gain matrix is $\textbf{K}_I = [18, -69, 0; 0, -69, 62; -44, 24, 0; 0, 1, -40; -54, 0, 0; 0, -30, 0;0, 0, -51]$, $n_{i,start} = 0.99 \hspace{0.1cm }\hat{n_i}$ and  $n_{i,stop} = 0.93 \hspace{0.1cm } \hat{n_i}$, for every region $i = 1,2,3$. In all cases, activation of the PI controller happens when at $n_i \geq n_{i,\textrm{start}}$ for at least 2 regions $i = 1,2,3$, while deactivation happens when $n_i < n_{i,\textrm{stop}}$ for all 3 regions. PC application requires inputs of aggregated regional accumulations $n_i$ for all regions $i=1,2,3$ for the PI controller, while for the phase of applicable green time calculation (optimization problem \ref{eq:feasibleGreensPC}), queue measurements for all approaches of primary and secondary phases of PC intersections, $Q_{m,p}$ and $Q_{m,s}$ respectively, are required, together with the latest applied signal plan. Similar to MP case, all required real-time traffic information are considered given and accurate and for the scope of this work are collected directly from the simulator. The process is repeated every 90 seconds.

\section{Results}

\subsection{Single Max Pressure}

\begin{figure}[tb]%
	\centering
    \subfloat[]{\includegraphics[scale=0.33]{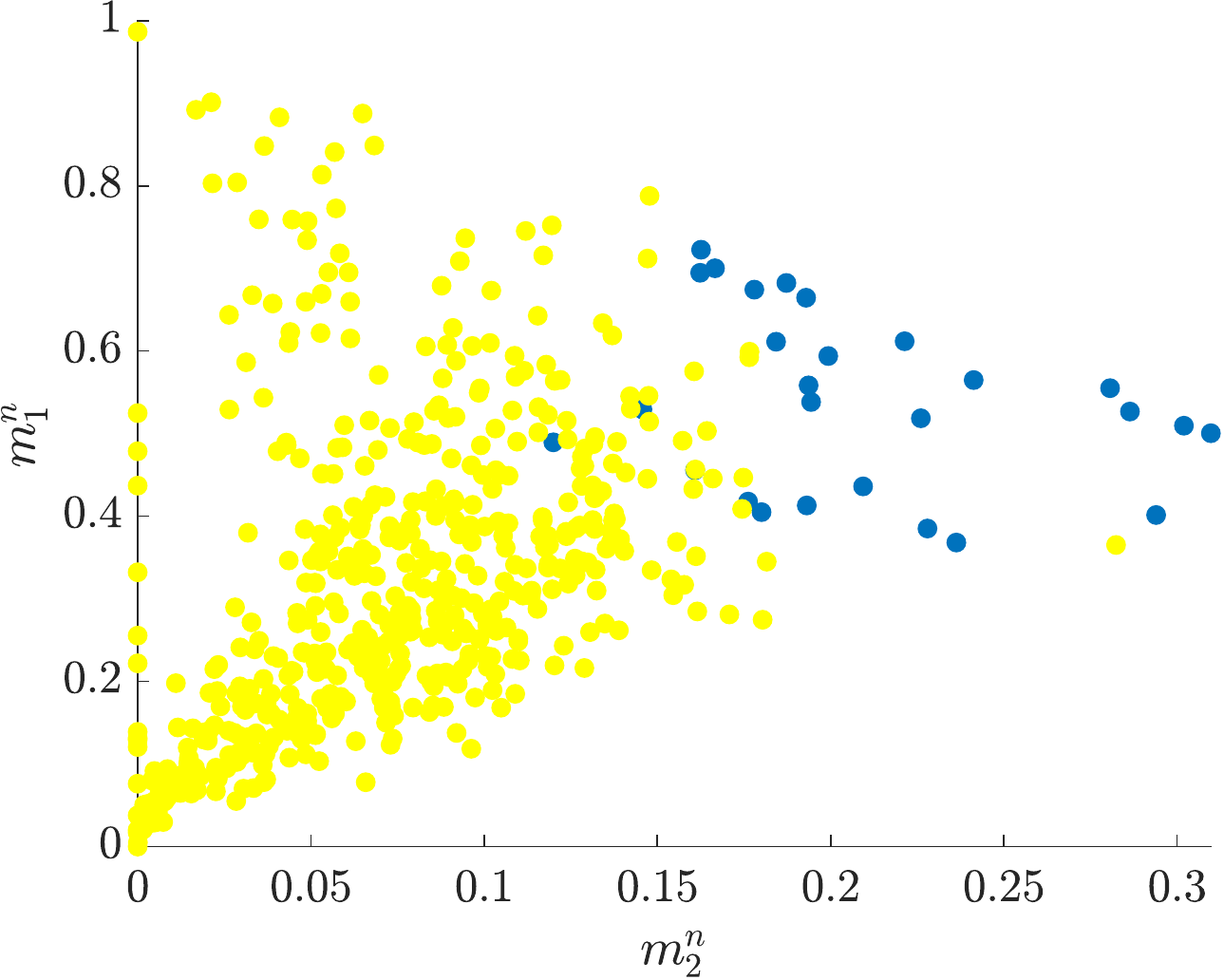}}%
    \qquad
    \subfloat[]{\includegraphics[scale=0.33]{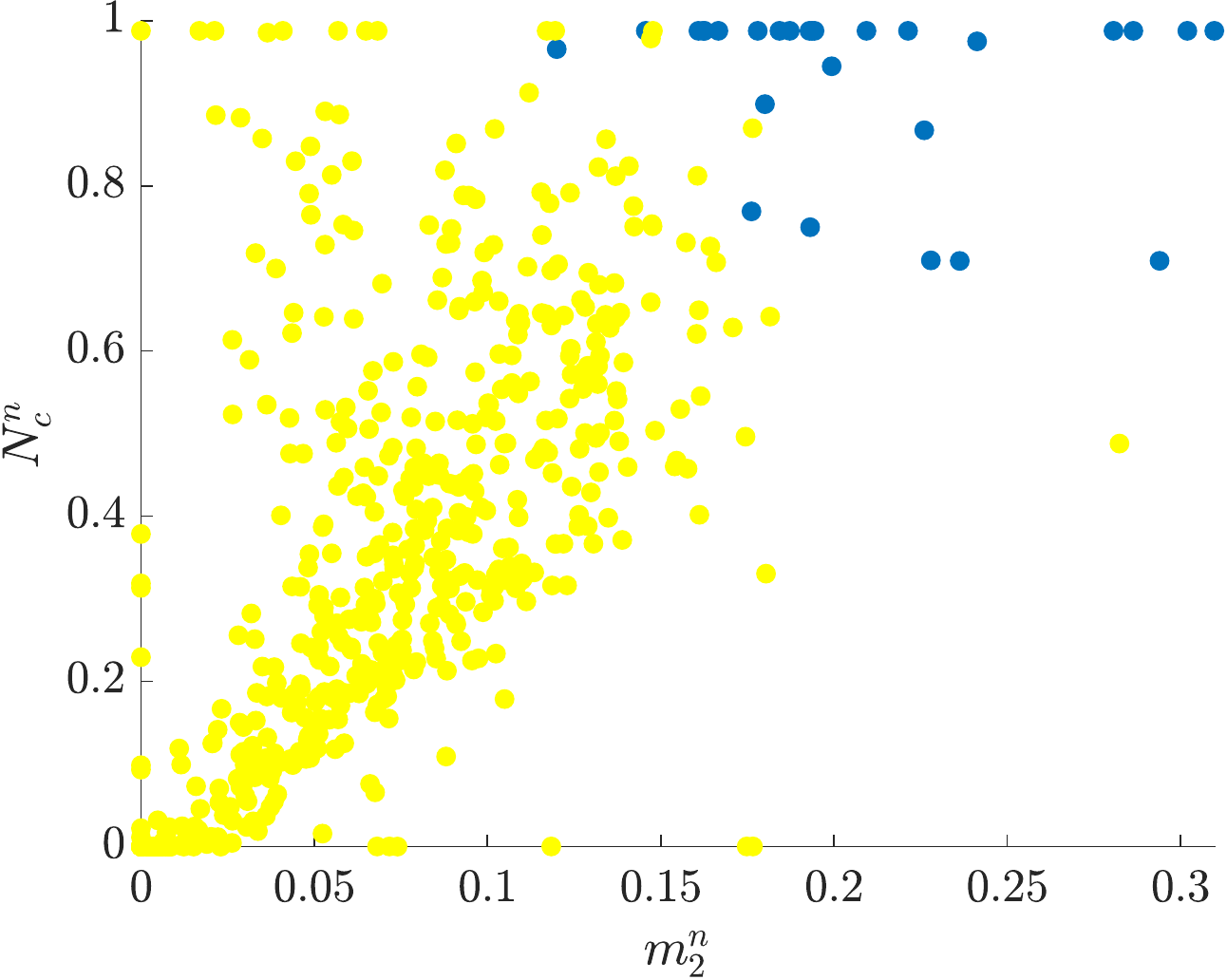}}%
    \qquad
     \subfloat[]{\includegraphics[scale=0.33]{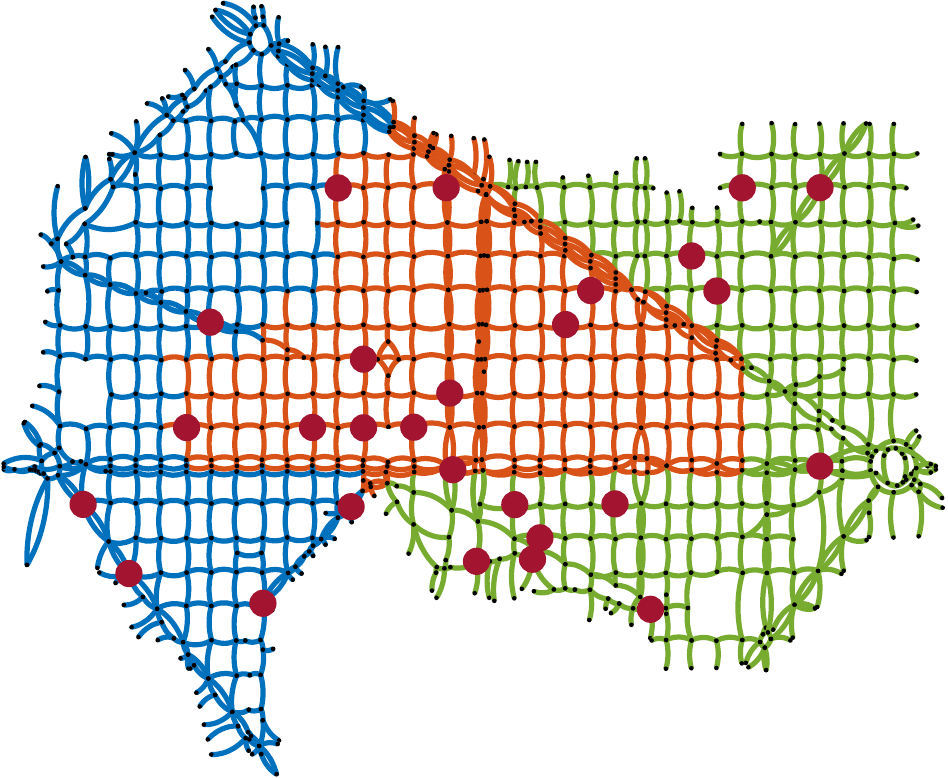}}%
    \qquad
    \subfloat[]{\includegraphics[scale=0.33]{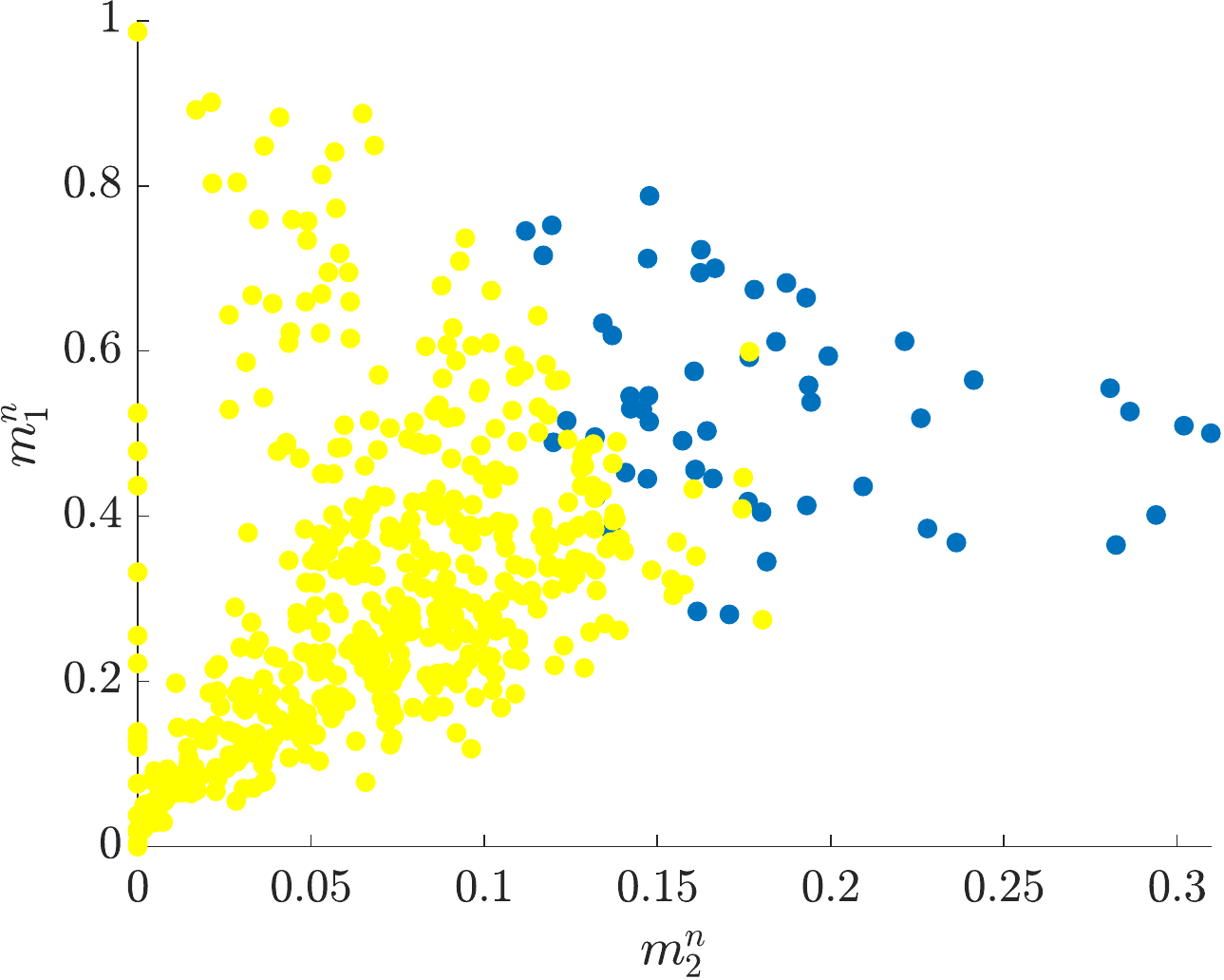}}%
        \qquad
    \subfloat[]{\includegraphics[scale=0.33]{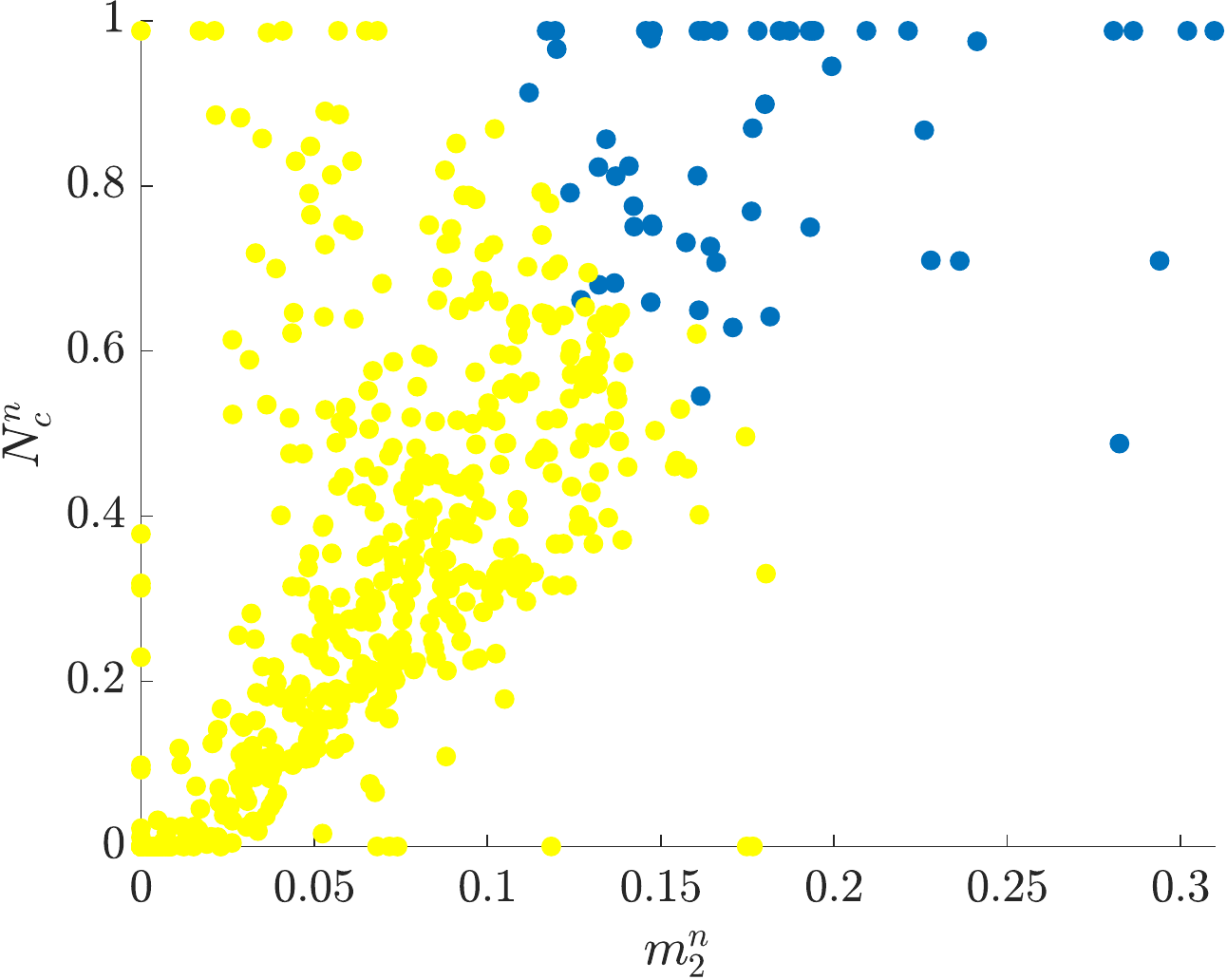}}%
        \qquad
    \subfloat[]{\includegraphics[scale=0.33]{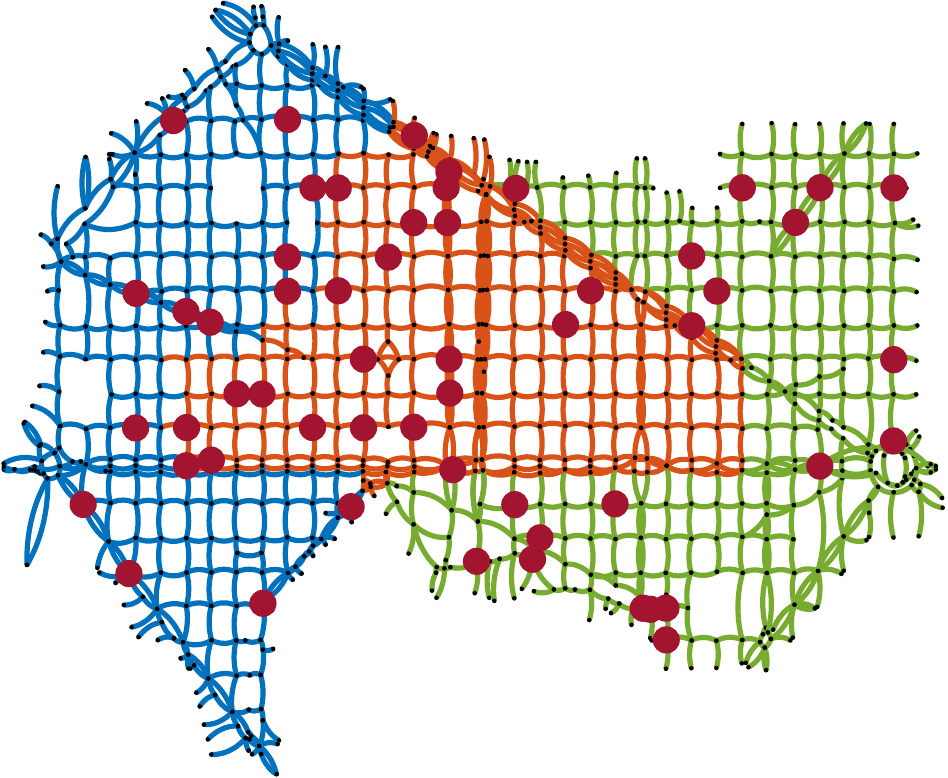}}%
    \qquad
     \subfloat[]{\includegraphics[scale=0.33]{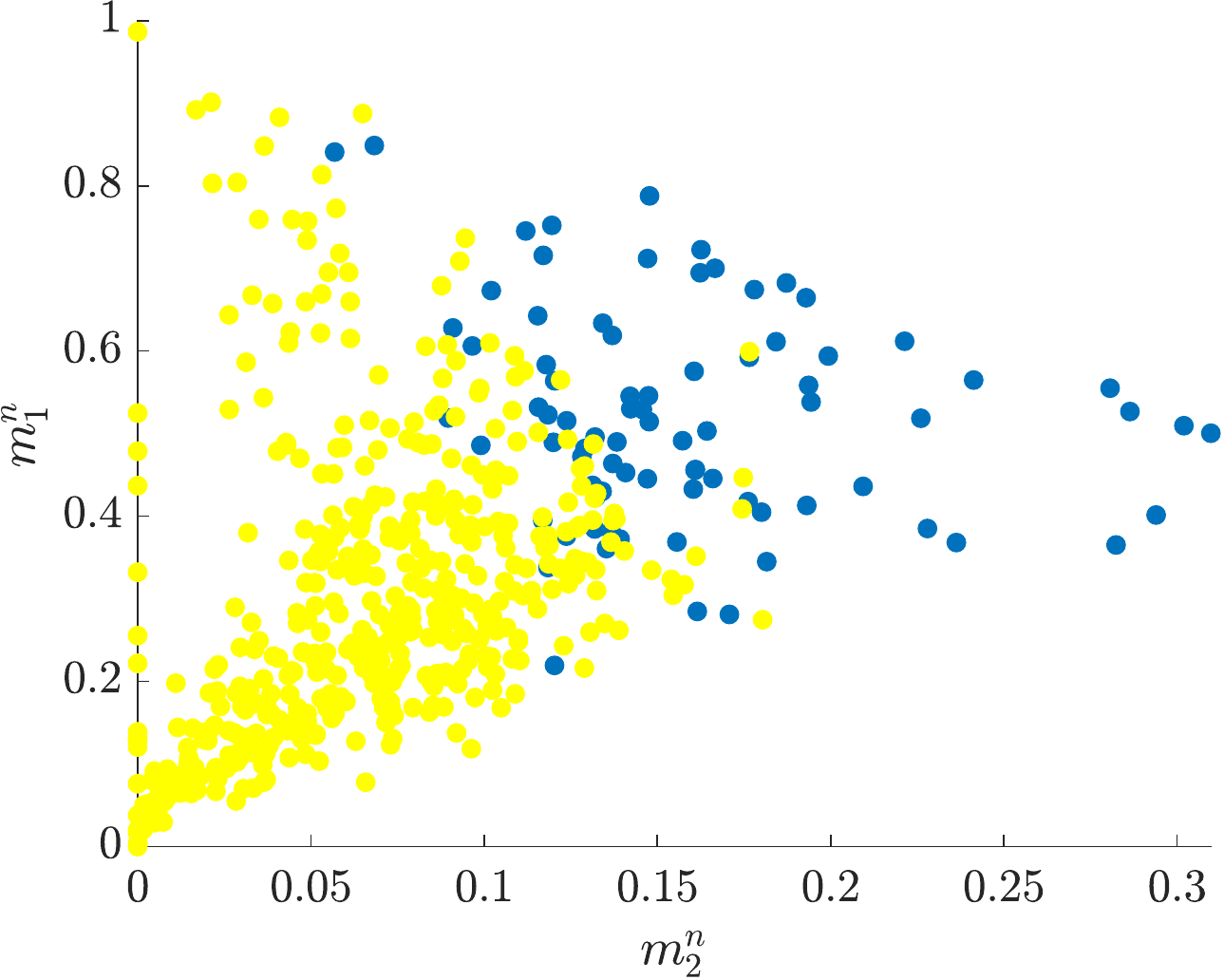}}%
    \qquad
    \subfloat[]{\includegraphics[scale=0.33]{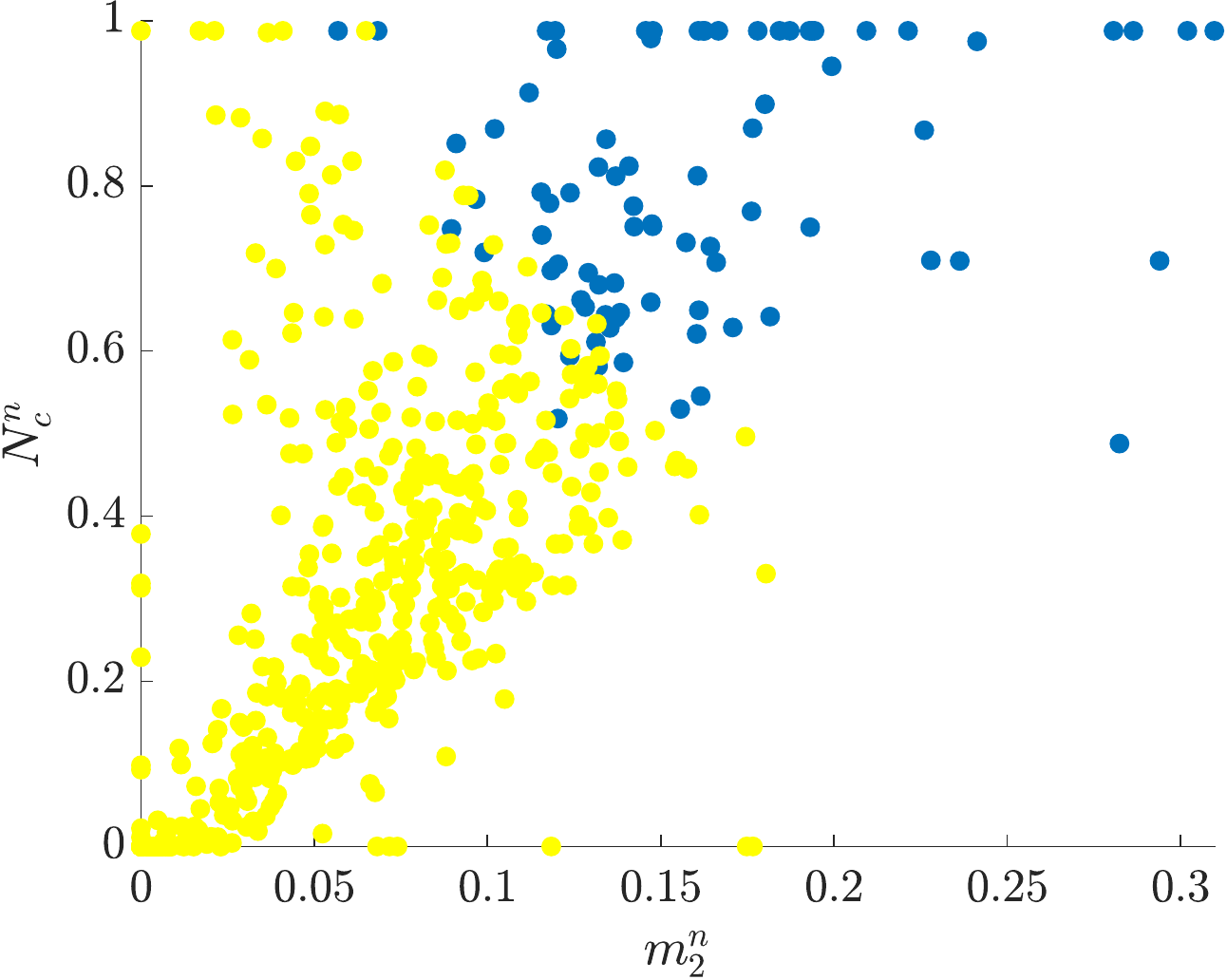}}%
        \qquad
    \subfloat[]{\includegraphics[scale=0.33]{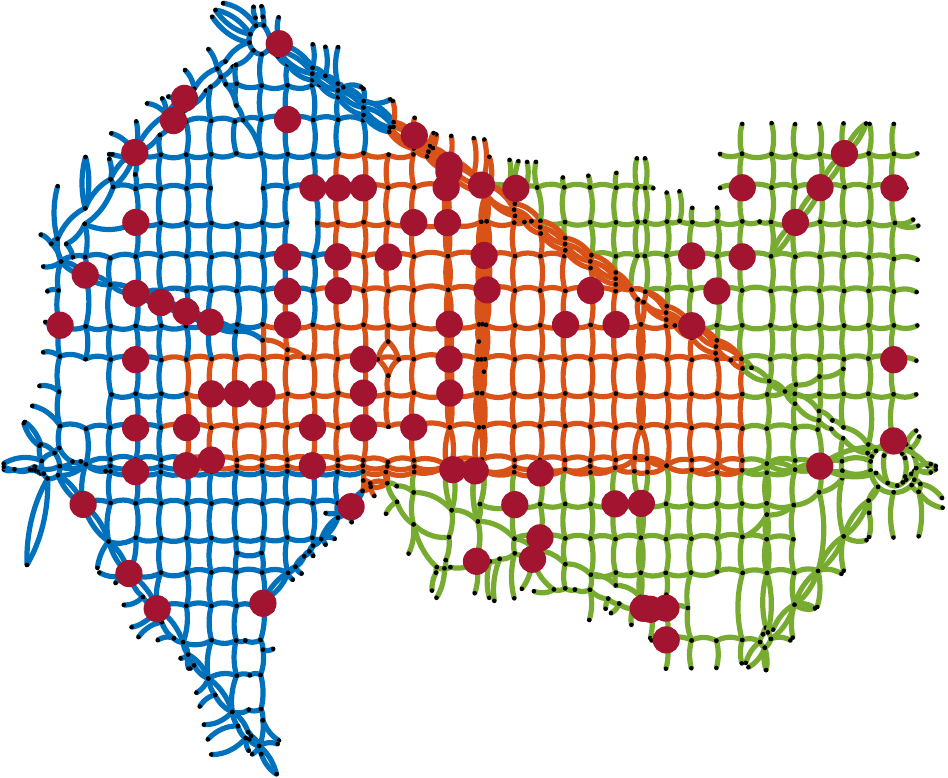}}%
    \caption{Visualization of the MP node selection process for the case of medium demand, according to the proposed method. Each row refers to different penetration rate (5\%, 10\% and 15\%, from top to bottom). First and second column graphs show relations between selection variables $m_2$ - $m_1$ and $m_2$ - $N_c$, respectively, for all network nodes, for the benchmark case of FTC. Blue dots represent the selected nodes for MP control. Third column figures show the plan of the studied network, partitioned in 3 regions, with the spatial distribution of the selected MP intersections shown as red dots.}%
    \label{fig:nodeSelection_med}%
\end{figure}

Results of simulation experiments concerning single Max Pressure schemes are presented in this section. The case of full-network MP implementation, i.e. MP regulator assigned to all eligible nodes, is compared to the FTC case (no adaptive control) as well as to different scenarios of partial MP implementation, in different fractions of network nodes, selected both randomly and by the methodology described in section \ref{sss:nodeSelection}. For the process of node selection, the considered peak-period is 2-hour long (starting from 0.5 and ending at 2.5 h) and consists of $T_P = 80$ control cycles of 90 seconds. FTC case is simulated and results are used for the calculation of $m_1$, $m_2$ and $N_c$. After performing a trial-evaluation test as described in section \ref{sss:nodeSelection} for the medium demand scenario, the best performing values are $\alpha = 0.6$, $\beta = -1.8$ and $\gamma = -1$, and selection is done starting from nodes with lowest $R$. In this way, the algorithm prioritizes selection of nodes with relatively high queue length variance and spill-back occurrence during peak time but with moderate mean queue lengths. In figure~\ref{fig:nodeSelection_med} the node selection process for the case of medium demand is pictured  for penetration rates $15\%$, $20\%$ and $25\%$, where dots represent all network signalized nodes. First column graphs (a, d, and g) show the relation between $m_1$ and $m_2$, while second column graphs (b, e and h) show the relation between $N_c$ and $m_2$, all calculated based on simulation results of FTC case. Blue dots represent nodes that are selected to receive MP controller according to the proposed algorithm. Third column graphs (c, f and i) visualize the spatial distribution of the selected MP nodes, depicted as red dots. 

\begin{figure}[tb]%
	\centering
   \includegraphics[scale=0.75]{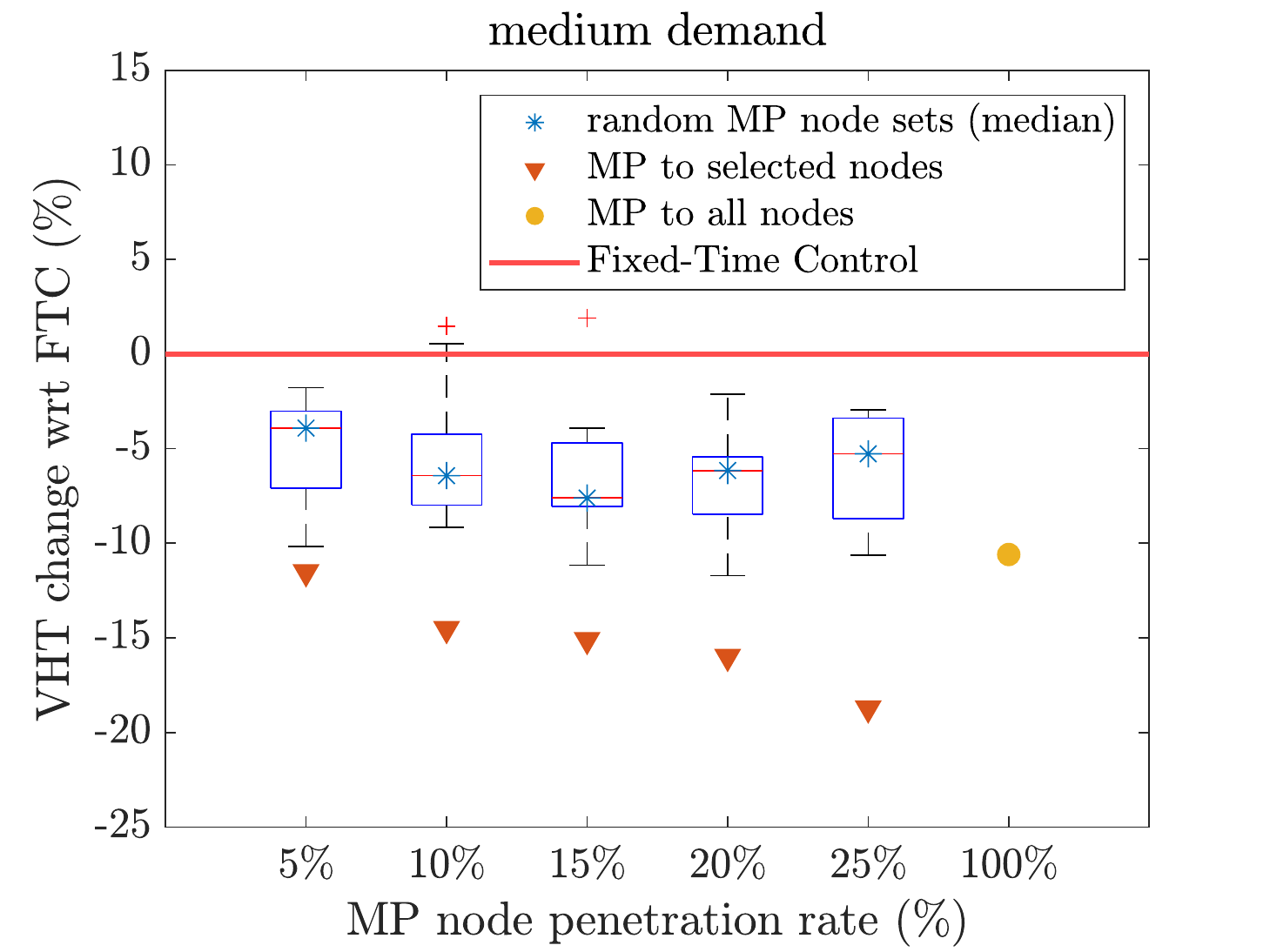}
    \caption{Comparison of Total Travel Time improvement, with respect to FTC  scenario, of single Max Pressure application, for medium demand scenario. Boxplots refer to 10 randomly created MP node sets for every penetration rate, red triangles refer to node selection based on the proposed method and yellow dot represents the full-network implementation.}%
    \label{fig:boxplotsMP_TTT_med}%
\end{figure}

The performance of single MP network control for the medium demand scenario, for different fractions of MP controlled nodes, as well as for the standard full network implementation (penetration rate of 100\%), is shown in figure~\ref{fig:boxplotsMP_TTT_med}. On the vertical axis the percentile improvement of total travel time (vehicle-hours travelled or VHT) with respect to FTC scenario is shown. Each boxplot refers to 10 cases of random selection of MP controlled intersections, corresponding to the respective penetration rate. The case of 100\% rate refers to full MP network implementation. Red triangles represent scenarios where controlled nodes are selected according to the method described in section~\ref{sss:nodeSelection}. Firstly, we observe that for the case of medium demand, almost all MP scenarios lead to improved total travel time, even those with randomly selected MP nodes. However, most cases of node selection made by the proposed method significantly outperform random assignment. In fact, we notice that the higher the number of controlled nodes, the larger the difference between random and targeted selection performance. These observations indicate that the proposed selection process is successful in identifying critical intersections for MP control. Interestingly, the case of installing MP regulator to all network nodes leads to smaller improvement than those including only a fraction of controlled nodes according to the proposed algorithm. More specifically, with only 10\% of critical nodes the system travel time improves by 14.5\% and with 25\% of critical nodes, it improves by 18.8\%, while in the case of controlling all nodes, it improves only by 10.6\%. This remark indicates not only significant cost reduction can be achieved by reducing the number of controlled nodes through the proposed selection process, but also system performance can be increased. However, this behaviour is observed for moderate demand, where the network does not reach highly congested states in the FTC case.

\begin{figure}[tb]%
	\centering
    \subfloat[]{\includegraphics[scale=0.33]{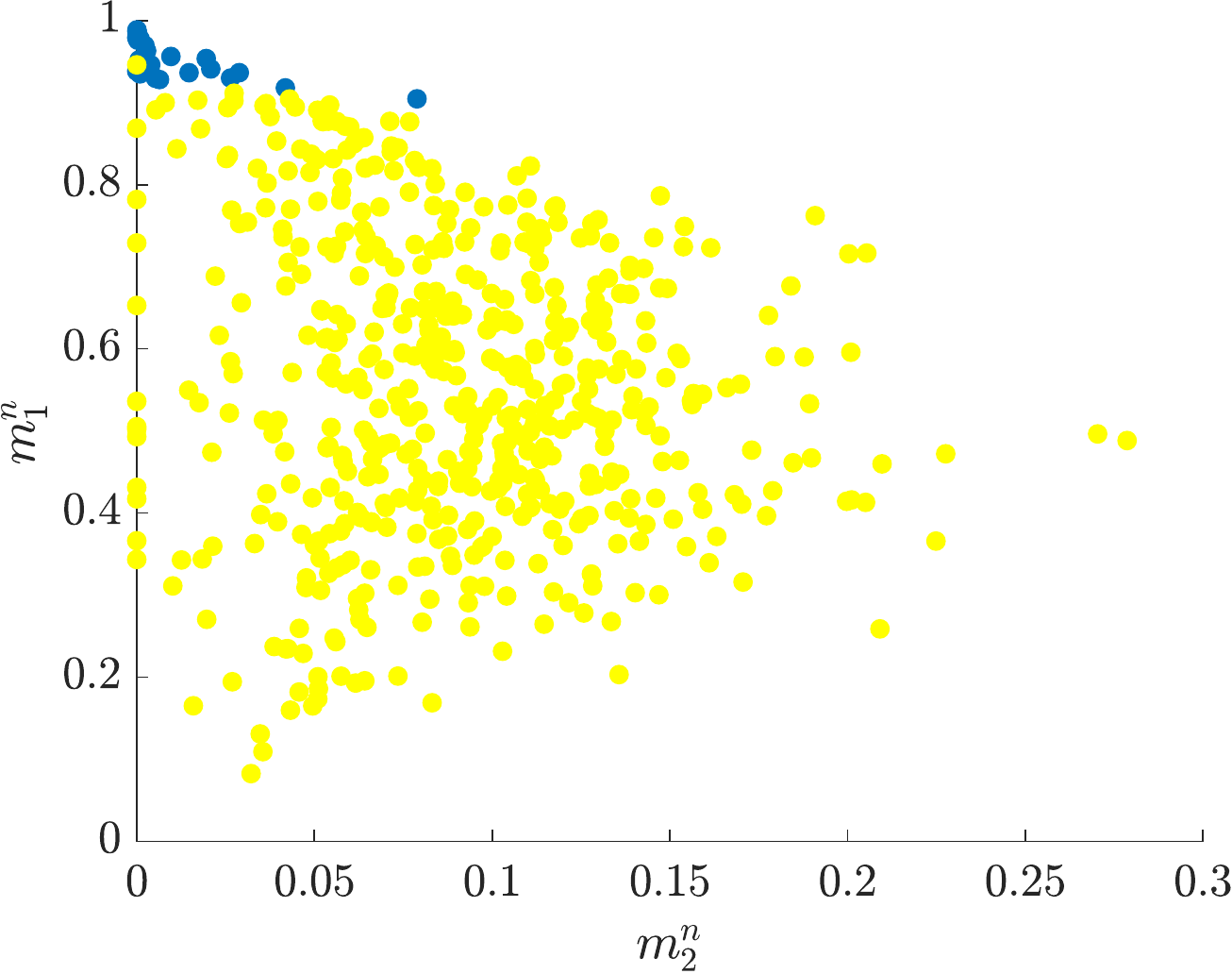}}%
    \qquad
    \subfloat[]{\includegraphics[scale=0.33]{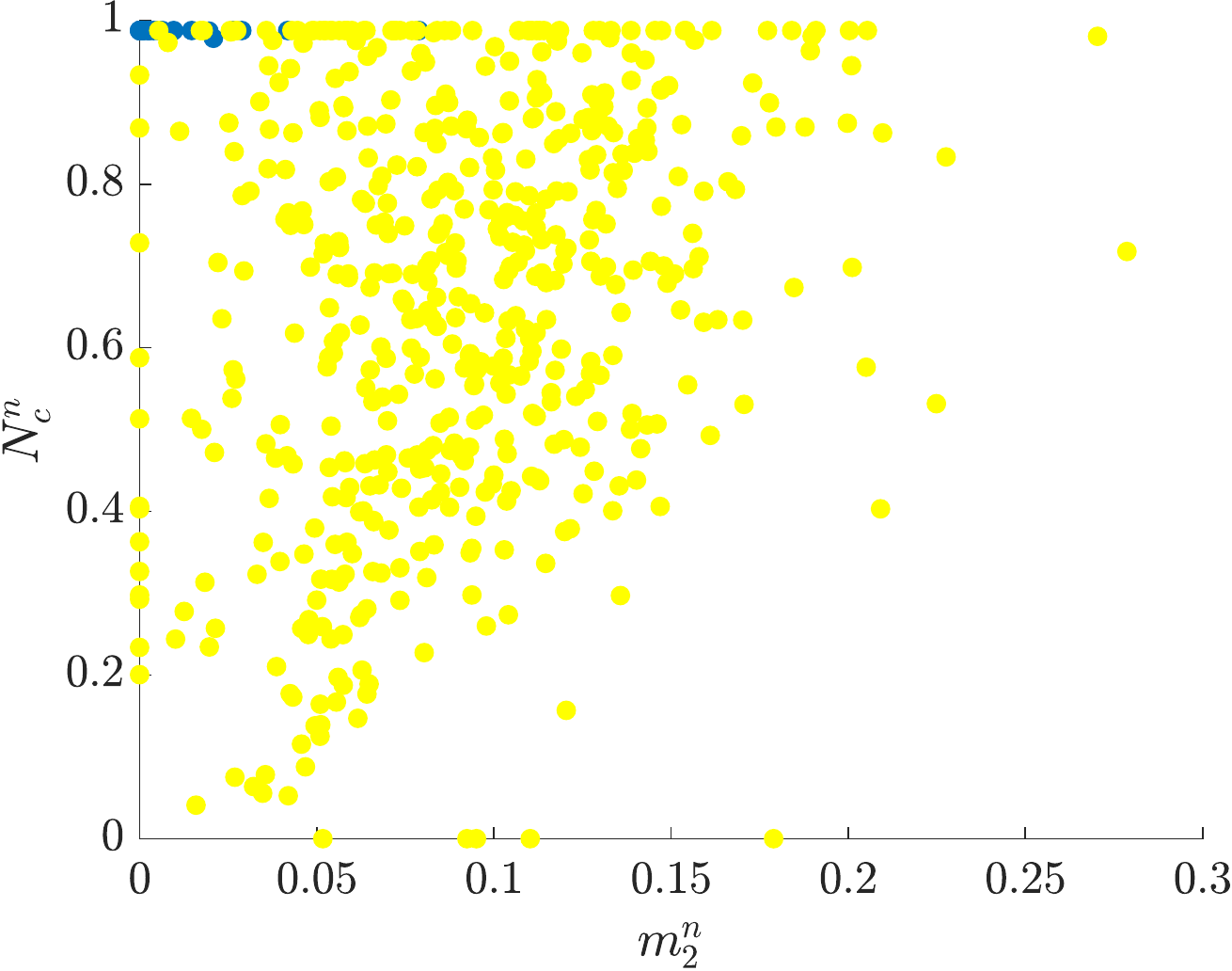}}%
    \qquad
     \subfloat[]{\includegraphics[scale=0.33]{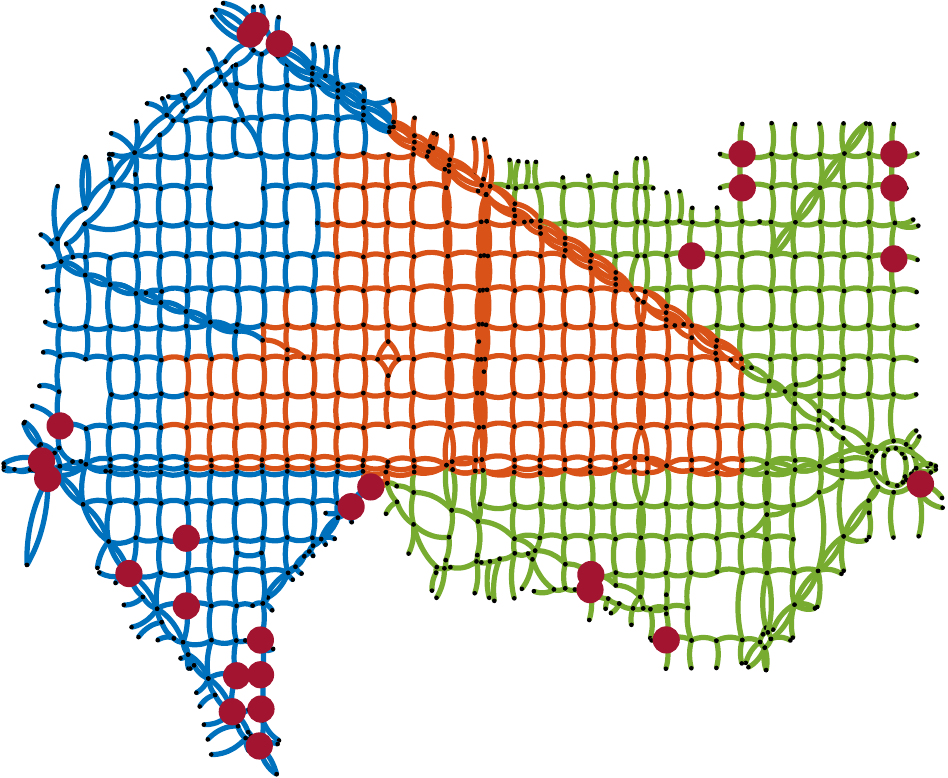}}%
    \qquad
    \subfloat[]{\includegraphics[scale=0.33]{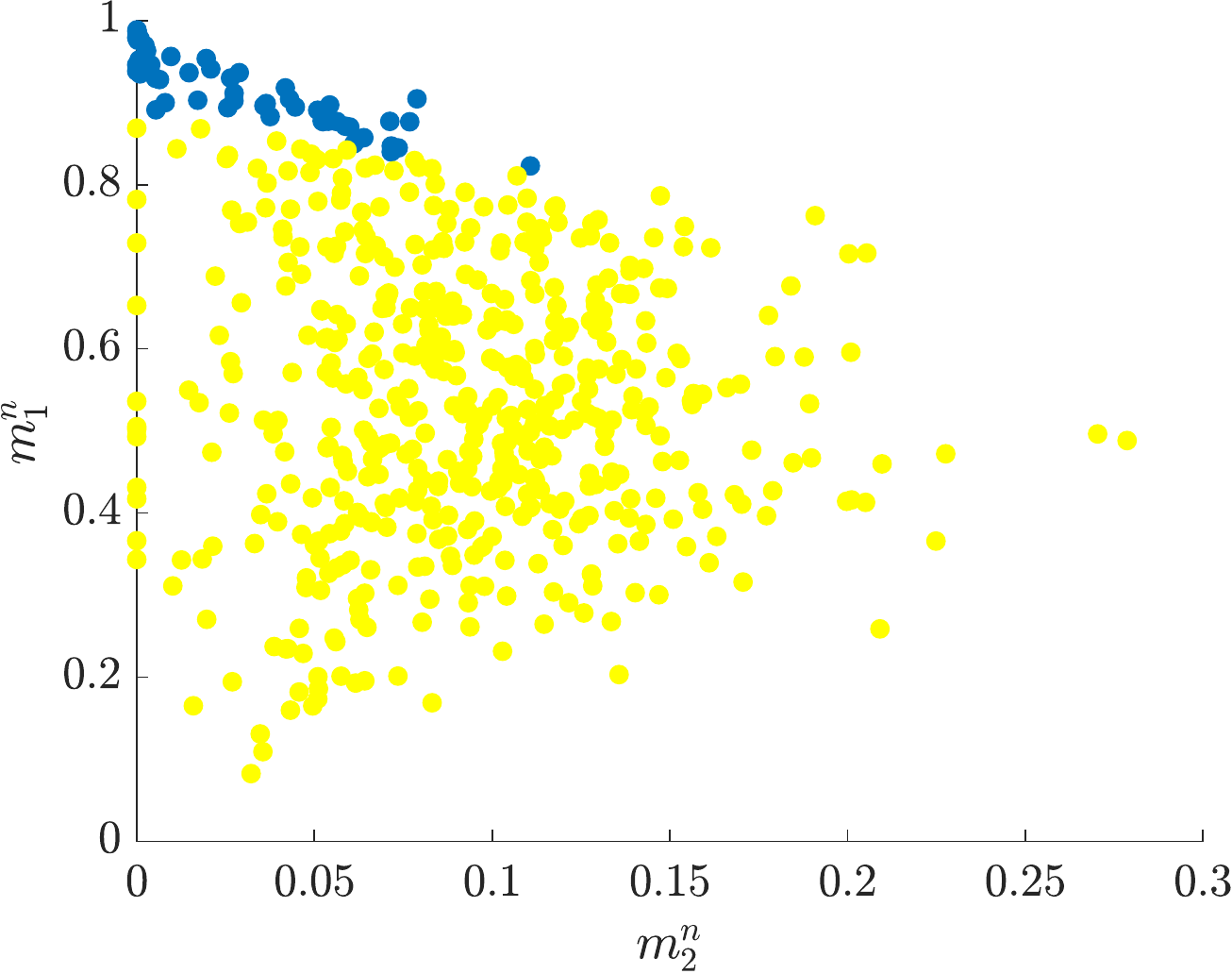}}%
        \qquad
    \subfloat[]{\includegraphics[scale=0.33]{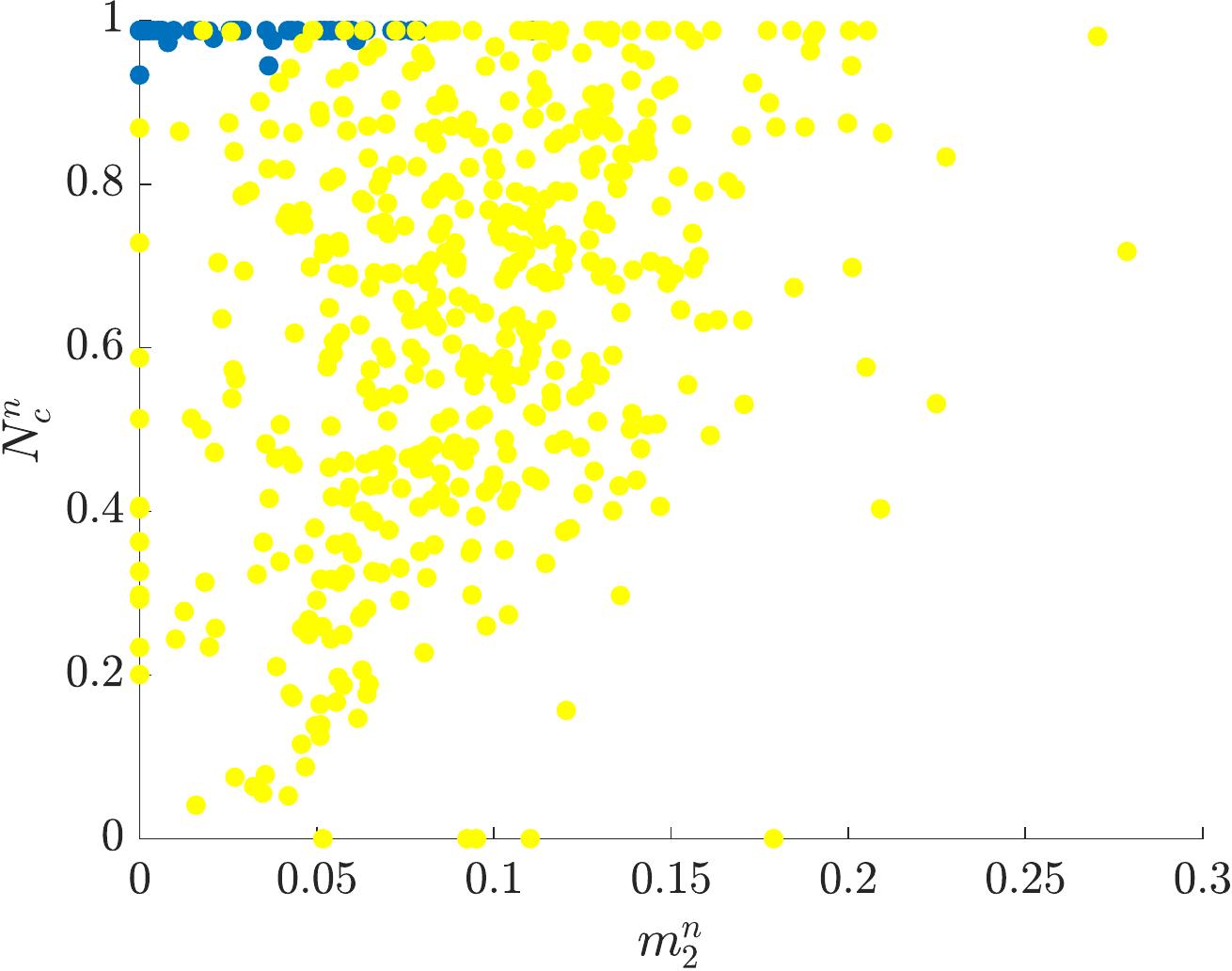}}%
        \qquad
    \subfloat[]{\includegraphics[scale=0.33]{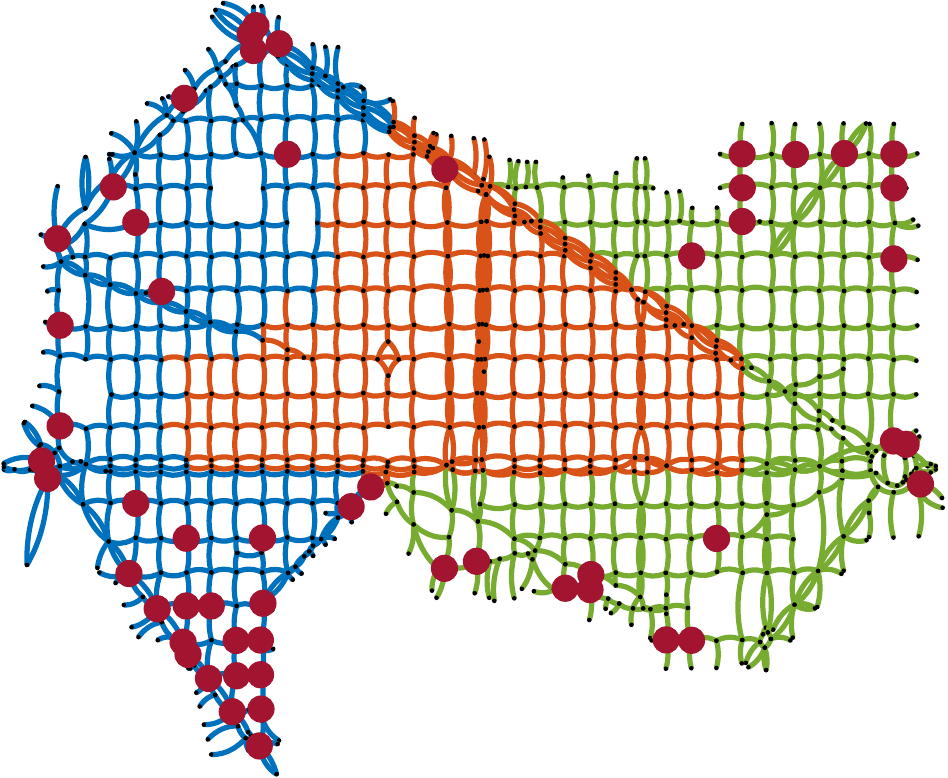}}%
    \qquad
     \subfloat[]{\includegraphics[scale=0.33]{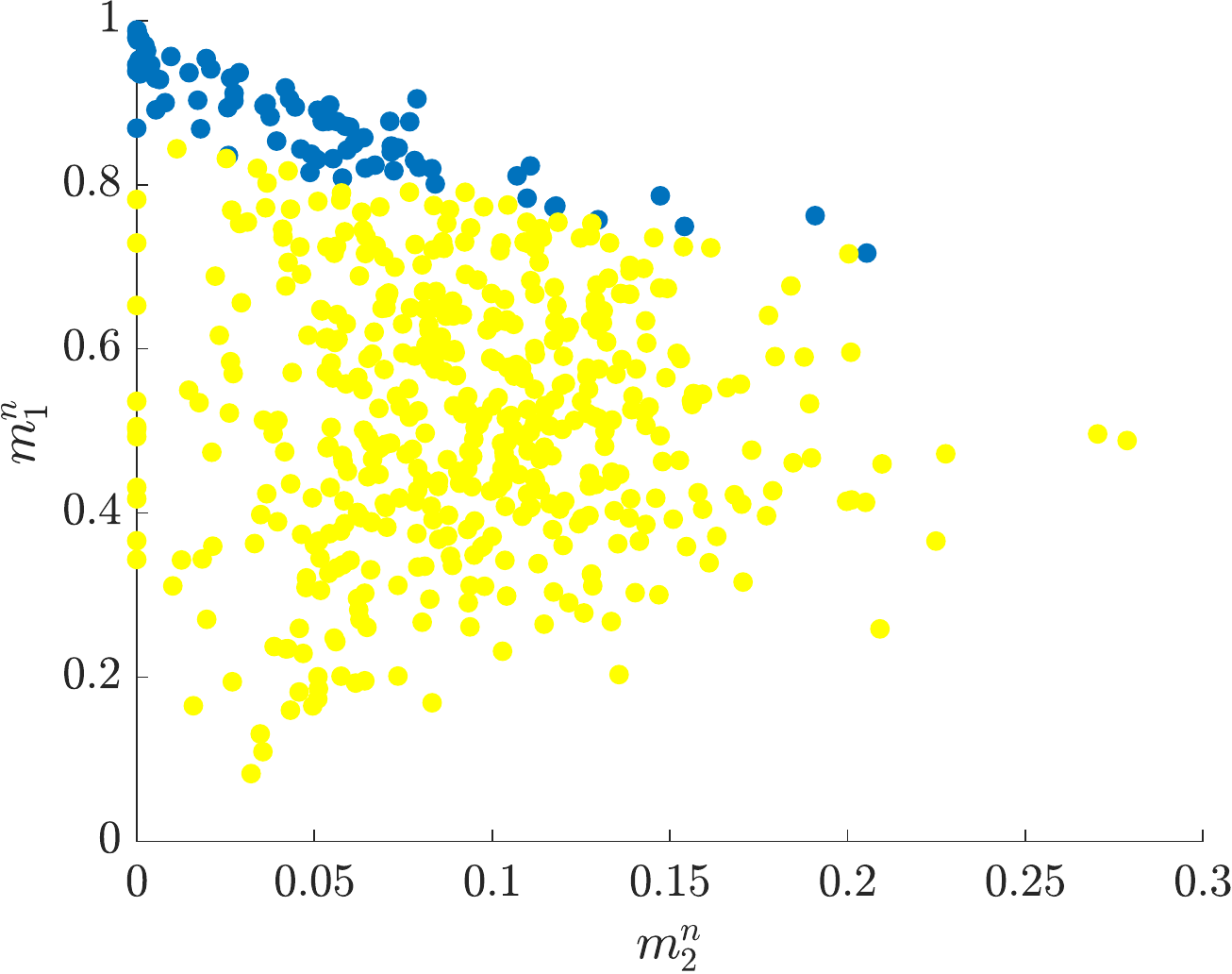}}%
    \qquad
    \subfloat[]{\includegraphics[scale=0.33]{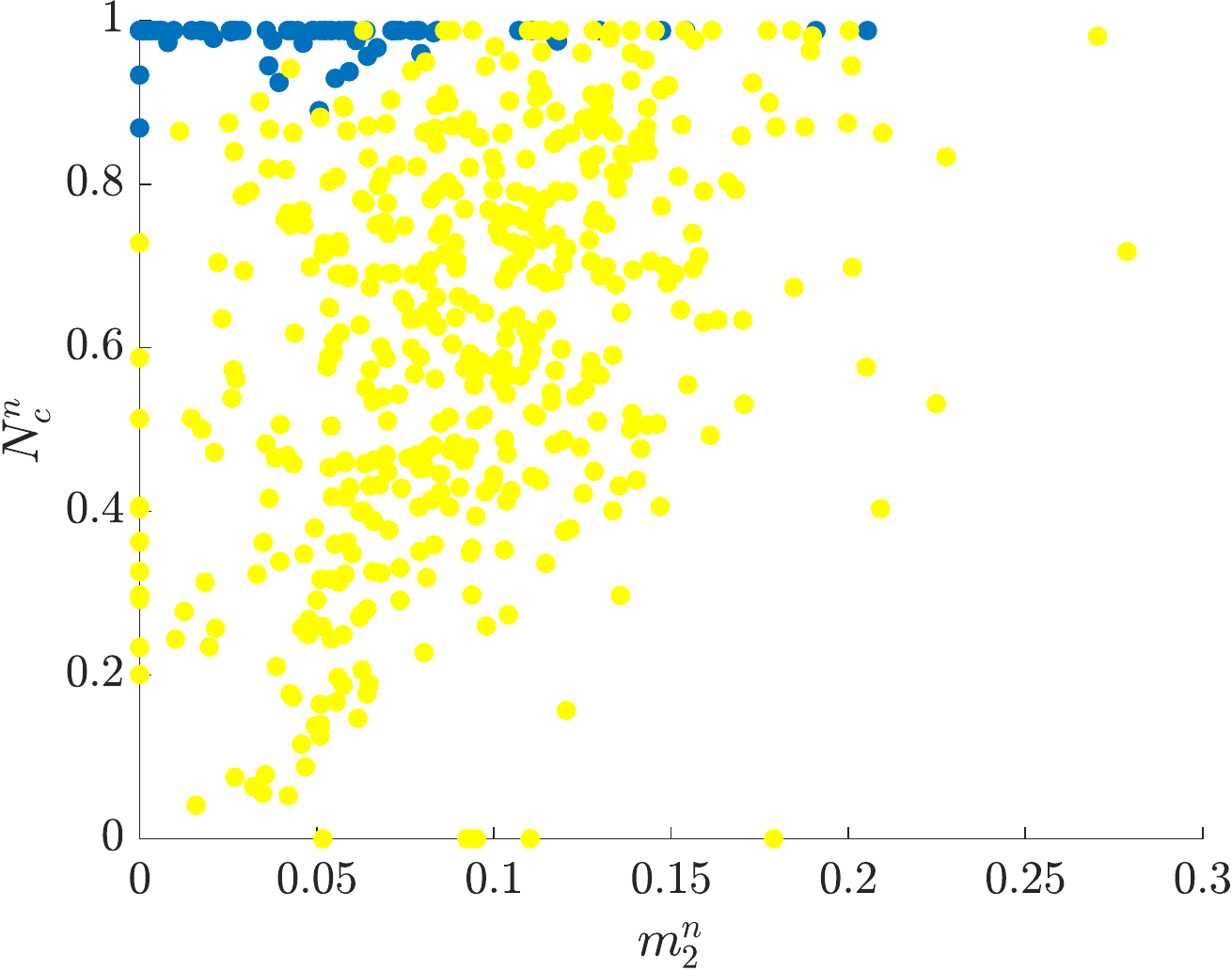}}%
        \qquad
    \subfloat[]{\includegraphics[scale=0.33]{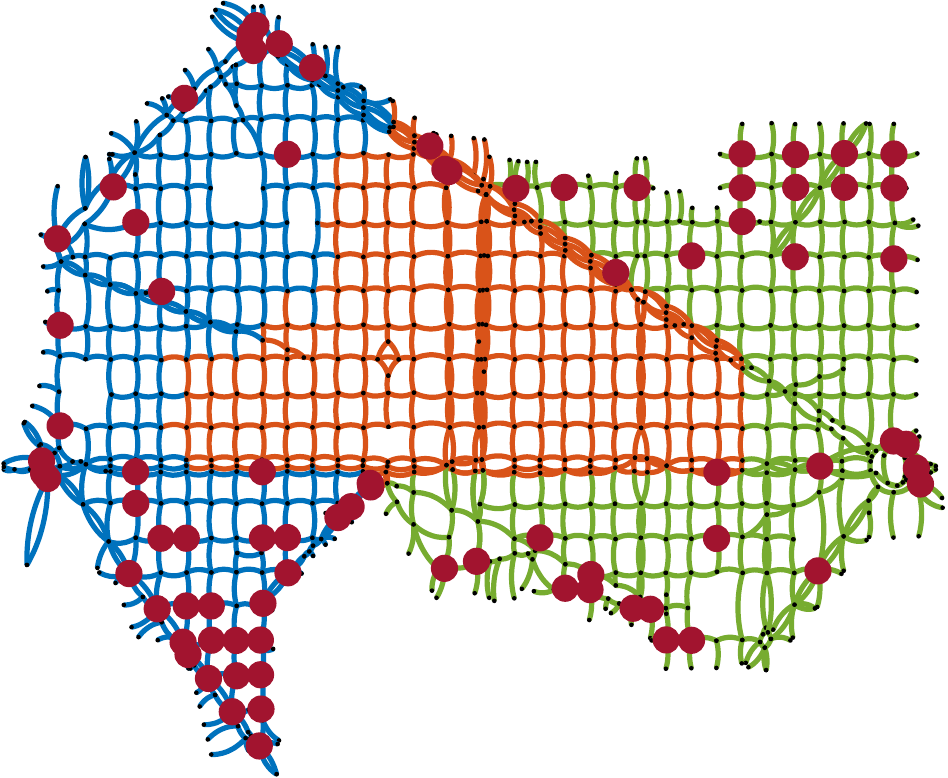}}%
    \caption{Visualization of the MP node selection process for the case of high demand, according to the proposed method. Each row refers to different penetration rate (5\%, 10\% and 15\%, from top to bottom). First and second column graphs show relations between selection variables $m_2$ - $m_1$ and $m_2$ - $N_c$, respectively, for all network nodes, for the benchmark case of FTC. Blue dots represent the selected nodes for MP control. Third column figures show the plan of the studied network, partitioned in 3 regions, with the spatial distribution of the selected MP intersections.}%
    \label{fig:nodeSelection_high}%
\end{figure}

\begin{figure}[tb]%
	\centering
     \includegraphics[scale=0.75]{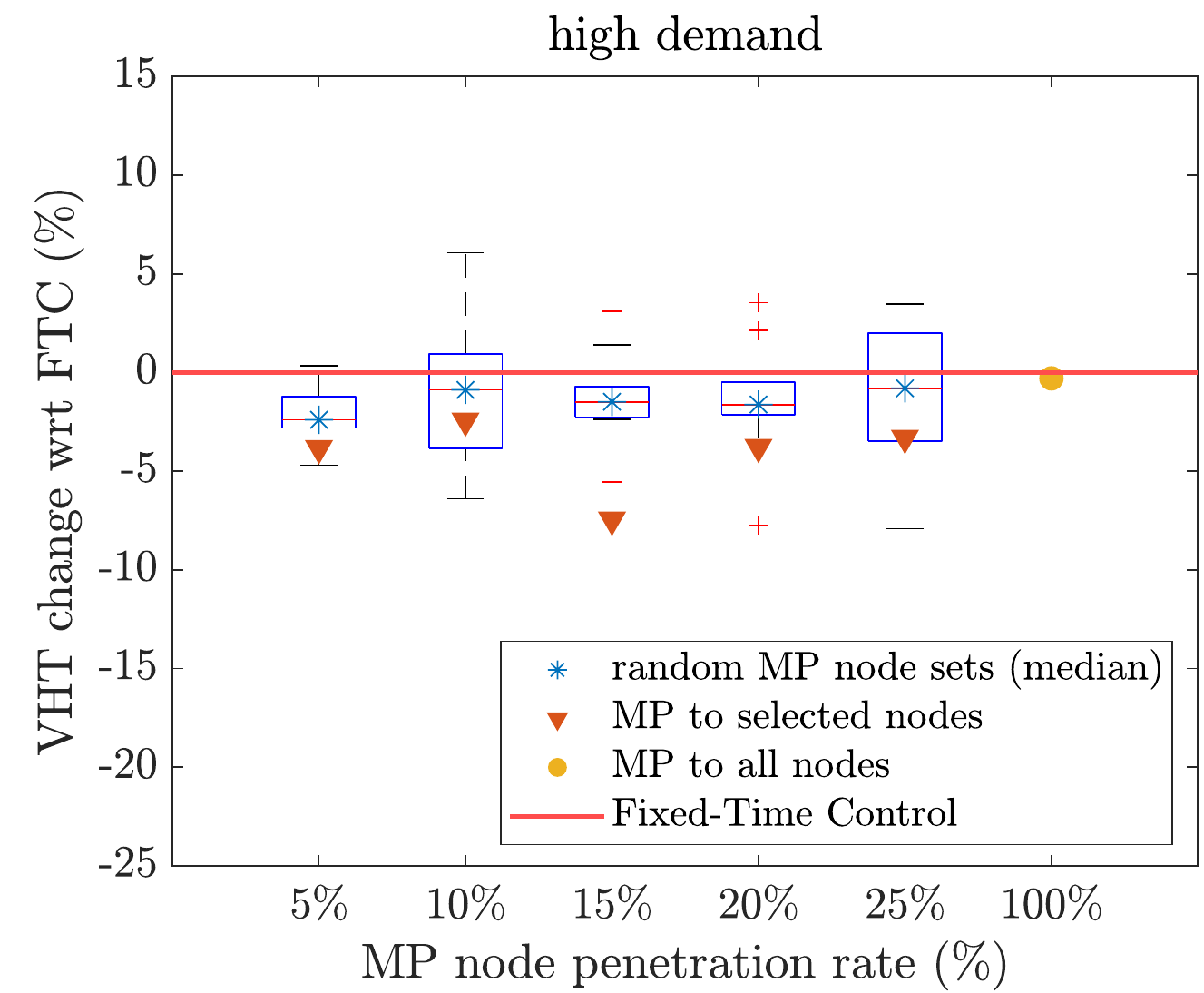}
    \caption{Comparison of Total Travel Time improvement, with respect to FTC  scenario, for single Max Pressure application for high demand scenario.  Boxplots refer to 10 randomly created MP node sets for every penetration rate, red triangles refer to node selection based on the proposed method and yellow dot represents the full-network implementation.}%
    \label{fig:boxplotsMP_TTT_high}%
\end{figure}

A different behaviour is observed in the case of high travel demand, both in the node selection pattern and in the corresponding network performance. Firstly, we observed that using the same values for parameters $\alpha, \beta, \gamma$ as in moderate demand case and following the same selection pattern leads to poor system performance, very similar to the one of random selection. Therefore, parameter optimization is done again, by performing a new trial-evaluation test, specifically for the high demand scenario. The new values found are $\alpha = -0.72$, $\beta = -0.4$, $\gamma = -0.2$. In this case, we observe a change of sign in $\alpha$, which directs the selection towards nodes with high mean queues ($m_1$) while high queue variance ($m_2$) and spill-back duration ($N_v$) are given lower weights. In other words, in this case, the best found selection pattern prioritizes nodes with high mean queues. However, as we can see in figure~\ref{fig:nodeSelection_high} which is made based on FTC results of the high demand scenario, there is correlation between $m_1$, $m_2$ and $N_c$. For instance, in the high demand scenario, a significant number of nodes experience complete gridlock ($m_1$ values close to 1), which corresponds to zero queue variance ($m_2$) and maximum gridlock duration ($N_c$ equal or close to 1). As figure~\ref{fig:nodeSelection_high} shows, the selection process in this case prioritizes highly congested/gridlocked nodes for MP selection and, as we will see, leads to significant performance gains. Node selection is visualized for penetration rates of 5\%, 10\% and 15\%, in the same format as in figure~\ref{fig:nodeSelection_med} for medium demand. 

Regarding system performance of single MP in high demand scenario, for which the network reaches more congested states in the FTC case, results show relatively smaller improvement with respect to FTC than in the case of medium demand. This can be seen in figure~\ref{fig:boxplotsMP_TTT_high}, where percentile performance improvement with respect to FTC is shown on the vertical axis, for different MP node penetration rates. Again, boxplots refer to 10 cases of randomly selected node sets for MP control per penetration rate. Firstly, it is interesting that case of full-network MP control (yellow dot) leads to practically zero improvement compared to FTC. However, smaller MP penetration rates in best case result in improvement between 3\% and 7\%. The effectiveness of node selection method seems to drop, even with re-optimized parameters of function $R$, since we observe a few random sets performing better than the ones of the proposed method. Performance of targeted selection is always better than the median of the random set though, and in the case of 15\%, for which parameter optimization was performed, performance is significantly better than random cases. Based on these remarks, we can infer that partial MP implementation, except for being less costly, can also lead to improved performance, especially in highly congested networks with single MP control, which implies that significant spatial correlation exist between performance of MP controlled nodes that can act detrimentally to system performance. In other words, even if performance generally improves in the proximity of a controlled intersection, adding MP controllers to all network intersections does not necessarily mean that system performance will globally improve as a result. Hence, optimizing spatial node distribution for MP control requires further investigation. 

\begin{figure}[tb]%
\centering
\includegraphics[width=0.9\textwidth]{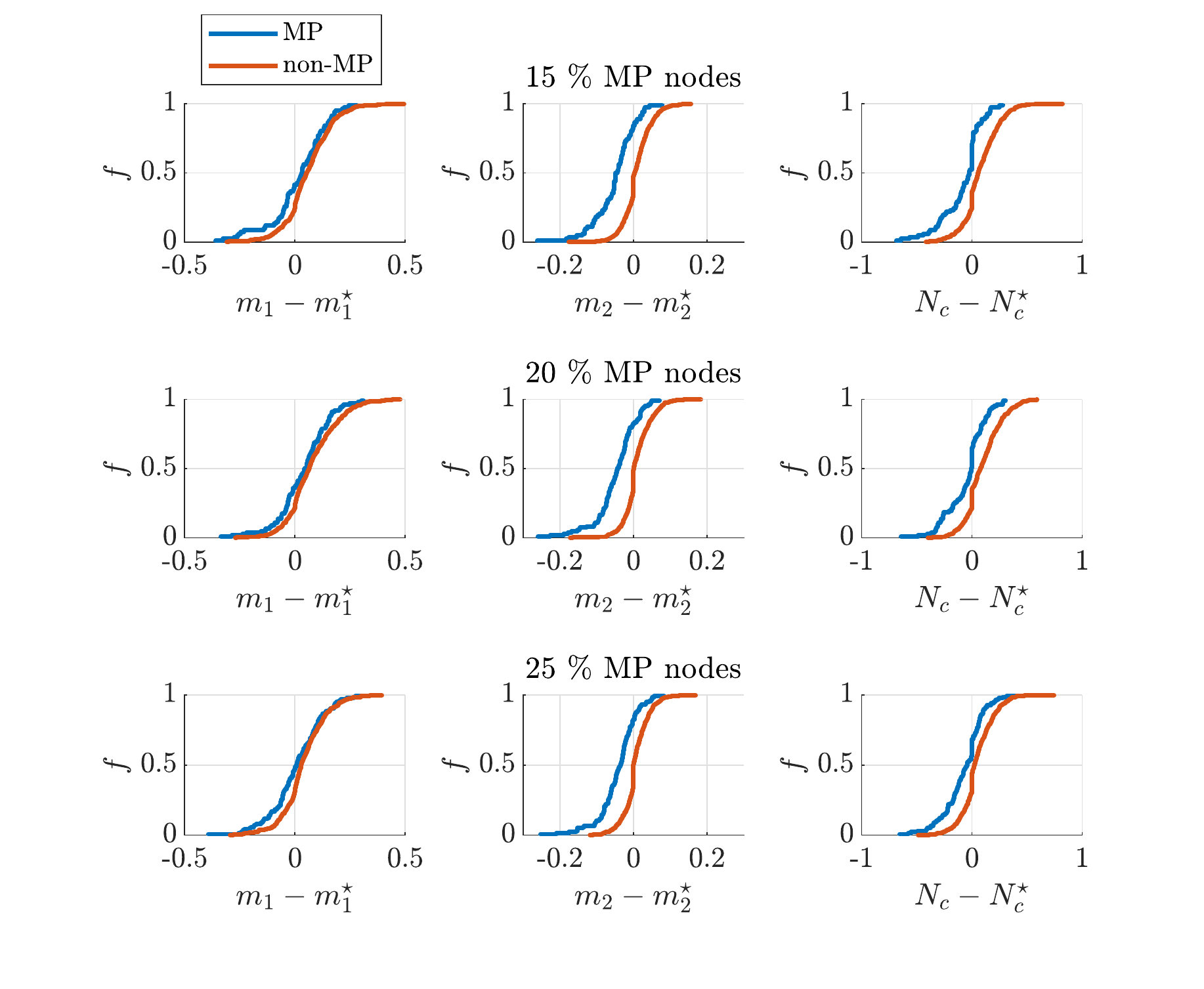}%
\caption{Cumulative distribution functions for the changes in node selection criteria $m_1$, $m_2$, $N_c$ after MP control implementation for medium demand scenario. Start superscript refers to the FTC case. Blue lines correspond to MP nodes and red lines to the rest. Each row represents a different MP node penetration rate (15\%, 20\% and 25\%).}\label{fig:CDFs_med}%
\end{figure}

\begin{figure}[tb]%
\centering
\includegraphics[width=0.9\textwidth]{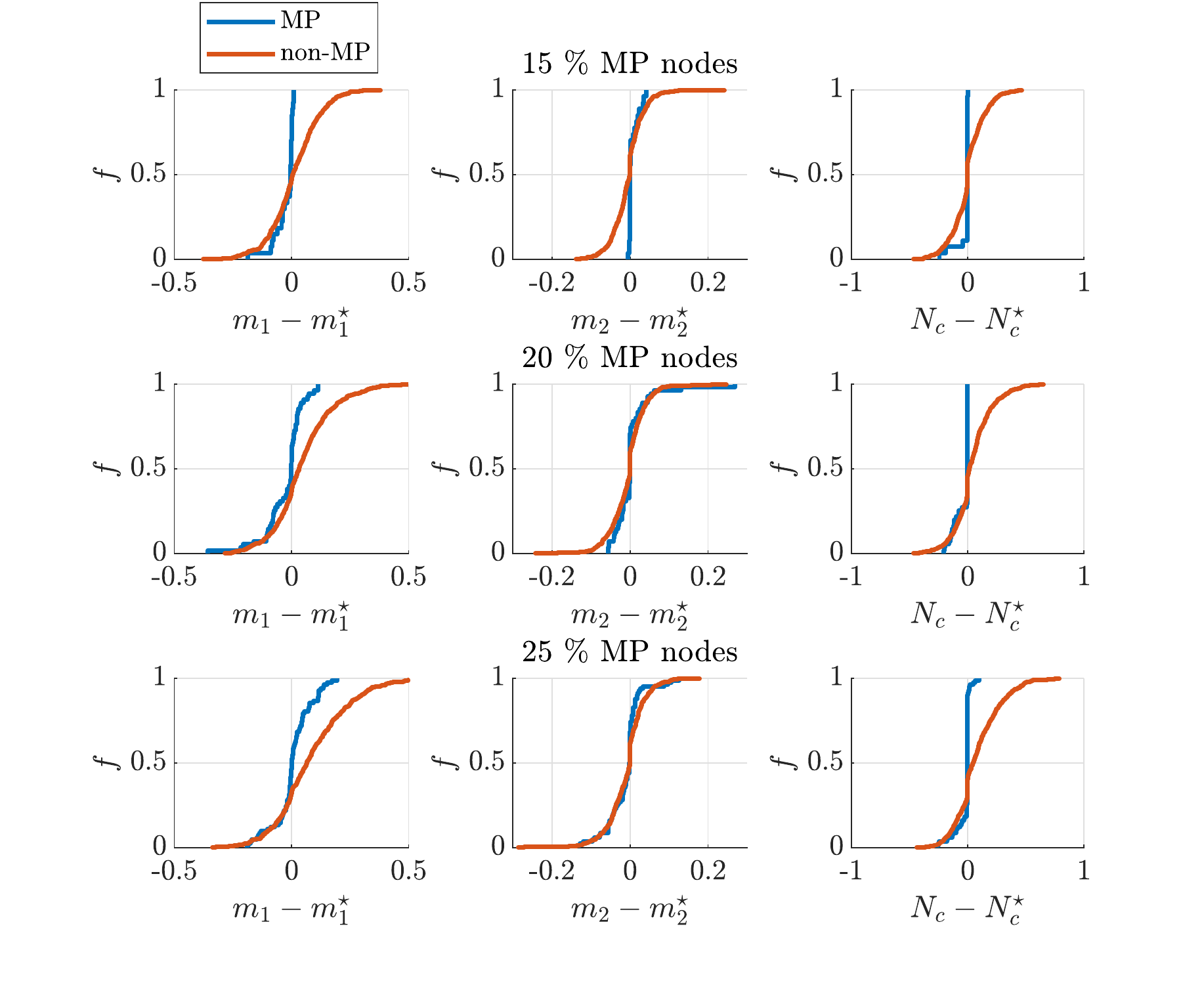}%
\caption{Cumulative distribution functions for the changes in node selection criteria $m_1$, $m_2$, $N_c$ after MP control implementation for high demand scenario. Start superscript refers to the FTC case. Blue lines correspond to to MP nodes and red lines to the rest. Each row represents a different MP node penetration rate (15\%, 20\% and 25\%).}\label{fig:CDFs_high}%
\end{figure}

In the scope of investigating in detail the effectiveness of partial MP control in targeted node sets, for both demand scenarios, we plot cumulative distribution functions of selection variable $m_1$, $m_2$ and $N_c$ changes, after implementing MP control, with respect to their values in FTC, for three MP node sets selected by the proposed method (15\%, 20\% and 25\%), for medium demand case in figure~\ref{fig:CDFs_med} and for high demand case in figure~\ref{fig:CDFs_high}. Blue lines correspond to MP node sets while red lines to the remaining nodes, where FTC applies. First, second and third column figures show changes in values of $m_1$, $m_2$ and $N_c$, respectively, where $m_1^*$, $m_2^*$ and $N_c^*$ refer to the FTC case, before MP control application. In medium demand, it is evident from second column graphs of figure~\ref{fig:CDFs_med} that queue variance $m_2$ was decreased for 80\% to 90\% of nodes that received MP controller according to the proposed algorithm, for penetration rates of 15\% to 25\%, while for the remaining nodes, it was reduces for half and increased for the other half. Also, reduction to the non-MP nodes was smaller than the one to MP nodes. Moreover, spill-back duration during peak time $n_c$ was also decreased in the majority of selected MP nodes in all three cases shown, though mean queue length was mostly increased. The increase in mean queue length in cases of both MP and non-MP nodes is justified, as a result of more efficient road space use that is achieved due to MP control, leading to more vehicles entering the network at the same time and increasing node occupancy. Different pattern of improvement is observed in the case of high demand as shown in figure~\ref{fig:CDFs_high}, where we observe a significant decrease in spill-back occurrence $N_c$ in almost all MP nodes (blue curve in third column graphs), and a decrease in mean queue length $m_1$, while no significant difference can be seen between MP and non-MP nodes in terms of queue variance $m_2$. This behaviour can be explained with reference to the node selection pattern of the high demand case, where MP is assigned by priority to capacitated nodes with very high queues, where by definition variance is very low and spill-back occurrence very high. Therefore, in both cases selected MP nodes are improved with respect to either spill-back occurrence and either queue variance or queue length, in moderate and high congestion scenarios, respectively.

\subsection{Two-layer framework: Perimeter control combined with Max Pressure}

\begin{figure}[tb]%
	\centering
    \subfloat[]{\includegraphics[scale=0.55]{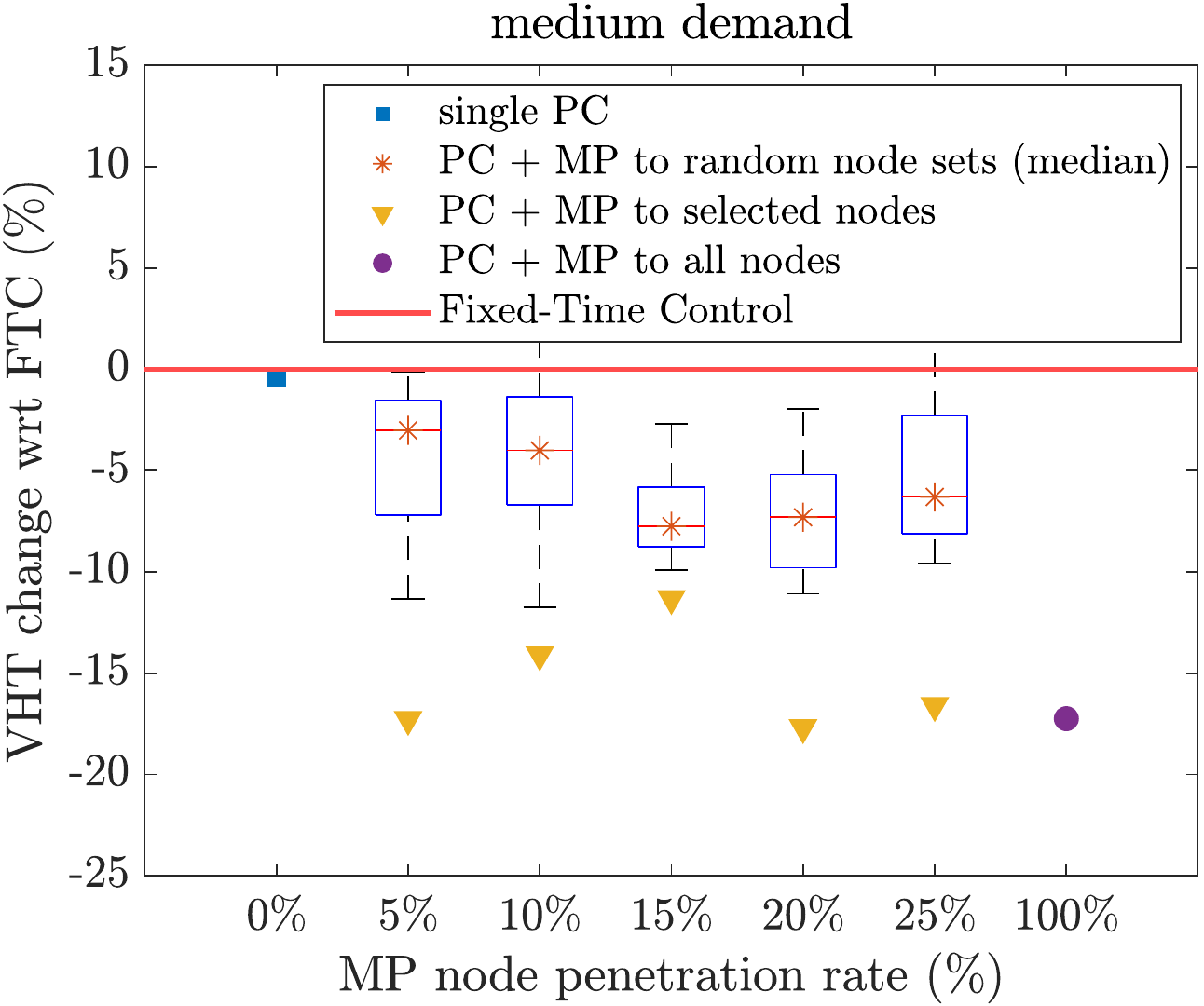}}%
    \qquad
    \subfloat[]{\includegraphics[scale=0.55]{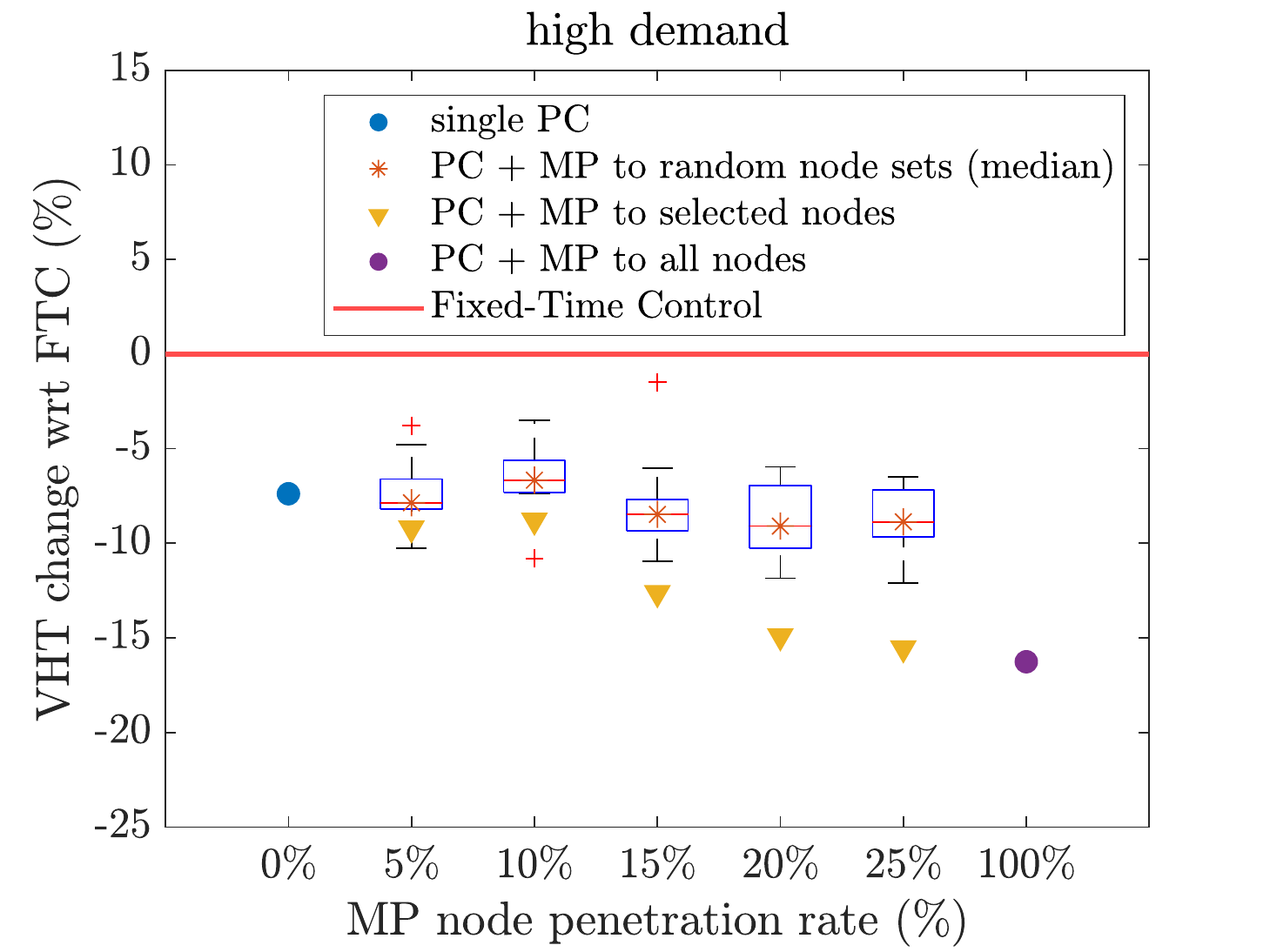}}%
    \caption{Comparison of Total Travel Time improvement, with respect to FTC scenario, for single PC and combined PC plus MP implementation, for (a) medium, and (b) high demand. Graphs show the performance for different MP node penetration rates (0\% is single PC) for the two-layer PC+MP framework. Boxplots refer to 10 random MP node selections per rate while yellow triangles refer to node selection based on the proposed method.}%
    \label{fig:boxplotsPC_TTT}%
\end{figure}

Results of the two-layer framework combining PC with MP schemes are discussed in this section. Simulation results in terms of performance improvement with respect to FTC case are shown in figure \ref{fig:boxplotsPC_TTT}, where (a) refers to medium and (b) to high demand. The case of 0\% penetration rate (square) corresponds to typical PC application without any MP control (for comparison), while the remaining cases refer to combined control of PC and MP in different node penetration rates. Again, boxplots aggregate results of 10 scenarios of random MP node selection of the corresponding penetration rate, combined with the same PC scheme. Triangles refer to the combined scheme where MP nodes are selected according to the proposed methodology, while the case of 100\% (dot) refers to the combined scheme with MP installed in all nodes. Selection of MP nodes is based on FTC results and is done with the same parameter values $\alpha, \beta, \gamma$ as in single MP, for every demand scenario respectively, while PI controller parameters are the same for all cases of the same demand scenario. 

For medium demand, we observe that single PC does not lead to considerable improvement with respect to FTC (only 0.5\%). This is not surprising, since there is not enough demand to drive the network to heavily congested states, where PC would get activated for longer and would have a higher impact, and production MFD does not drop significantly for FTC case, as we will see also in figure \ref{fig:resMPsoloBest} below. However, in all combined scheme cases, we observe significant travel time improvement with respect to the FTC case. Especially in the cases of 5\% and 20\% MP nodes selected by the proposed method, performance is slightly higher than the best performing single MP scheme of the respective penetration rate. Moreover, similar to the single MP cases, we observe that the proposed node selection algorithm leads to higher performance gains compared to random selection for most cases. The highest improvement for the two-layer controller, which is about 17.7\% , is recorded for the case of 20\% MP nodes, while both 100\% and 5\% penetration rates achieve about 17.2\%. Therefore, in multiple cases, as in 5\%, 20\% and 100\%, adding PC on top of MP increases network performance from 3\% to 7\% with respect to single MP.

\begin{table}[tb]
\centering
\caption{Total travel time for all control scenarios, for medium and high demand, in vehicle hours travelled (VHT). Under $\Delta$VHT, the percentile change of VHT with respect to FTC case is shown. For random MP node selection, the median VHT of 10 random node set replications is reported.}\label{tab:VHT}
\footnotesize
\addtolength{\tabcolsep}{-2pt}
\begin{tabular}{@{}lllllllll@{}}
\toprule
                 & \multicolumn{4}{c}{Medium demand}                                                             & \multicolumn{4}{c}{High demand}                                                               \\ \midrule
MP node selection & \multicolumn{2}{c}{Targeted} & \multicolumn{2}{c}{Random} & \multicolumn{2}{c}{Targeted} & \multicolumn{2}{c}{Random} \\ \midrule
  Control scenario               & VHT                & $\Delta$VHT(\%)                 & VHT            & $\Delta$VHT(\%)            & VHT                    & $\Delta$VHT(\%)             & VHT             & $\Delta$VHT(\%)            \\
FTC              & 221400             & -                         & -                     & -                    & 484000                 & -                     & -                     & -                    \\
MP 5\%           & 195580             & -11.7                     & 212743                & -3.9                 & 464390                 & -4.1                  & 472394                & -2.4                 \\
MP 10\%          & 189250             & -14.5                     & 207173                & -6.4                 & 471680                 & -2.5                  & 479732                & -0.9                 \\
MP 15\%          & 187860             & -15.1                     & 204558                & -7.6                 & 447360                 & -7.6        & 476813                & -1.5                 \\
MP 20\%          & 185940             & -16.0                     & 207777                & -6.2                 & 465280                 & -3.9                  & 476145                & -1.6                 \\
MP 25\%          & 179850             & -18.8           & 209726                & -5.3                 & 467480                 & -3.4                  & 480113                & -0.8                 \\
MP 100\%         & 197960             & -10.6                     & -                     & -                    & 483000                 & -0.2                  & -                     & -                    \\
PC               & 220380             & -0.5                      & -                     & -                    & 447000                 & -7.6                  & -                     & -                    \\
PC + MP 5\%       & 183000             & -17.3                     & 214725                & -3.0                 & 438540                 & -9.4                  & 445948                & -7.9                 \\
PC + MP 10\%      & 190070             & -14.2                     & 212544                & -4.0                 & 441060                 & -8.9                  & 451741                & -6.7                 \\
PC + MP 15\%      & 196190             & -11.4                     & 204242                & -7.8                 & 422550                 & -12.7                 & 442860                & -8.5                 \\
PC + MP 20\%      & 182230             & -17.7            & 205216                & -7.3                 & 411510                 & -15.0                 & 440004                & -9.1                 \\
PC + MP 25\%      & 184650             & -16.6                     & 207452                & -6.3                 & 408430       & -15.6       & 441069                & -8.9                 \\
PC + MP 100\%     & 183220             & -17.2                     & -                     & -                    & 405300        & -16.3       & -                     & -                    \\
                 &                    &                           &                       &                      &                        &                       &                       &                      \\ \bottomrule
                
\end{tabular}

\end{table}

For high demand, results of the combined scheme are more promising than single MP, as we can see in figure figure~\ref{fig:boxplotsPC_TTT}(b). While single MP application, as well as single PC, only improve traffic performance by around 8\% compared to FTC in the best case, the two-layer framework manages to improve up to 15\%, in the case of 25\% MP nodes selected by the proposed algorithm, and up to 17\% in the case of full-network MP implementation. Therefore, in high demand scenario, PC strategy can be significantly enhanced by the additional distributed MP layer, even with only a fraction of properly selected, controlled MP nodes, which leads to performance very similar to the one of 100\% controlled nodes, but with only 25\% of the respective cost. 
Detailed performance of all evaluated controlled schemes can be found in Table \ref{tab:VHT}. 

\begin{figure}[tb]%
	\centering
    \subfloat[]{\includegraphics[scale=0.5]{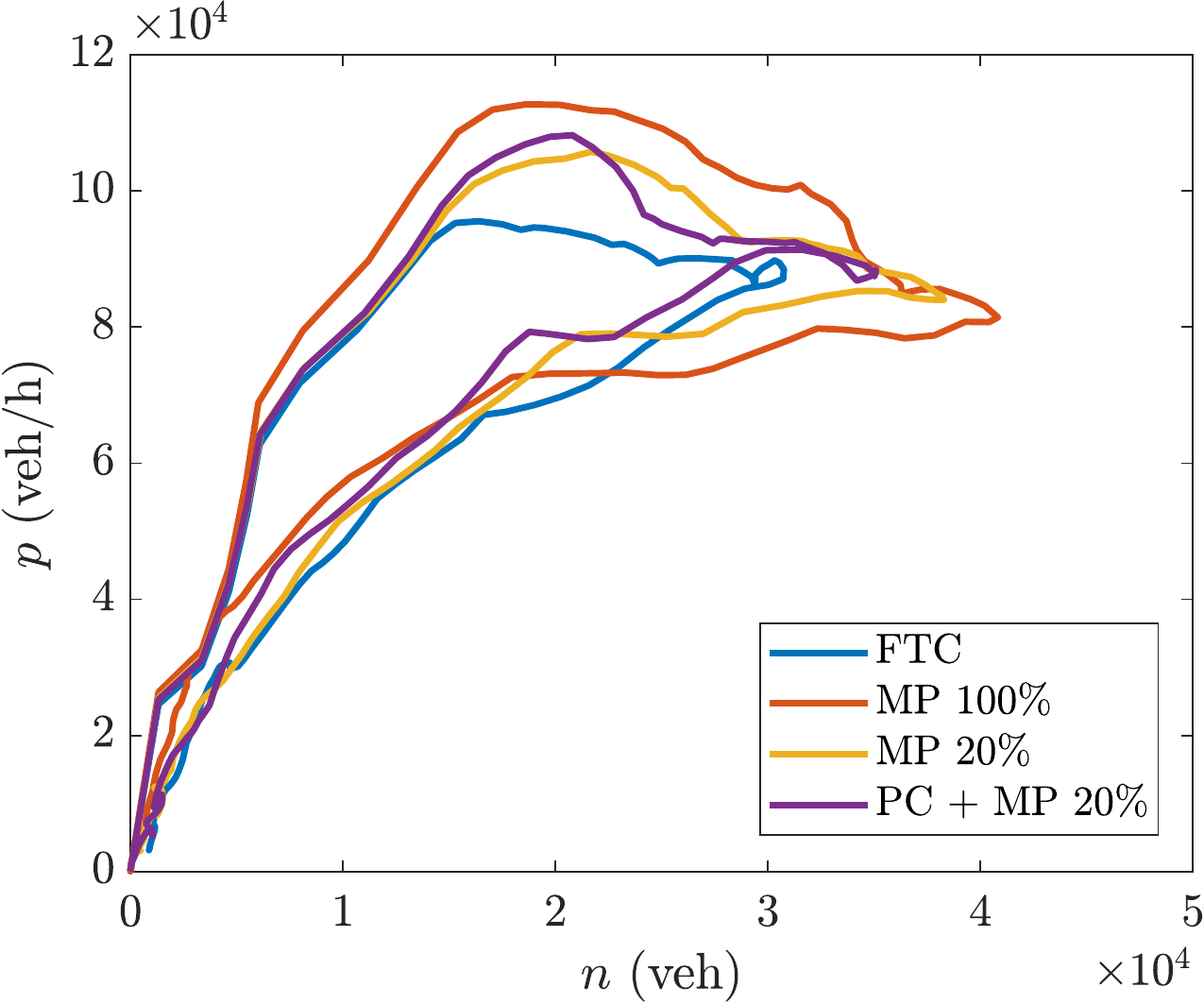}}%
    \qquad
    \subfloat[]{\includegraphics[scale=0.5]{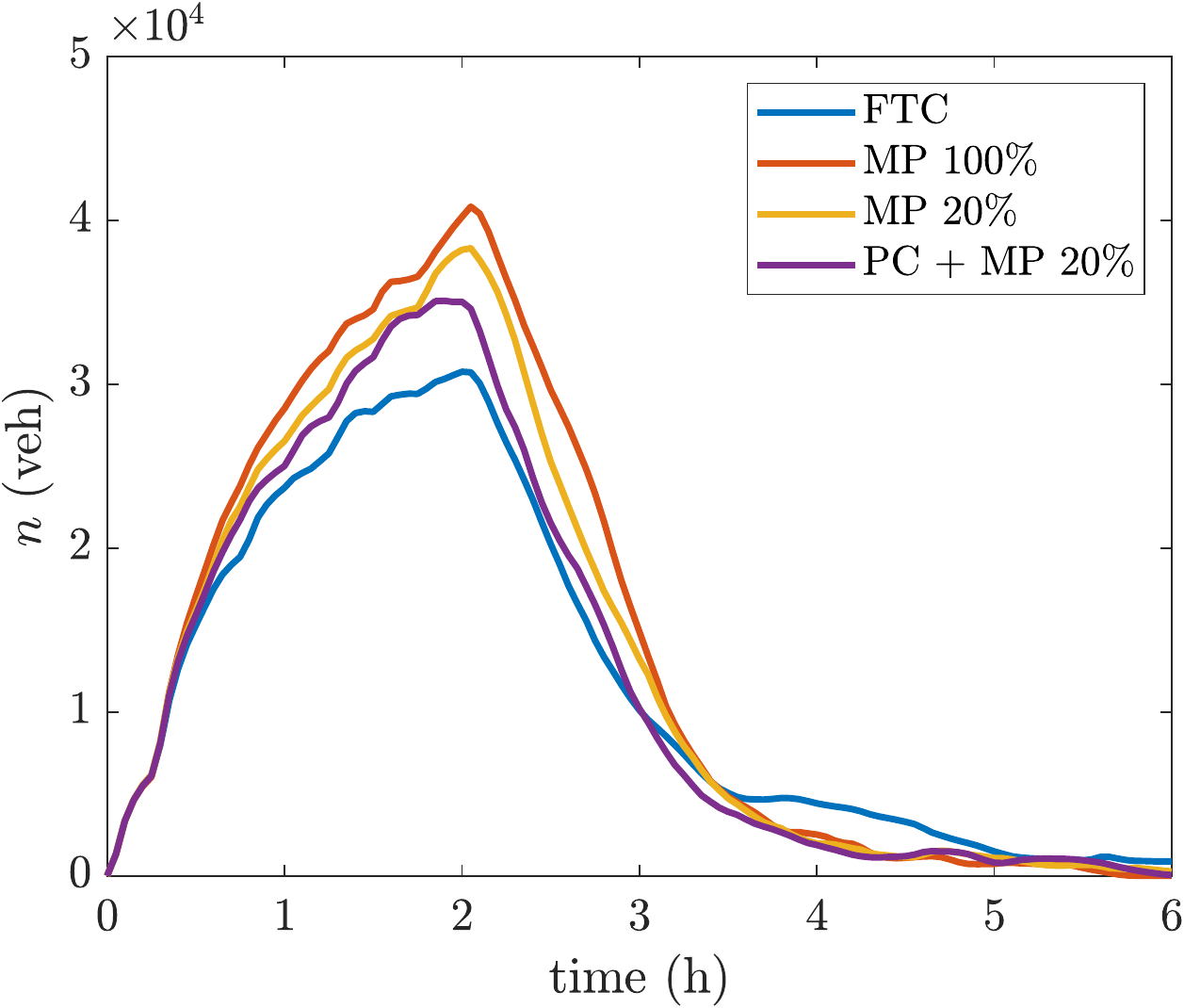}}%
    \qquad
     \subfloat[]{\includegraphics[scale=0.5]{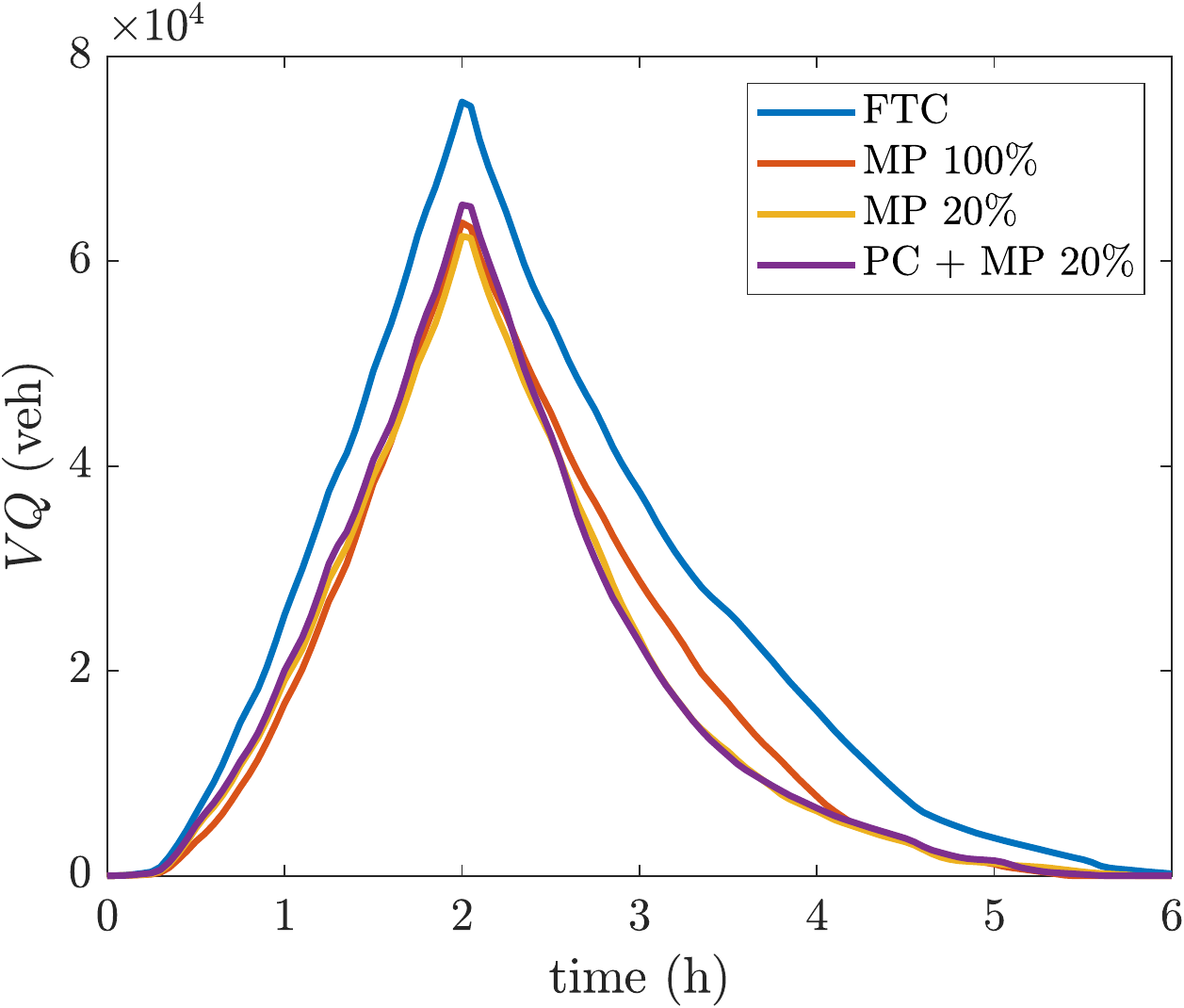}}%
    \qquad
    \subfloat[]{\includegraphics[scale=0.5]{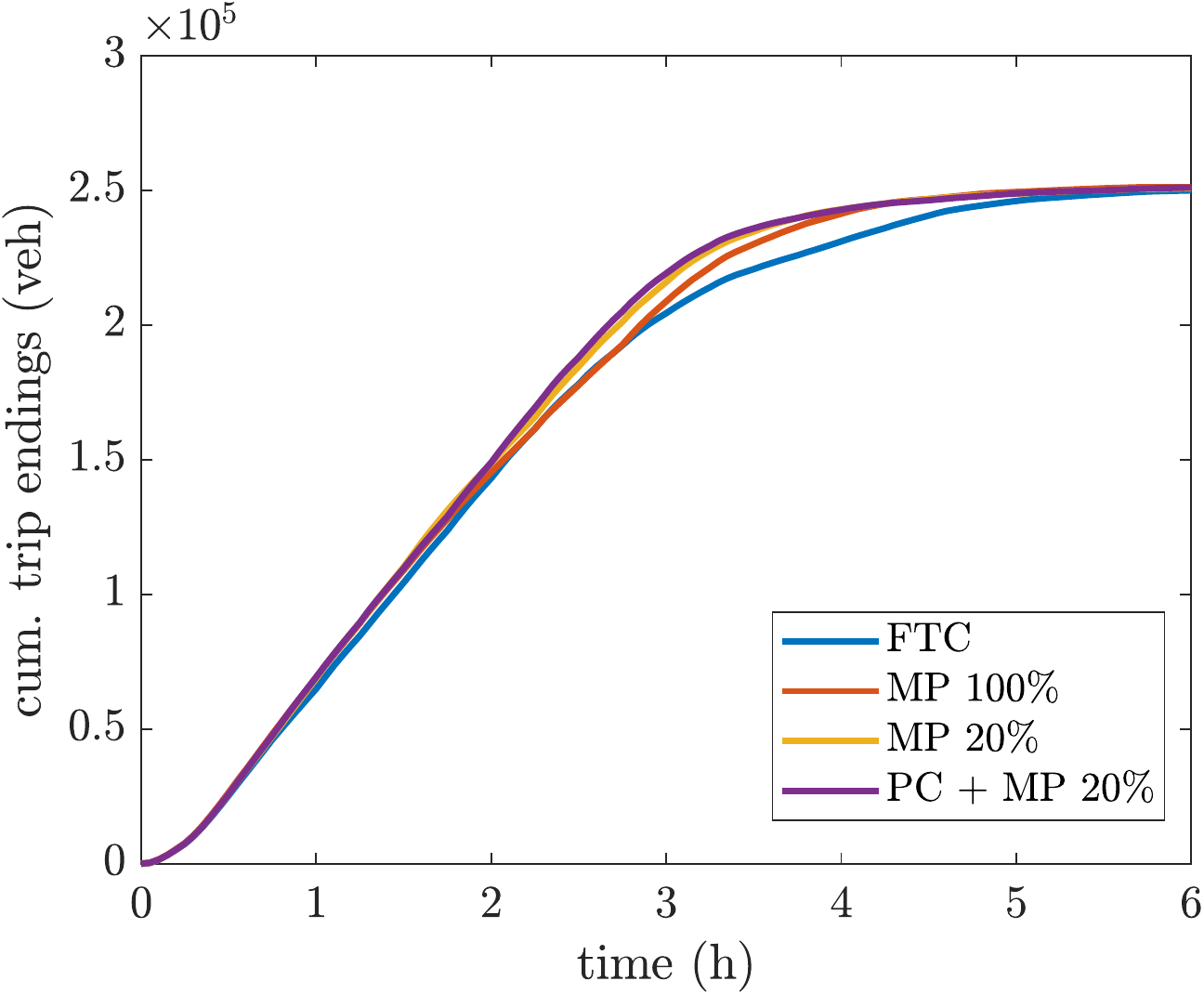}}%
    \caption{Simulation results for medium demand scenario. Comparison between FTC, MP to all network nodes (100\%), MP to only 20\% nodes selected by the proposed method, and combined PC with MP to 20\% selected nodes. Figures refer to the entire network: (a) MFD of accumulation vs. production; (b) time-series of accumulation; (c) time-series of total virtual queue; (d) time-series of cumulative trip endings.} 
    \label{fig:resMPsoloBest}%
\end{figure}

Figure~\ref{fig:resMPsoloBest} depicts simulation results of the benchmark case of FTC, single MP case for all eligible nodes (`MP 100\%'), single MP controlling 20\% of nodes selected according to the proposed method (`MP 20\%'), and the combined PC with MP to 20\% of selected nodes (`PC+MP 20\%'), all for the case of medium demand. In \ref{fig:resMPsoloBest}(a) MFDs of accumulation versus production are shown for the four cases. We notice that all scenarios involving MP significantly increase the maximum production, compared to the FTC case, and therefore, increase both critical and maximum observed vehicle accumulation. This remark indicates that MP strategy can increase system serving capacity in conditions of moderate congestion, and by balancing queues around controlled nodes, it leads to better road space utilization. As a result it allows a higher number of vehicles to be in the system at the same time, which was not possible in FTC due to local gridlocks that were forcing excess demand to stay in virtual queues. This is evident in (b), where total network accumulation of all scenarios is higher than in FTC, as well as in (c), where total virtual queues are remarkably lower. Moreover, by comparing `MP 20\%' and `PC+MP 20\%' MFD curves in (a), we see that the latter leads to slightly lower maximum accumulation and, thus, smaller capacity drop and hysteresis loop in the unloading part. In this case, combined PC+MP performs slightly better than MP by approximately 2\%. However, the opposite is observed in the respective cases of 25\% MP nodes, which is probably due to traffic correlation among additional MP controlled nodes. We should note here that some change of MFD curve in presence of MP is to be expected, especially with respect to critical accumulation, and this should be considered in the process of parameter tuning for PC. Another interesting remark is that, between the two single MP scenarios, `MP 100\%' results in higher increase in system serving capacity, but on the other side, it introduces more vehicles in the network and thus, reaches higher congestion levels and capacity drop in peak time than FTC. Or, production rises significantly, but also drops during peak, causing some delays and heterogeneity non-existent in FTC, as shown by the unloading part of the MFD curves. This effect can explain why MP installed in all nodes performs worse than partial installation to 20\% of nodes, and is closely related to the shape of MFD and how fast production drops when network enters the congested regime. However, in this case, minimum recorded production is not much lower than the one in FTC, while capacity increase is significant, thus, despite the importance of the production drop in MP 100\% case, VHT savings compared to FTC are still significant. This effect is less intense in the `MP 20\%' and `PC+MP 20\%', where a slightly lower maximum production is recorded, but lower maximum congestion and capacity drop are observed as well. For medium demand, the highest delay savings are achieved for the case of PC + MP 25\%.

\begin{figure}[tb]%
	\centering
    \subfloat[]{\includegraphics[scale=0.5]{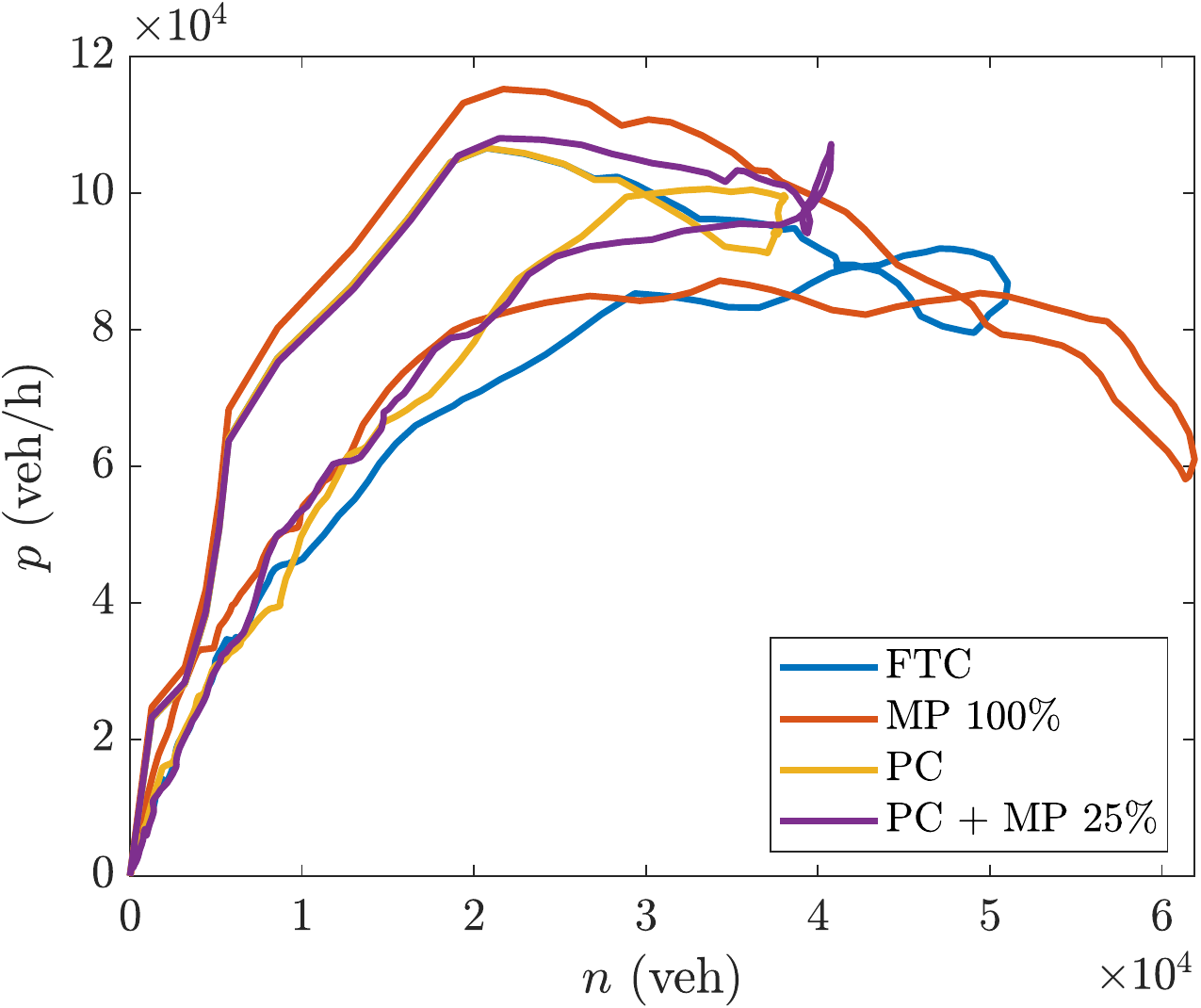}}%
    \qquad
    \subfloat[]{\includegraphics[scale=0.5]{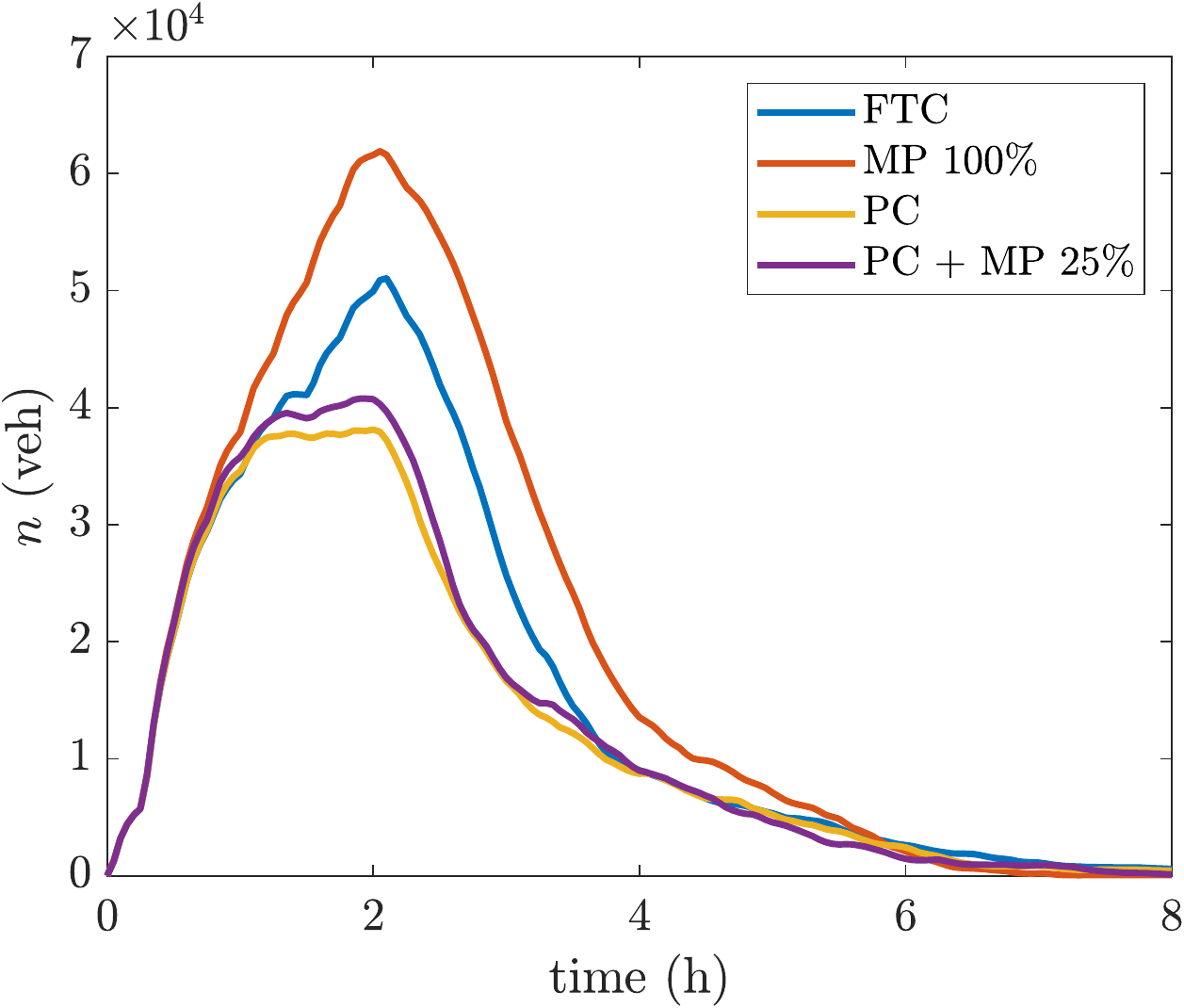}}%
    \qquad
     \subfloat[]{\includegraphics[scale=0.5]{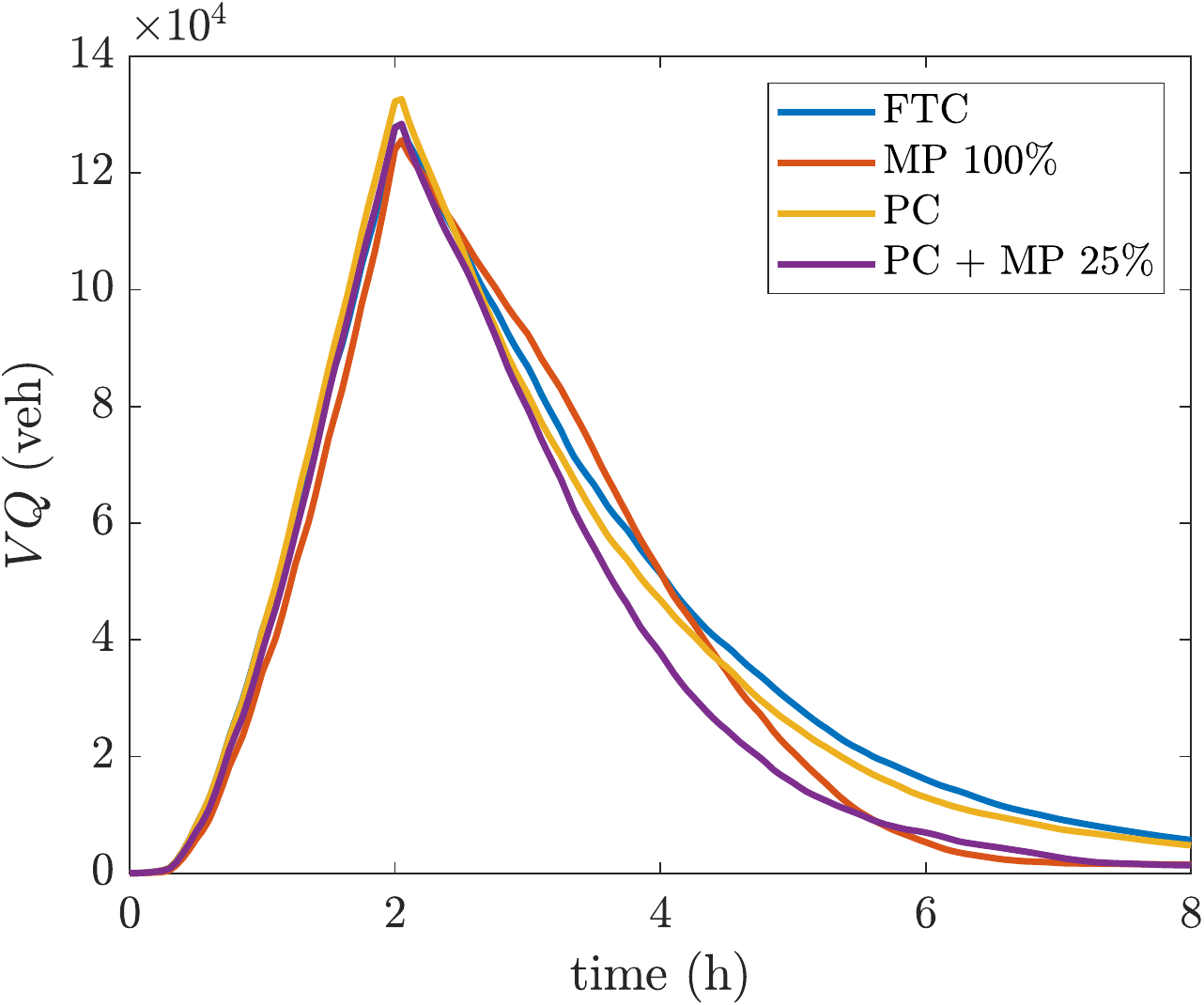}}%
    \qquad
    \subfloat[]{\includegraphics[scale=0.5]{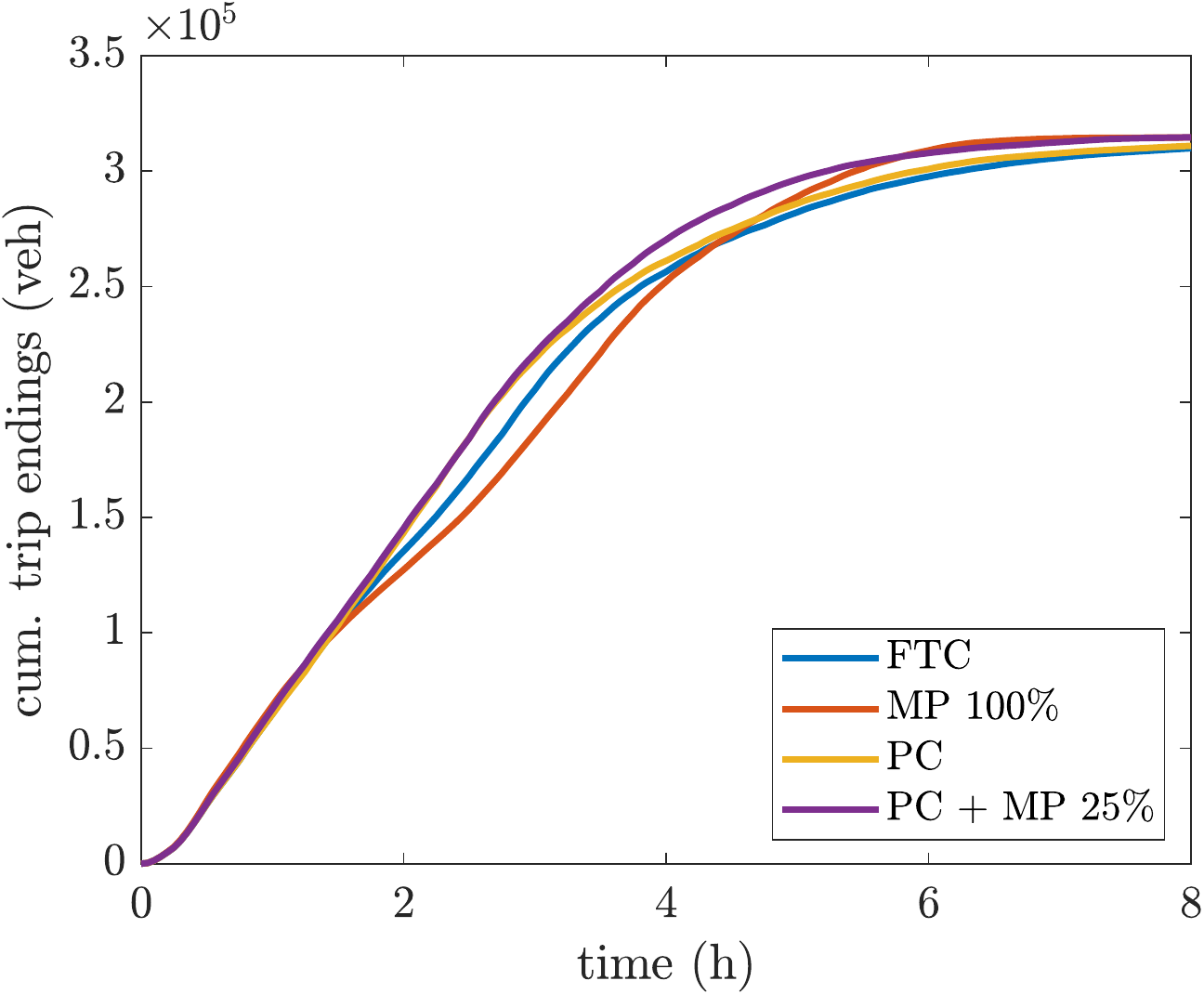}}%
        \qquad
    \caption{Simulation results for high demand scenario. Comparison between FTC, single PC, single MP in 100\% of nodes and combined PC with MP in 25\% of nodes selected by the proposed method. Figures refer to the entire network: (a) MFD of accumulation vs. production; (b) time-series of accumulation; (c) time-series of total virtual queue; (d) time-series of cumulative trip endings} 
    \label{fig:resMPPCBest}%
\end{figure}

Similarly, figure~\ref{fig:resMPPCBest} shows simulation results of four, best-performing control scenarios, for the high demand scenario. FTC case is compared to the case of single MP with full-network control (`MP 100\%'), the case of single PC, and the case of combined PC with distributed MP in subset of 25\% of eligible nodes, selected according to the proposed method. Regarding single MP scheme, a behaviour similar to medium demand case is also observed for the high demand, although in the latter, capacity increase is relatively smaller compared to FTC (about 6.5\%), while production drop in peak time is significantly higher (around 27\%), which can be   due to reaching more congested state by allowing more vehicles inside the network at the same time, as we can see in \ref{fig:resMPPCBest}b. Overall, MP 100\% case performs almost similar to FTC in terms of VHT but significant differences are observed between MFD curves. However, smaller hysteresis is recorder during network unloading in the case of MP 100\%, thus reducing the damage made by the production drop. Interestingly this effect is eliminated in the case of the two-layer framework, where PC plays an important role in prohibiting the system from reaching highly congested states. Therefore, in the combined case of `PC + MP 25\%', the network reaches slightly higher production in peak period compared to single PC case, which drops with a smaller rate as accumulation increases above critical, due to MP control. Also, PC impedes the excessive increase of vehicle accumulation in the system and prevents highly hysteretic behaviour owed to heterogeneity. Among the four cases shown, the combined framework leads to shorter total travel time, reduced by almost 15\% with respect to FTC, when single PC achieves a decrease of around 7.5\%. In short, adding a MP layer significantly improves single PC performance, while properly selected MP nodes allow for a smaller network penetration rate that leads to comparable performance as in full-network MP implementation. 

\begin{figure}[tb]%
	\centering
     \subfloat[]{\includegraphics[scale=0.20]{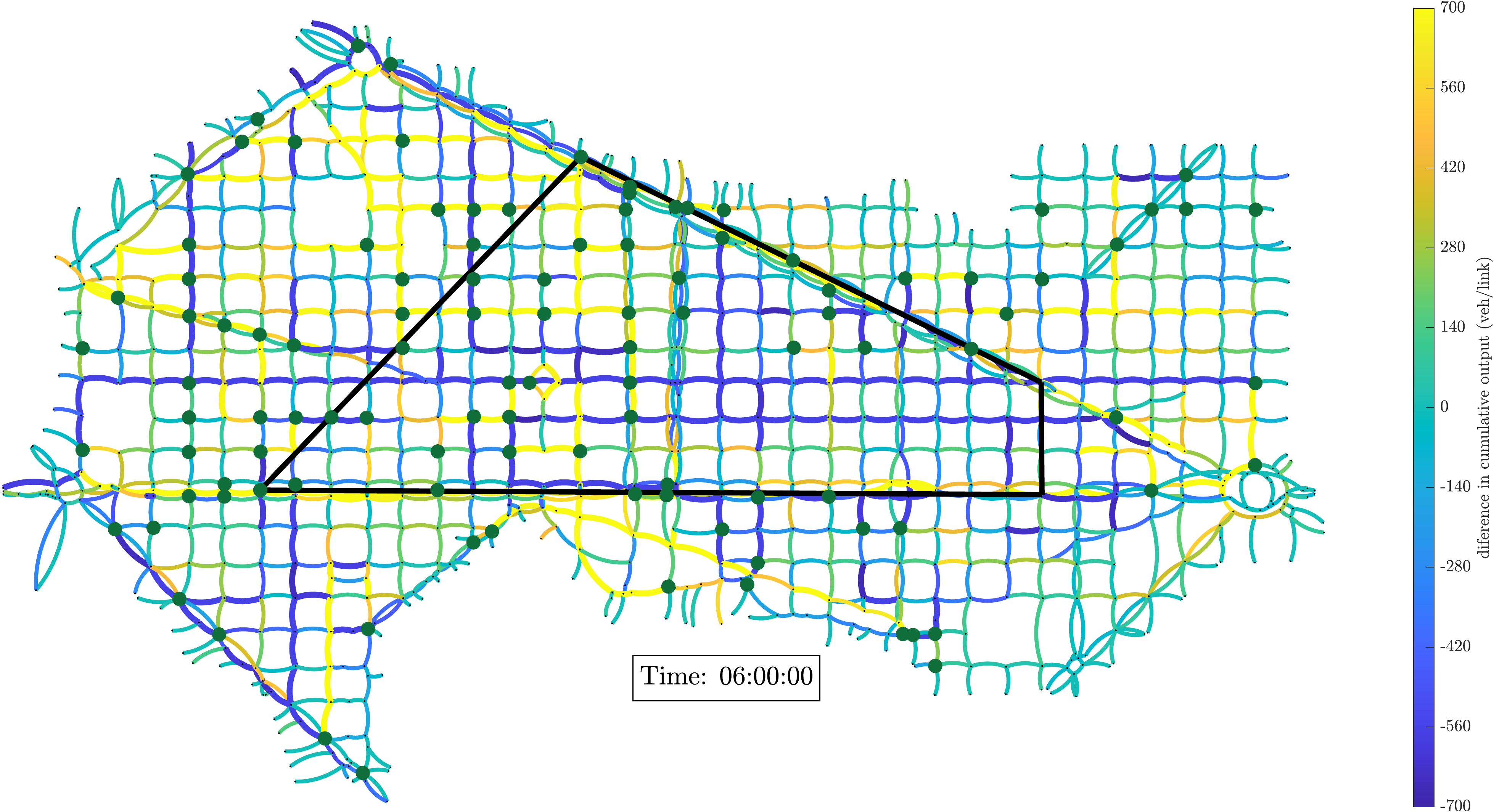}}%
    \qquad
    \subfloat[]{\includegraphics[scale=0.20]{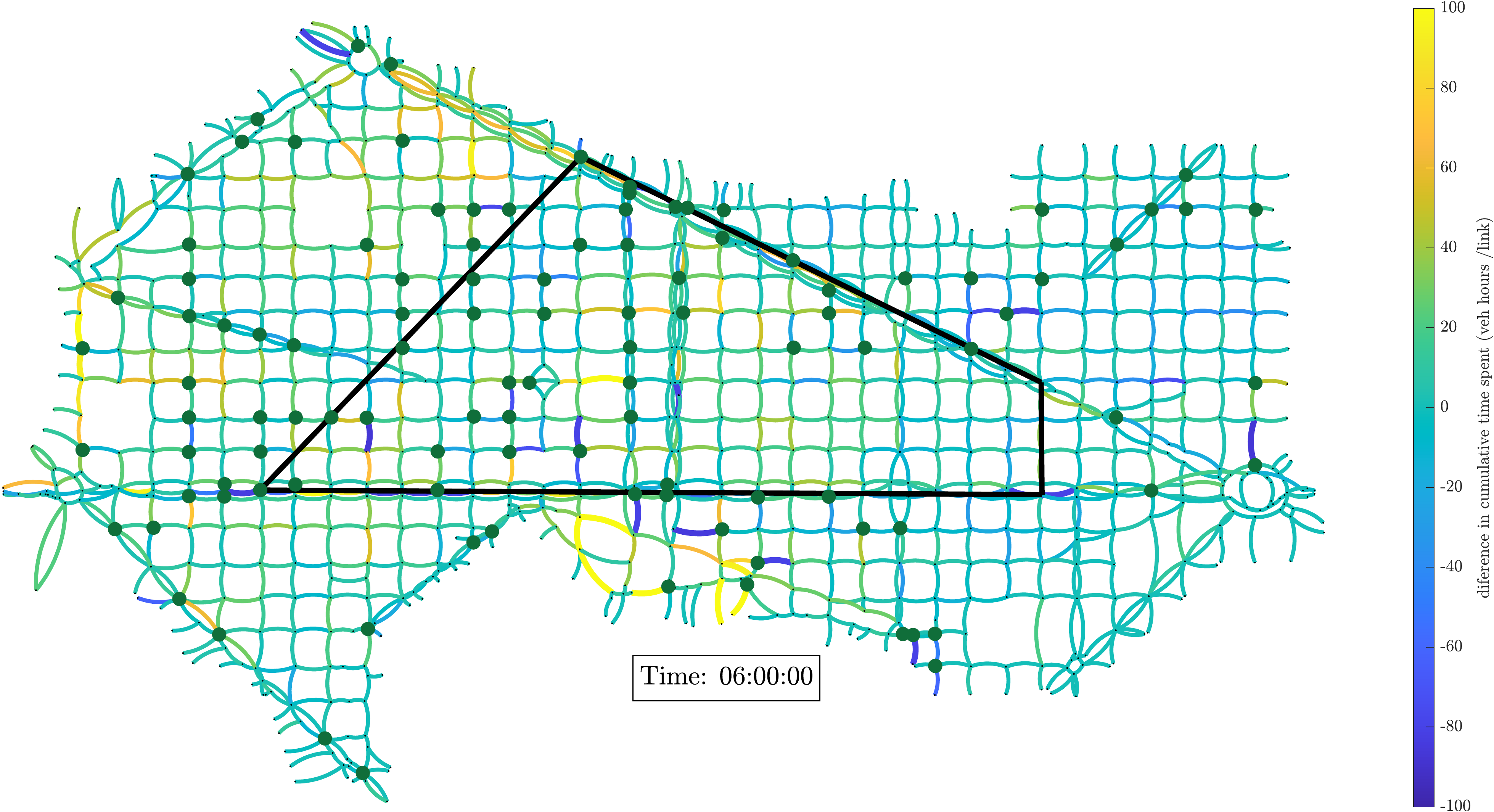}}%
    \qquad
   \subfloat[]{\includegraphics[scale=0.20]{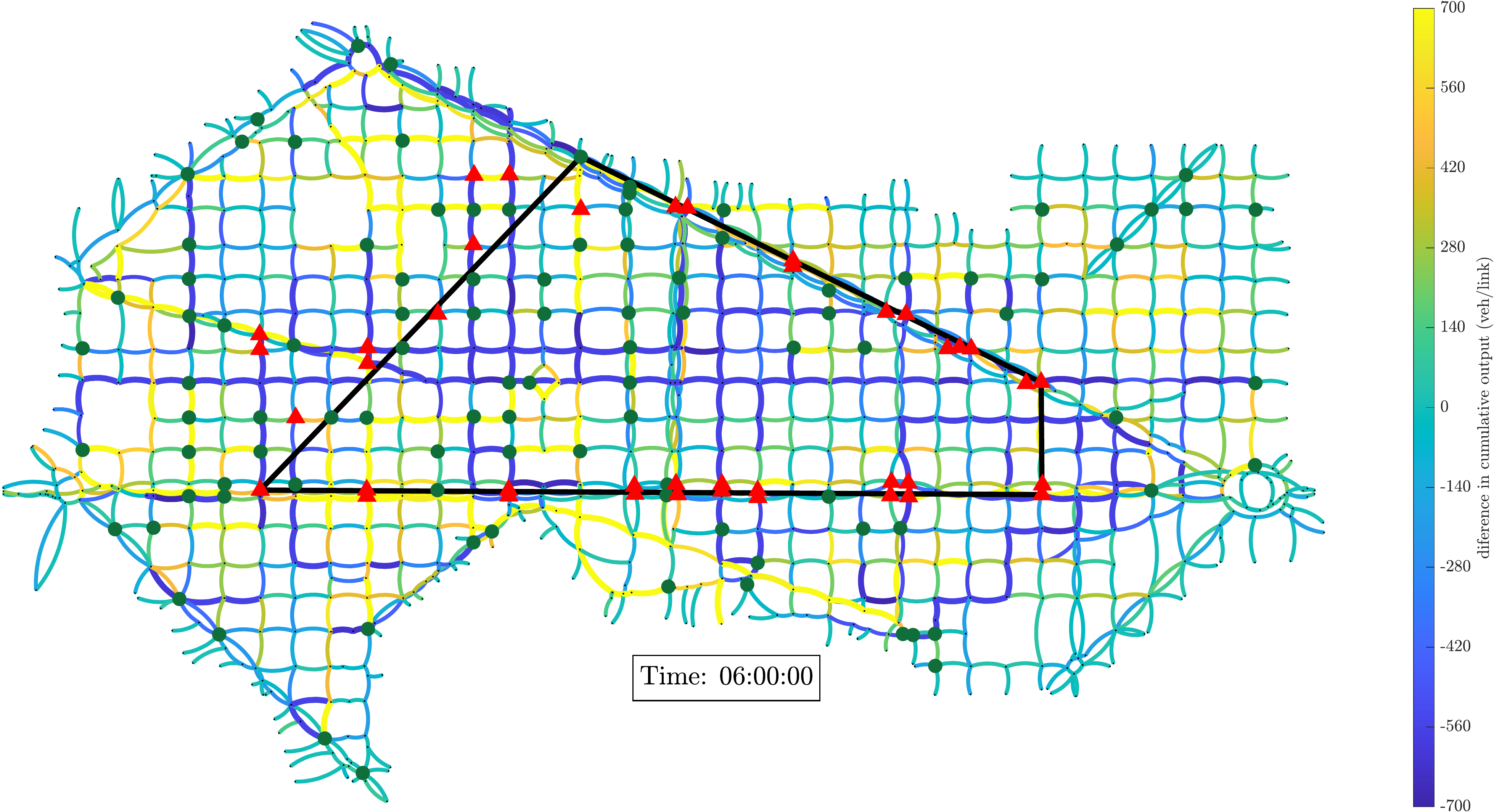}}%
    \qquad
     \subfloat[]{\includegraphics[scale=0.20]{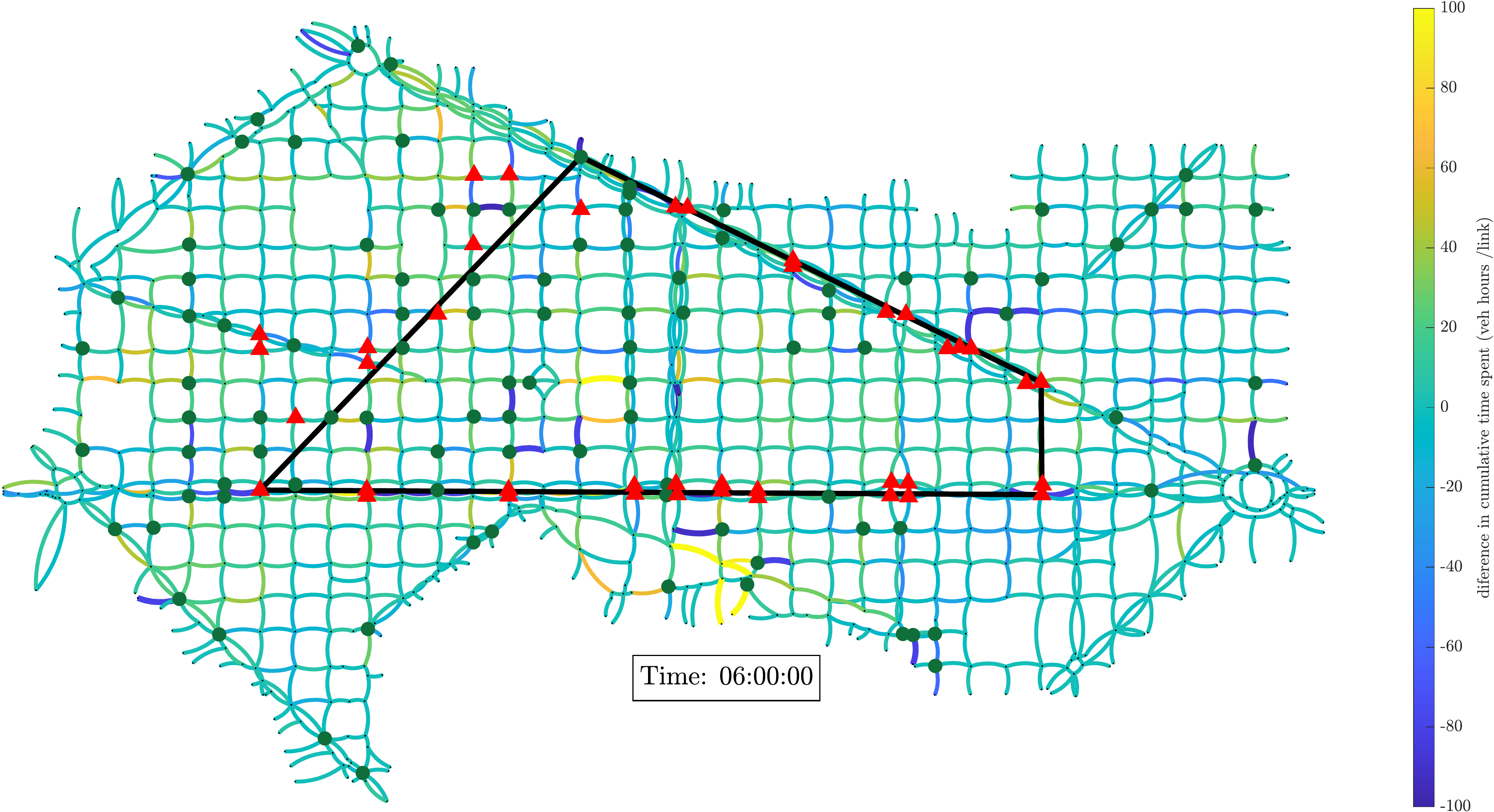}}%
    \caption{Difference of cumulative link throughput and cumulative time spent with respect to FTC, in medium demand scenario, at the end of simulation (6:00 h): (a)-(b) Case MP in 20\% critical nodes; (c)-(d) Case MP in 20\% critical nodes combined with PC. Dots and triangles represent MP and PC node locations, respectively.} %
    \label{fig:snaps_time_med}%
\end{figure}

Figure \ref{fig:snaps_time_med} depicts the impact of two well-performing control schemes, in terms of link throughput and link total time spent difference with respect to FTC scenario, for medium demand scenario, for the case of single MP in 20\% of nodes selected by the proposed method (see a and b), and for the case of the combined control of PC and MP in 20\% of nodes, selected in the same way (see c and d). All four graphs show the network map where each link's color corresponds to a range of values, as displayed on the bar on the right. Links appear slightly thicker when values approach or overpass bar limits. Graphs a and c depict the difference in link cumulative throughput with respect to FTC case, for each of the above control cases, respectively. Similarly, graphs b and d depict the difference in cumulative time spent per link, with respect to FTC case. Green dots denote MP controlled nodes while red triangles denote PC nodes. The black lines represent the approximate boundaries of central region 2. From graphs a and c, we observe a similar pattern of traffic redistribution, where intense blue links indicate significant cumulative throughput reduction in the case of responsive control compared to FTC, while yellow links indicate the opposite. We notice that cumulative throughput is reduced along several links that are also connected, but it is increased in the majority of network links, as warmer colors (green and yellow) indicate. However, around most MP nodes, we see at least one or several incoming or ongoing links, where throughput is overall higher, at least to one direction, which means that a higher number of vehicles left those links sooner than FTC case. This observation supports the idea that throughput is mostly increased around MP nodes. Nevertheless, a throughput decrease can also result from rerouting of vehicles due to change in congestion distribution around the network. Interestingly, we do not observe significant increase in cumulative total time spent in links where cumulative throughput is lower (by comparing left and right figures of each row), which indicates that throughput drop is related to rerouting rather than congestion increase. In fact, control scheme of single MP in 20\% of critical nodes achieves 16\% lower VHT than FTC, while combined PC + MP in 20\% critical nodes achieves 17.7\% lower VHT. This reduction is not particularly obvious in graphs b and d, since a large amount of this time gain comes from decreasing the time spent in virtual queues, which are not shown in these figures. 

\begin{figure}[tb]%
	\centering
    \subfloat[]{\includegraphics[scale=0.20]{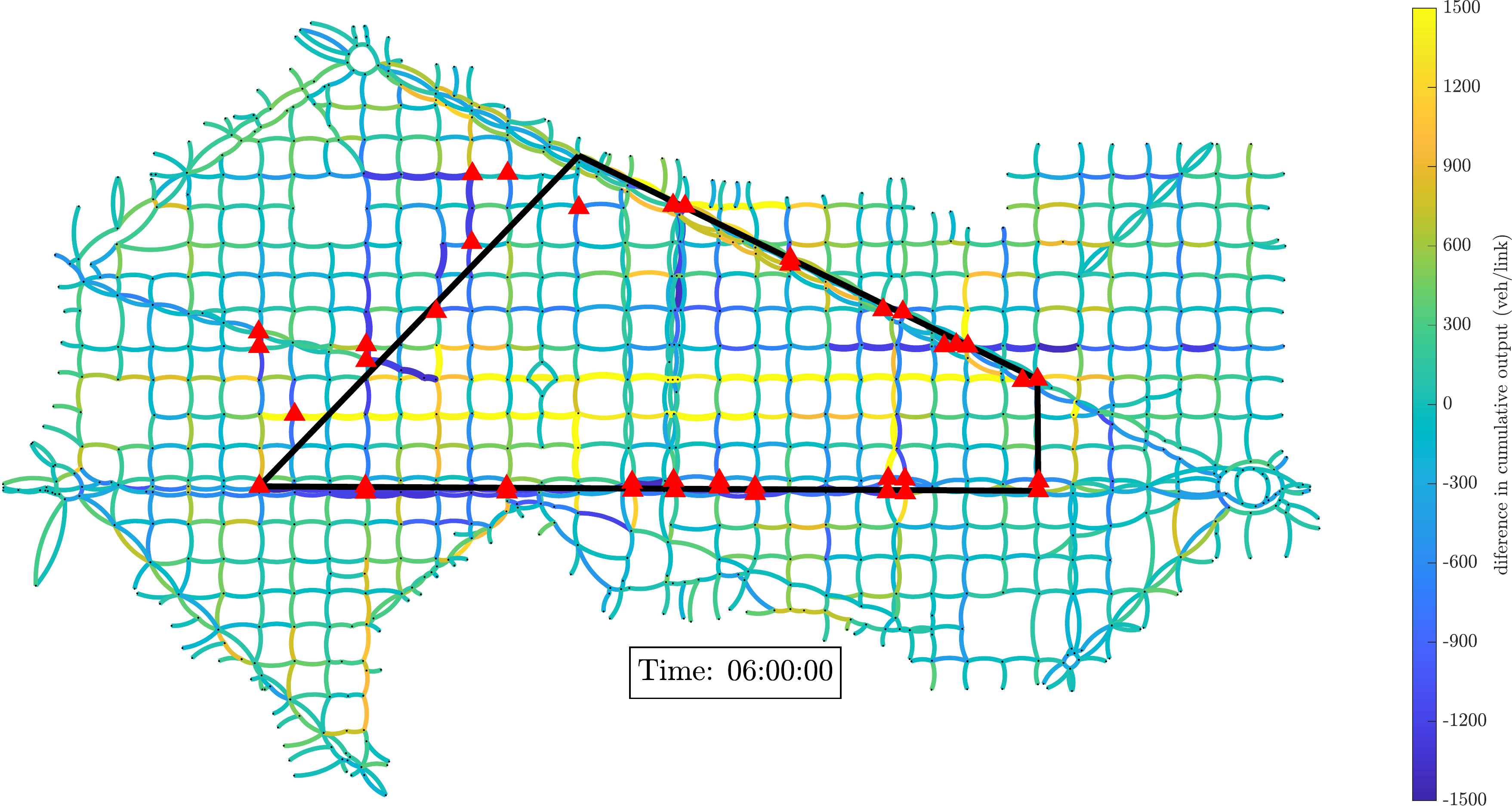}}%
    \qquad
     \subfloat[]{\includegraphics[scale=0.18]{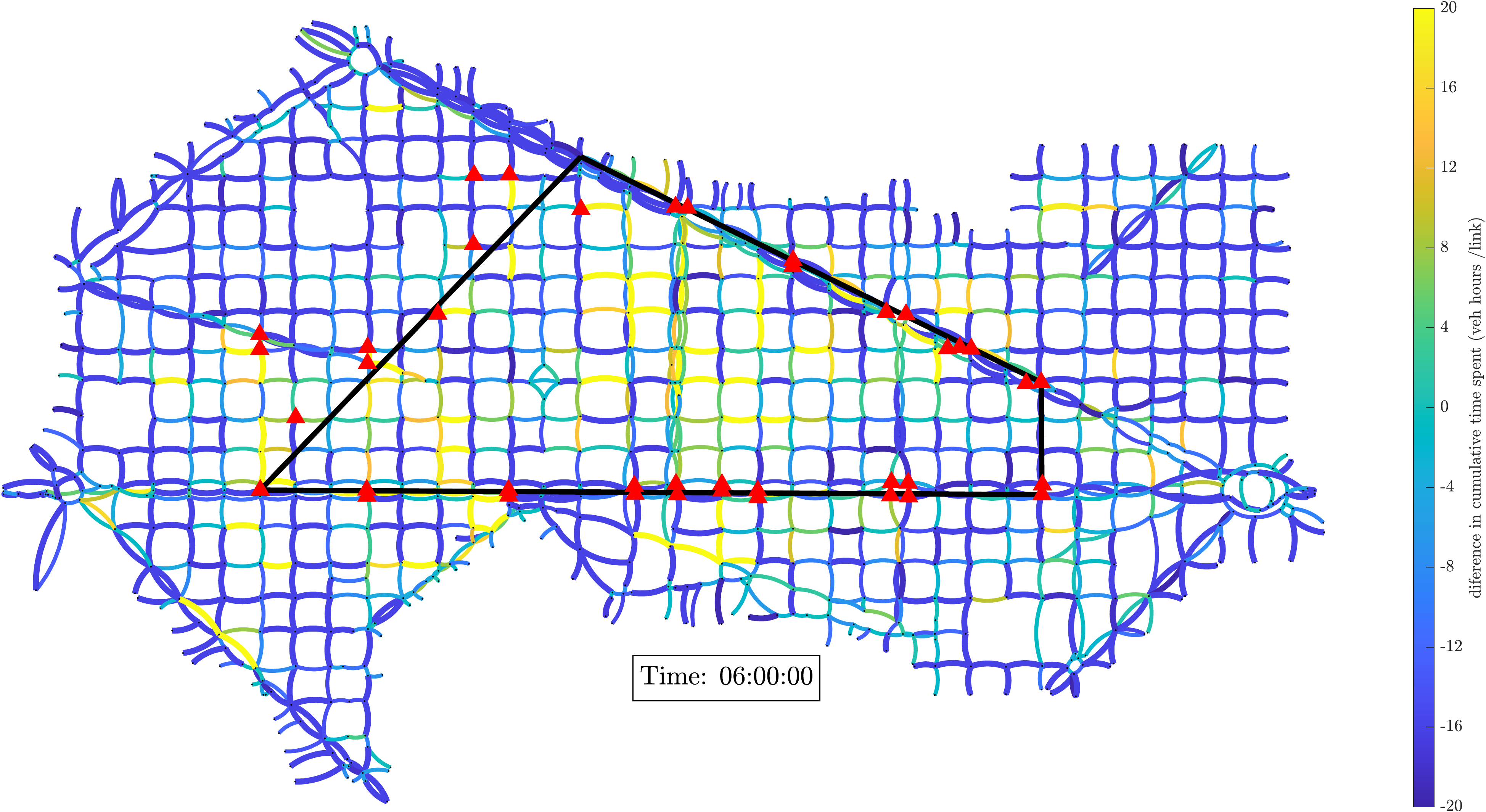}}%
    \qquad
    \subfloat[]{\includegraphics[scale=0.20]{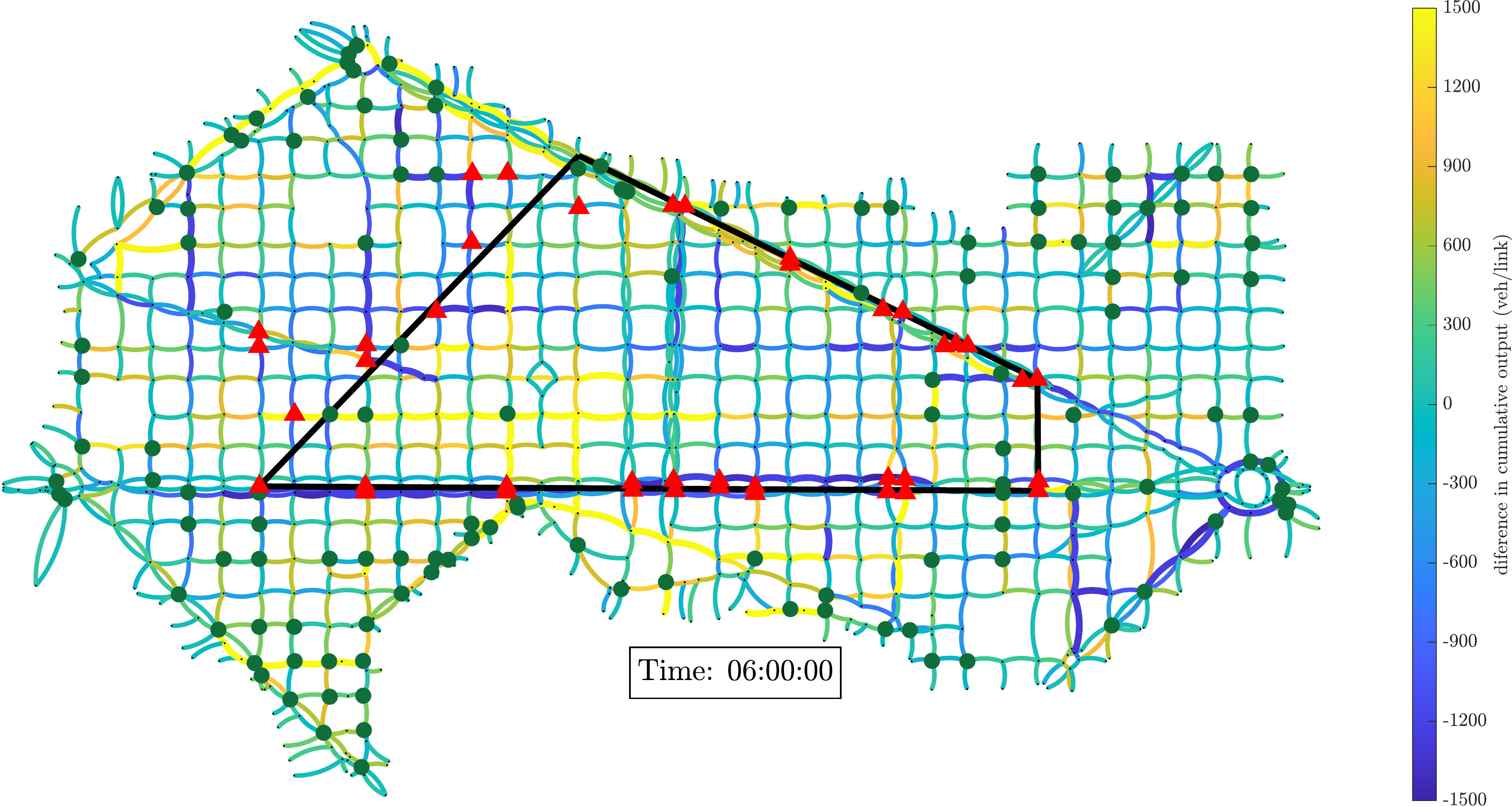}}%
    \qquad
    \subfloat[]{\includegraphics[scale=0.18]{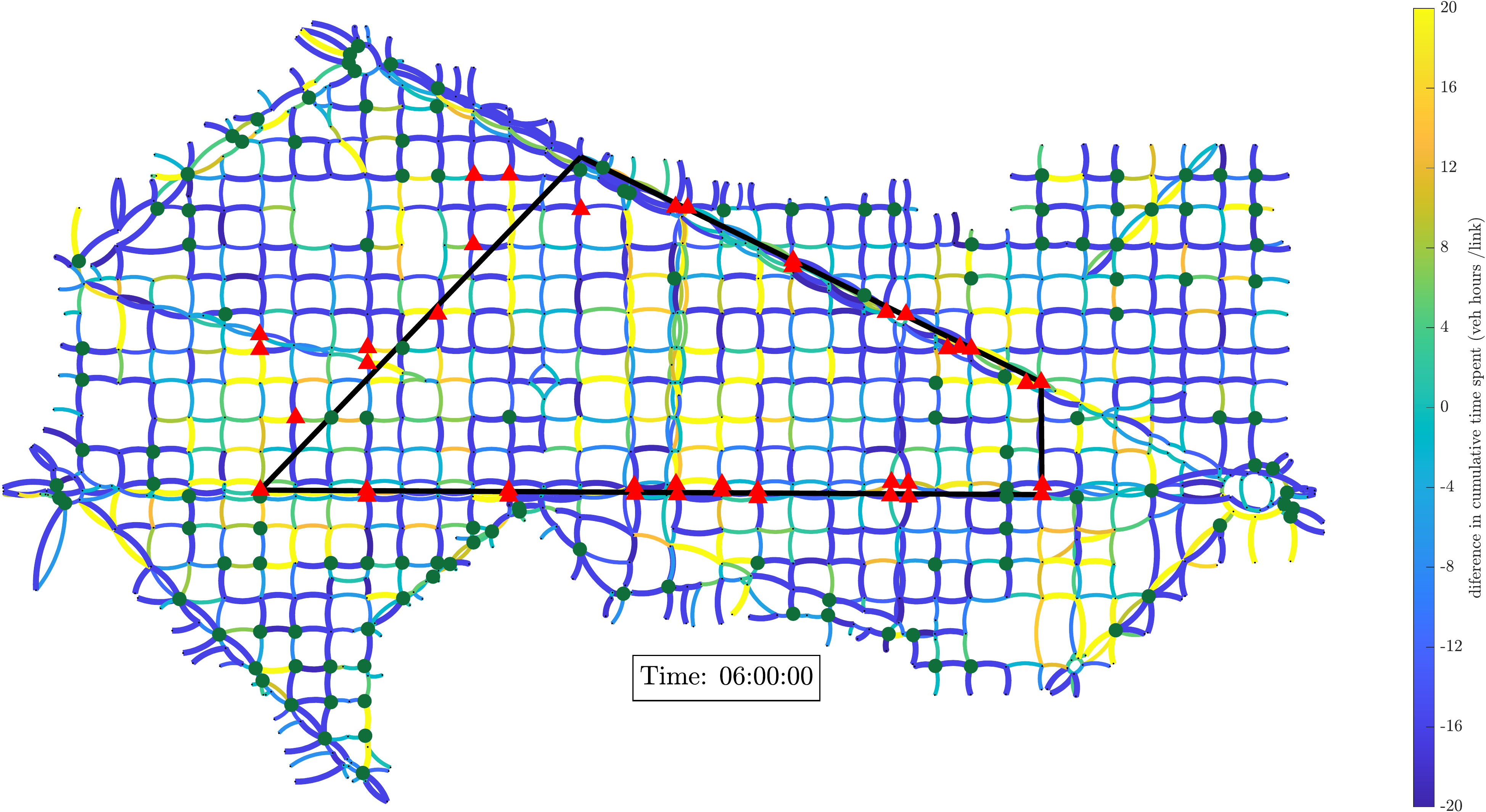}}%
    \caption{Difference of cumulative link throughput and cumulative time spent with respect to FTC in high demand scenario at the end of the simulation (6:00 h): (a)-(b) Case single PC; (c)-(d) Case PC combined with MP in 25\% of selected nodes. Dots and triangles represent MP and PC node locations, respectively.} %
    \label{fig:snaps_high}%
\end{figure}

The same type of information is shown in figure~\ref{fig:snaps_high}, but for the cases of single PC (see a and b) and combined PC with MP in 25\% of critical nodes selected by the proposed method, both for the high demand scenario. In a we observe significant increase in cumulative throughput along two main arterial roads in the interior of central region 2, which, for this demand scenario, attracts the majority of trips, while decrease is observed almost exclusively upstream or in the proximity of PC nodes (triangles), which is an expected effect of PC. However, throughput appears slightly higher in the majority of links also in peripheral regions, which is related to congestion drop, as shown in graph b, where most network links experience drop in cumulative time spent with respect to FTC case, with the exception of some links upstream PC nodes, which receive the queues that form due to gating. Performance appears even better for the combined scheme, as we see in c, where cumulative throughput is increased in most links (in green and yellow), and around the majority of MP nodes (dots). Alternatively, the decrease in cumulative time spent is obvious in most network links, as we see from graph d, where increase is recorded mostly around the boundaries with region 2. The improvement in VHT compared to FTC case is 7.6\% for single PC case (a and b), while for the combined case PC + MP 25\% (c and d) reaches 15.6 \%.  

\begin{figure}[tb]%
	\centering
    \subfloat[]{\includegraphics[scale=0.4]{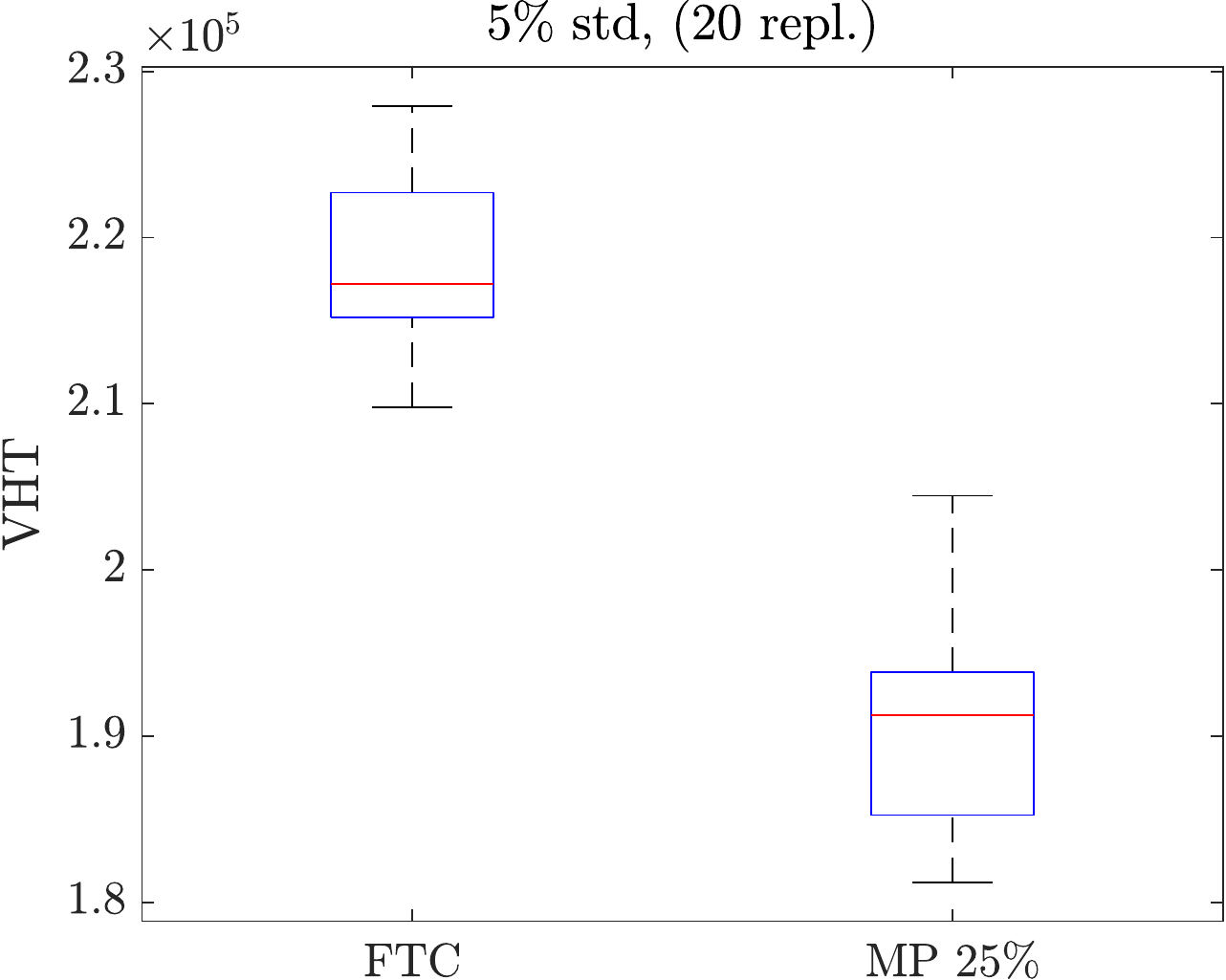}}%
    \qquad
    \subfloat[]{\includegraphics[scale=0.4]{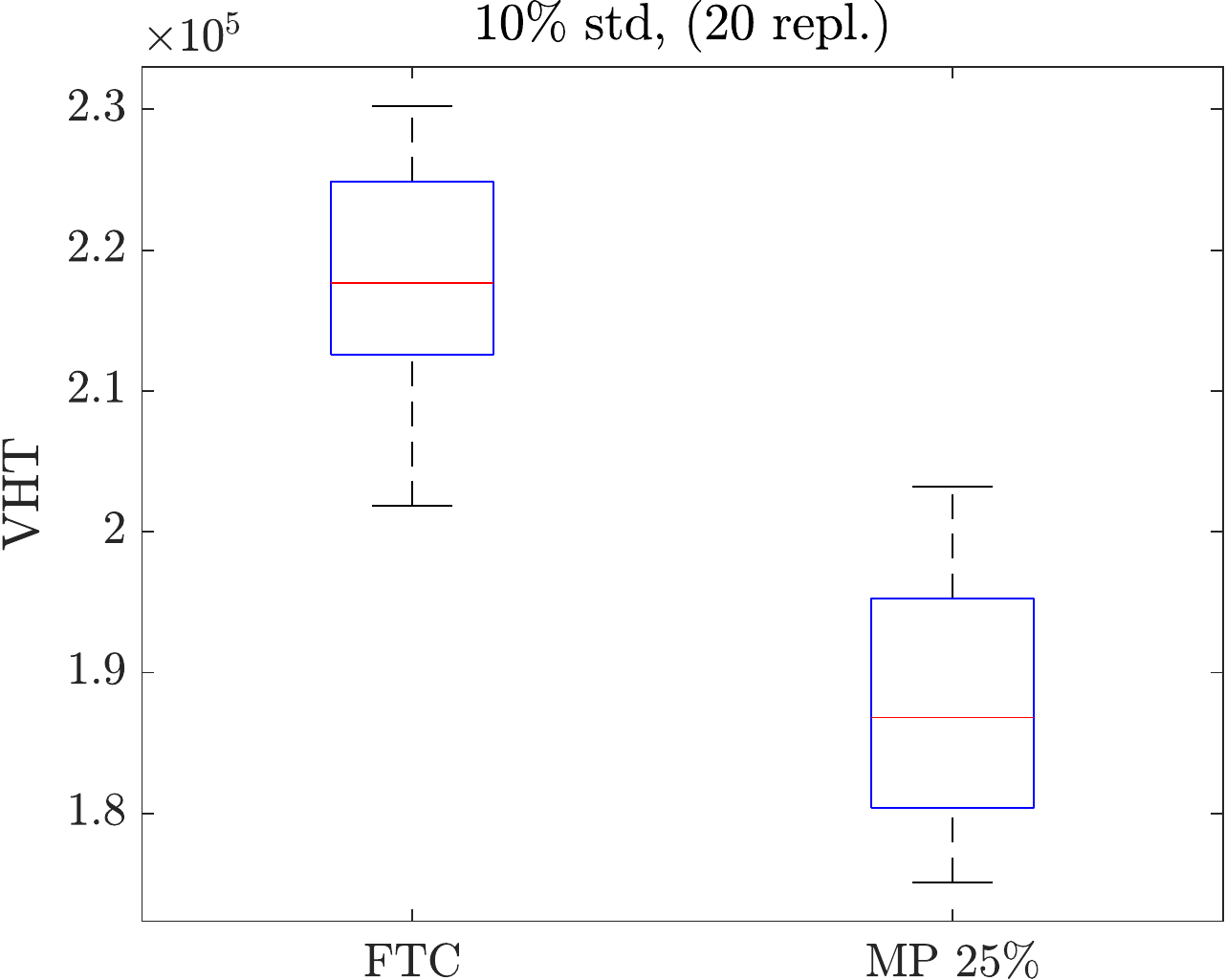}}%
    \qquad
     \subfloat[]{\includegraphics[scale=0.4]{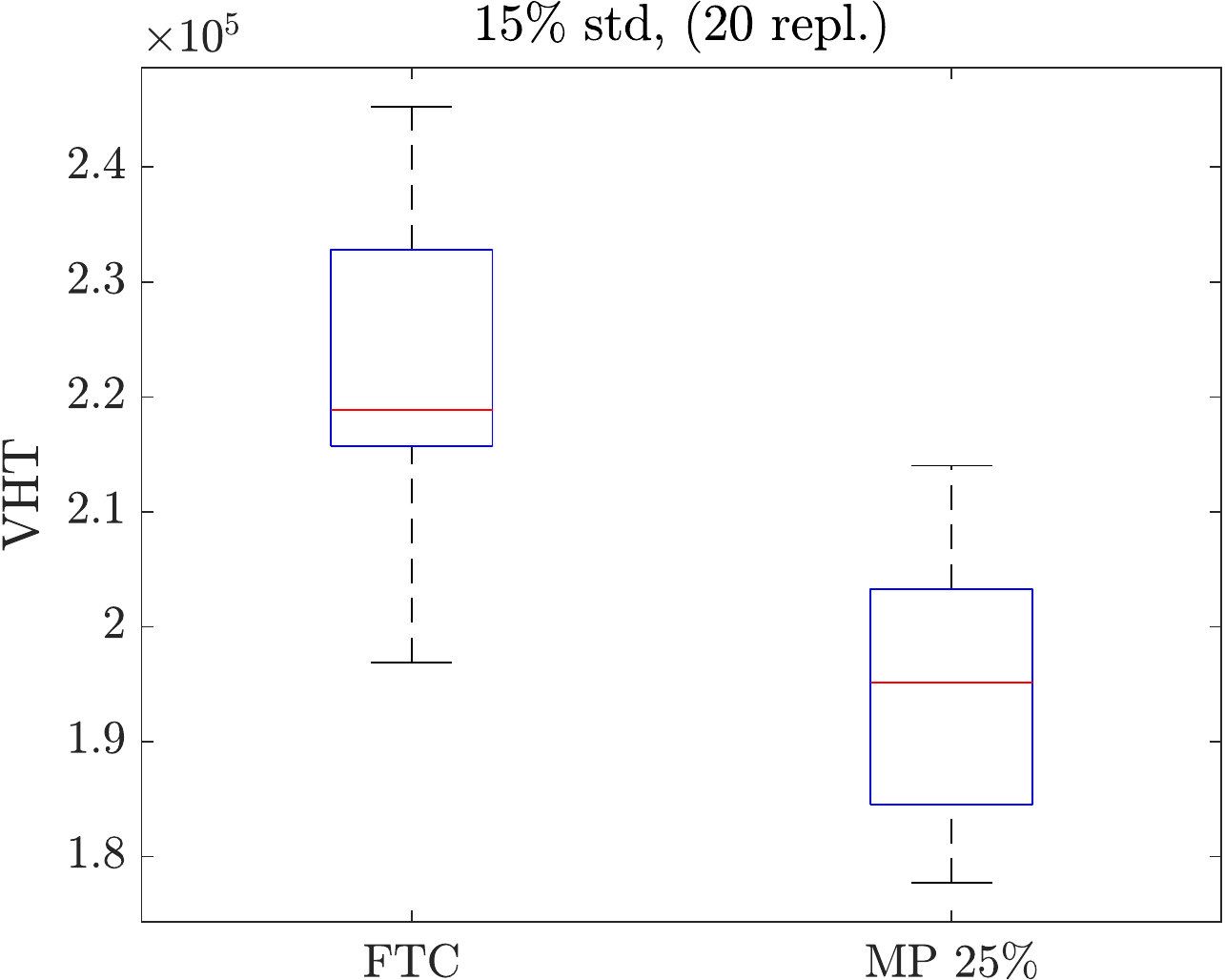}}%
    \qquad
    \subfloat[]{\includegraphics[scale=0.4]{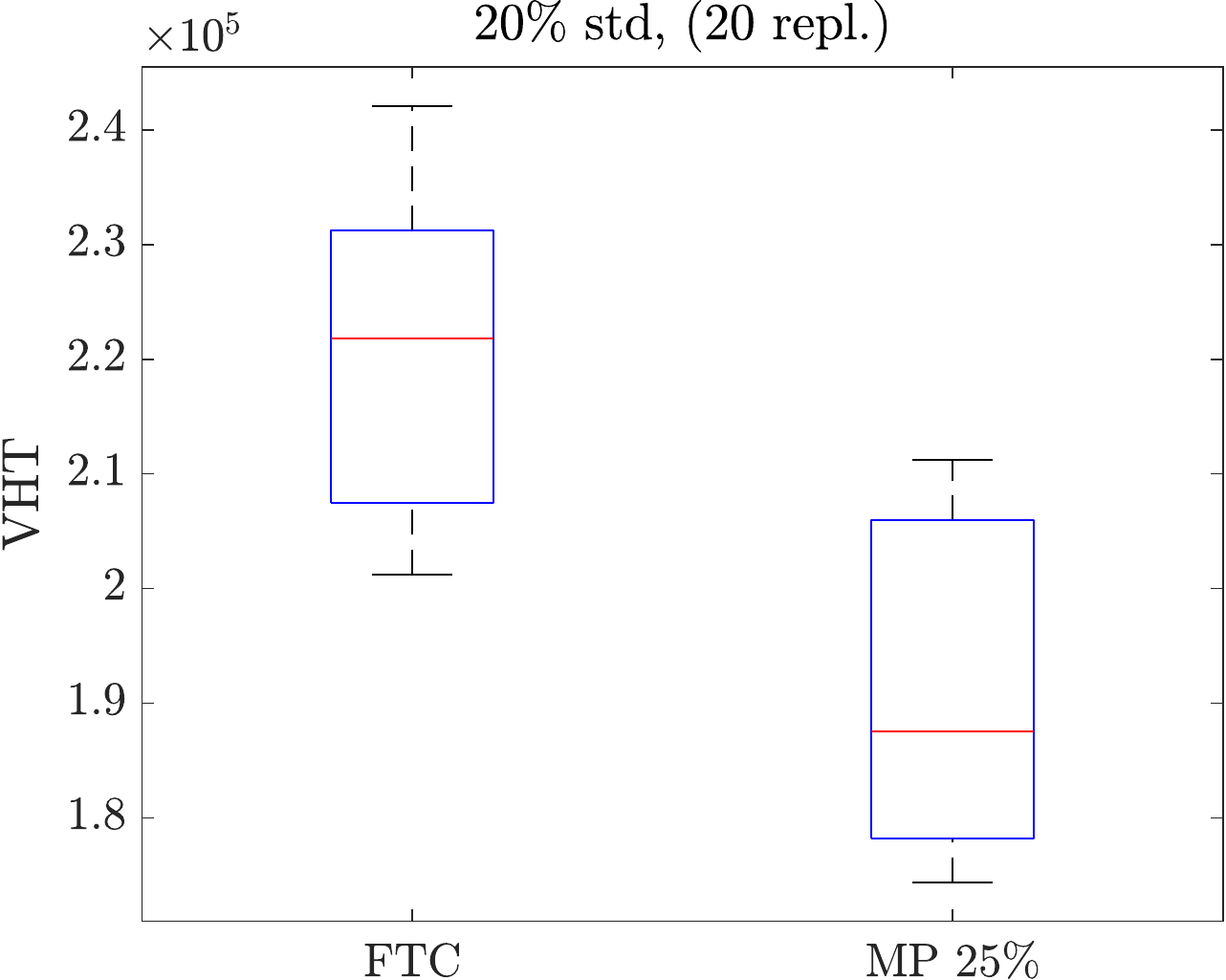}}%
        \qquad
    \subfloat[]{\includegraphics[scale=0.4]{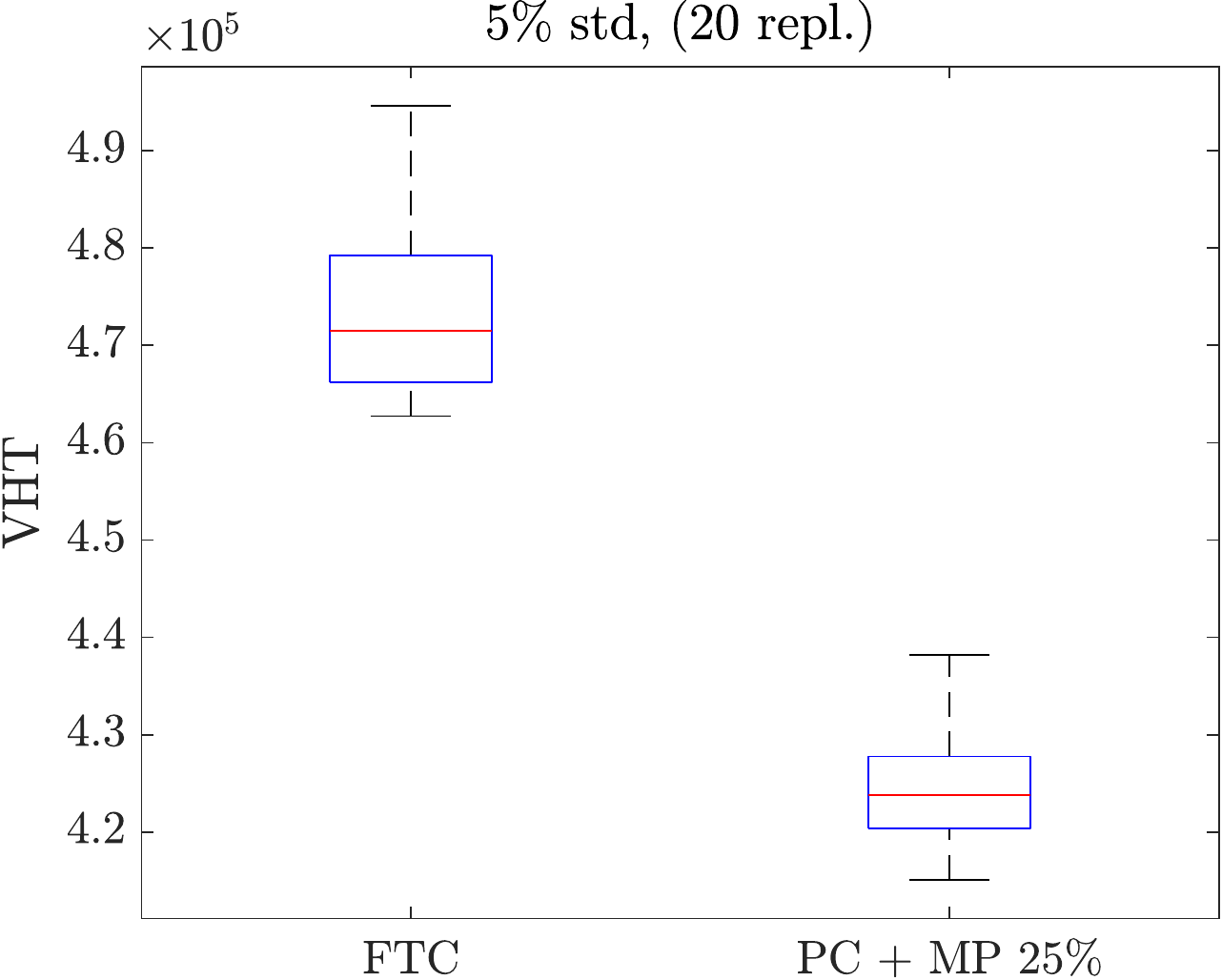}}%
    \qquad
    \subfloat[]{\includegraphics[scale=0.4]{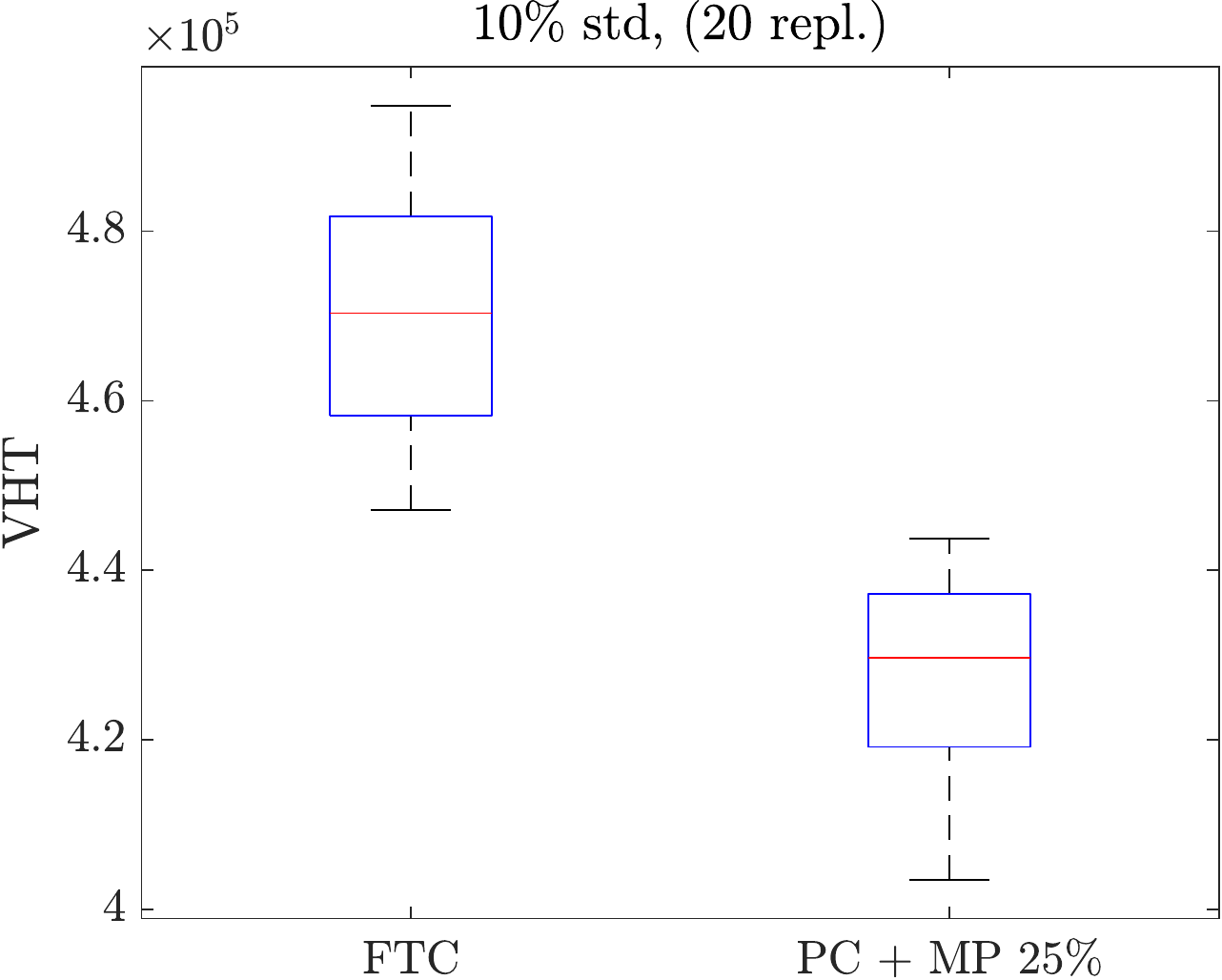}}%
    \qquad
    \subfloat[]{\includegraphics[scale=0.4]{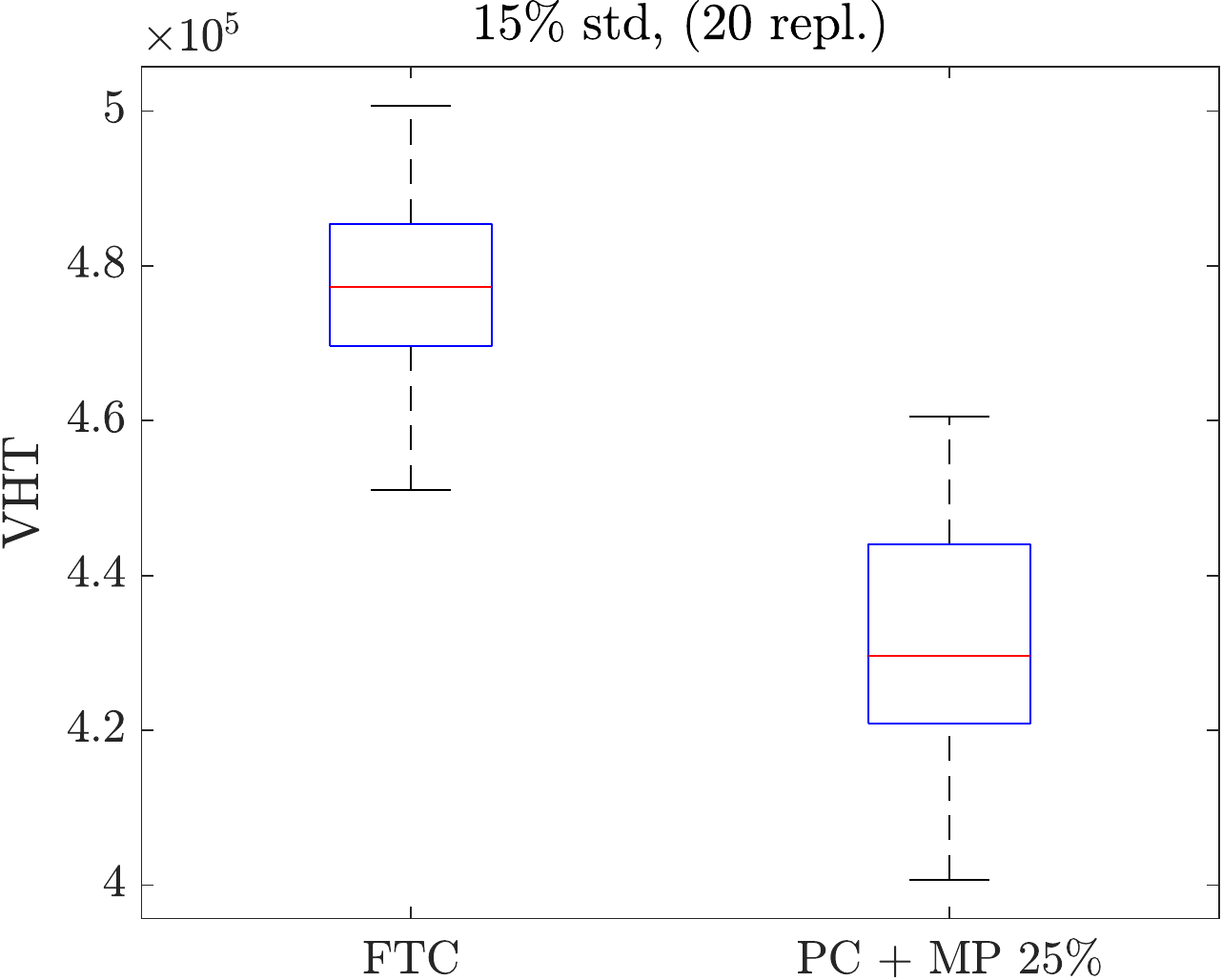}}%
    \qquad
    \subfloat[]{\includegraphics[scale=0.4]{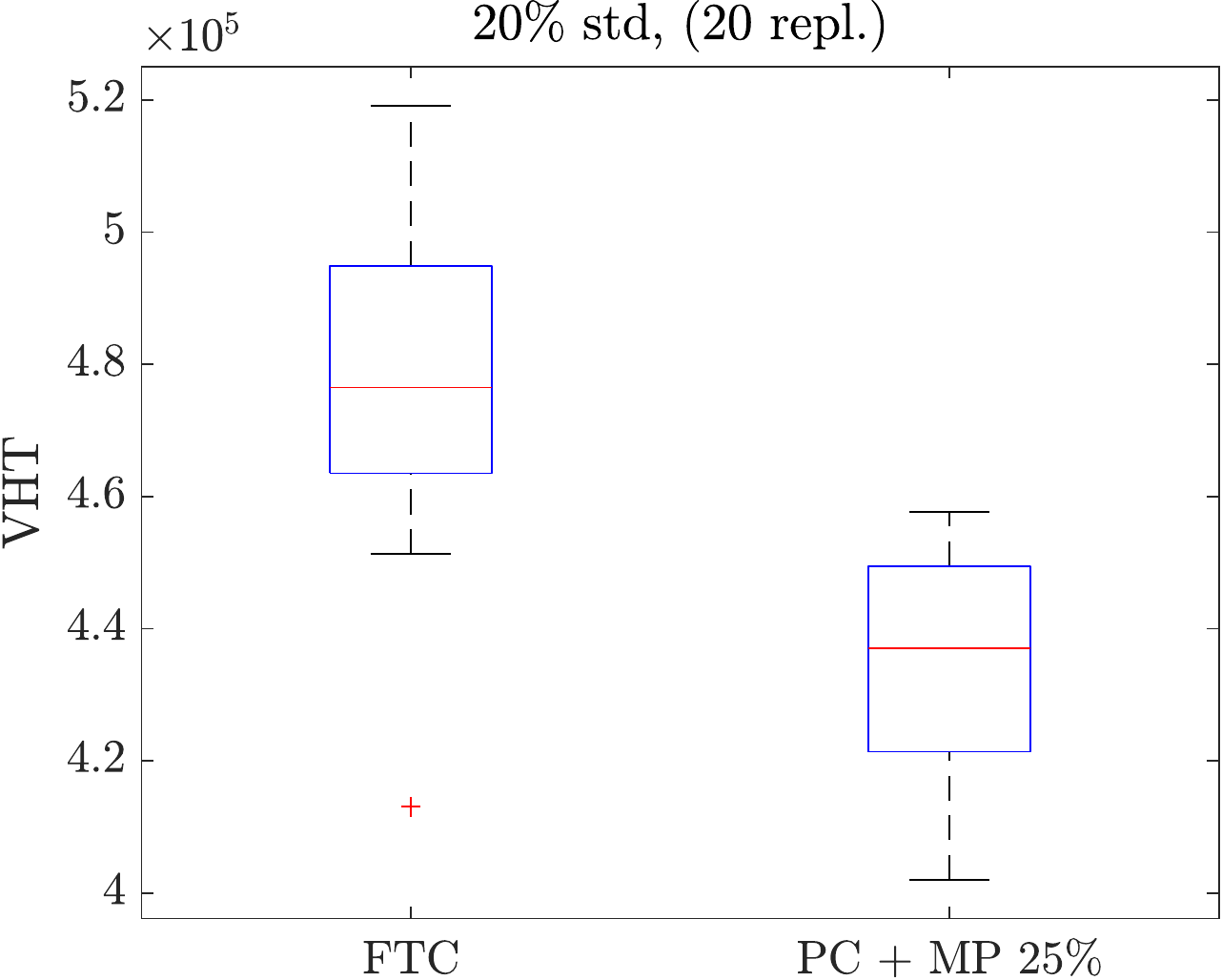}}%
    \caption{Sensitivity analysis of the two best performing control schemes under demand fluctuations: (a) to (d) refer to the medium demand case, and FTC is compared to single MP in 25\% of nodes selected by the proposed strategy; (e) to (h) refer to high demand case, and FTC is compared to combined scheme of PC and MP in 25\% of selected intersections. Boxplots correspond to 20 different OD matrices where each OD pair is a random variable with normal distribution, with standard deviation equal to 5, 10, 15 and 20\% of the mean, respectively.}%
    \label{fig:sensitivity}%
\end{figure}

In order to examine the sensitivity of the proposed MP node selection process, in terms of performance, against fluctuations of spatial traffic distribution, a set of additional simulation experiments are performed. Specifically, for two well-performing cases of single MP and combined MP-PC schemes with MP nodes selected through the proposed method, simulation tests are performed, where  instead of the mean demand value, for every origin-destination pair, a new value is considered generated by a stochastic process. More specifically, for the case of single MP with 25\% controlled nodes selected by the proposed method, based on the results of FTC scenario of the medium demand, 20 new demand matrices are generated, by replacing mean demand of every origin-destination pair, by a random draw from normal distribution, with mean equal to the initial value and standard deviation that corresponds to 5, 10, 15 and 20\% of the mean. The specific control scheme that was designed based on the mean demand, is tested for the 20  stochastically generated demand matrices for each of the aforementioned standard deviation cases, and performance is compared to the one of FTC case. Results are shown in figure~\ref{fig:sensitivity} (a) to (d), where each graph refers to a different standard deviation level of demand. Boxplots represent the distribution of total recorder travel time of FTC and MP-25\% scenarios for the same 20 modified demand matrices. We observe that MP-25\% case on average outperforms FTC for all standard deviation levels (median of MP-25\% is always less than median of FTC), although improvement decreases as deviation from mean increases. The same experiments are performed also for the combined MP-PC scheme with 25\% MP nodes selected by the proposed method, for the high demand scenario and results are shown in figure~\ref{fig:sensitivity} (e) to (h). Again, we observe that PC+MP 25\% outperforms FTC for almost all standard deviation levels, but improvement decreases with the deviation from the mean. This result indicates that the node selection process that is based on mean OD values is not highly sensitive to demand fluctuations of up to 20\% from the mean.



\section{Discussion}

The paper investigates the potential benefits of a two-layer signal control framework, combining centralized, aggregated perimeter control strategy, with partial and efficient distributed control, in critical intersections, based on the Max Pressure feedback controller. Cost-effective schemes of improved performance for network-wide MP control are produced by decreasing the number of controlled nodes, via selecting only nodes considered critical for MP implementation. A critical node identification method based on actual traffic characteristics is introduced and assessed. Simulation experiments performed with a modified version of Store-and-Forward model evaluate the effectiveness of PC and MP control schemes, both independently and combined in a two-layer framework. Spill-back effect and potential rerouting of drivers due to control-induced alteration of congestion patterns is taken into account. Regarding MP control, several scenarios are tested, including in the control scheme either subsets or the entire network node set. Subsets are selected both by the proposed method and randomly, for comparison. A sensitivity analysis of the best performing control schemes is done with respect to demand fluctuations. All scenarios are tested for two base origin-destination demand matrices, creating medium and high congestion, respectively. 

Analysis of simulation results provide interesting insights about single and combined implementation of MP and PC, as well as regarding partial/selective MP schemes. Single MP control proves more effective in moderate congestion conditions rather than in oversaturated states, while partial application to subsets of critical nodes, properly selected by the proposed algorithm, can result to similar or even better performance than full-network implementation. Therefore, not only application cost can be drastically reduced, but control effectiveness can increase as well. Furthermore, partial MP implementation to critical nodes is shown to increase link throughput in the proximity of controlled nodes and balance surrounding queues, while spill-back occurrence of surrounding links is also reduced, for both demand scenarios. Also, significant production capacity increase is observed in single MP application, though it can lead to increased link density (probably due to better road space use leading to less local gridlocks that keep vehicles in virtual queues) and thus increased congestion followed by production drop. Yet, recovery of the network can be smoother with smaller hysteresis shown by MFD curves.

The proposed node selection method seems effective in identifying critical node sets for MP control, since it outperforms random selection for all network penetration rates in the medium demand. Even though its effectiveness drops in the high demand for the single MP scheme, it remains effective in all combined schemes of PC and MP. However, it should be noted that the proposed selection method involves a parameter optimization step that resulted in different values for the medium and high demand, which indicates that the relative importance of the selection criteria can vary between moderate and highly congested conditions. Nevertheless, the proposed selection variables ($m_1, m_2, N_c$) seem to play an important role as indicators of node importance with respect to MP, while further research can help unravel the mechanism that relates selection variable importance to demand patterns, and thus determine optimal parameter values in a universal way by dropping parameter optimization requirement. Overall, it seems that significant correlation exists between controlled nodes, which affect each other in a way not necessarily beneficial for the system, since performance gains can decrease for penetration rates above 25\% and can even drop to zero in 100\% in highly congested scenarios. This phenomenon highlights the importance of partial MP implementation, especially in increased congestion, and the importance of spatial distribution of controlled nodes, which the proposed selection method tries to unravel.  

Regarding the two-layer combined scheme of PC and MP, results are promising in most tested cases, especially in the high demand scenario, where in our case study, adding MP in only 25\% of properly selected nodes leads to doubling the performance gains of single PC compared to FTC case, from 7.5\% to more than 15\%. Moreover, almost the same performance gain is achieved in the case of full-network MP implementation, proving the proposed selection method effective and, as a result, reducing implementation cost to one fourth, compared to full-network scheme. Furthermore, PC protects high-demand regions from reaching saturated states, and therefore from capacity drop, which seems to also increase MP efficiency, given that single MP shows zero improvement for full network implementation in high demand scenario, while combined MP with PC achieves twice the gain of single PC. Moreover, the node schemes generated by the proposed selection method do not appear sensitive to small demand fluctuations that alter traffic distribution in the network, and can still lead to performance improvement even if demand deviates up to 20\% from its mean value, which was used to generate the node scheme. However, as expected, improvement gains decrease as fluctuation increases. 

Future research can focus on including public transport priority polices both in MP and in PC intersections in the two-layer controller. Also, dynamic activation of subset of MP controlled nodes in real-time from a centralized approach would be an interesting research direction, especially with the arrival of connected and automated vehicles, where required traffic information will be provided to MP controller through vehicle-to-infrastructure communication and MP implementation cost related to node instrumentation will be zero, thus any node could be an MP node. On-line readjustment of PC parameters to address possible MFD alteration due to MP effects, as well as PC implementation through optimal MPC instead of PI regulator could prove even more beneficial, while robust control schemes that can address demand and input data noise in saturated conditions could be another interesting direction. Finally, system reaction under sudden incidents causing link or intersection closure, in presence of adaptive control based on PC and MP would be another relative topic to explore.

\clearpage

\section*{Acknowledgement}

This work is partially funded by Dit4Tram ``Distributed Intelligence \& Technology for Traffic \& Mobility Management'' project from the European Union's Horizon 2020 research and innovation program under Grant Agreement 953783. 

\section*{CRediT author statement}

\textbf{Dimitrios Tsitsokas:} Conceptualization, Methodology, Investigation, Validation, Software, Writing - original draft,  Writing - review \& editing. \textbf{Anastasios Kouvelas:} Conceptualization, Methodology, Investigation, Validation, Supervision,  Writing - review \& editing. \textbf{Nikolas Geroliminis:} Conceptualization, Methodology, Investigation, Validation, Writing - review \& editing, Supervision, Project administration, Funding acquisition.

\bibliographystyle{elsarticle-harv}
\bibliography{bibliography}

\end{document}